\documentclass[aps,prd,twocolumn,superscriptaddress,preprintnumbers,floatfix,longbibliography]{revtex4-1}

\usepackage{graphicx}
\usepackage{amsmath}
\usepackage[caption=false]{subfig}
\usepackage{siunitx}
\usepackage{placeins}
\usepackage{color}
\usepackage{standalone}
\usepackage{dcolumn}
\usepackage{bm}
\usepackage{microtype}
\usepackage{etoolbox}
\usepackage{amssymb}
\usepackage{mathrsfs}
\usepackage{accents}
\usepackage{gensymb}
\usepackage{float}
\usepackage[normalem]{ulem}
\usepackage[dvipsnames]{xcolor}
\usepackage[colorlinks,urlcolor=NavyBlue,citecolor=NavyBlue,linkcolor=NavyBlue,pdfusetitle]{hyperref}
\usepackage{subfig}

\usepackage{etoolbox}
\usepackage{xcolor}

\newtoggle{commentsoff}
\iftoggle{commentsoff}{
	\newcommand{\dhj}[1]{}
	\newcommand{\ls}[1]{}
}{
\newcommand{\dhj}[1]{{\color{red}~\textsf{[{\bf Dana}: #1]}}}
\newcommand{\ls}[1]{{\color{olive}~\textsf{[{\bf Lilli}: #1]}}}

}

\begin{document}

\title{Validating continuous gravitational-wave candidates \textcolor{black}{from a semicoherent search using Doppler modulation and an effective point spread function}}

\author{Dana Jones}
\affiliation{College of Arts and Sciences, University of Pennsylvania, Philadelphia, Pennsylvania 19104, United States of America}
\affiliation{OzGrav-ANU, Centre for Gravitational Astrophysics, College of Science,The Australian National University, Australian Capital Territory 2601, Australia}

\author{Ling Sun}
\affiliation{OzGrav-ANU, Centre for Gravitational Astrophysics, College of Science,The Australian National University, Australian Capital Territory 2601, Australia}

\author{Julian Carlin}
\author{Liam Dunn}
\author{Meg Millhouse}
\affiliation{School of Physics, University of Melbourne, Parkville, Victoria 3010, Australia}
\affiliation{OzGrav, University of Melbourne, Parkville, Victoria 3010, Australia}

\author{Hannah Middleton}
\affiliation{School of Physics, University of Melbourne, Parkville, Victoria 3010, Australia}
\affiliation{OzGrav, University of Melbourne, Parkville, Victoria 3010, Australia}
\affiliation{Centre for Astrophysics and Supercomputing, Swinburne University of Technology, Hawthorn, Victoria, 3122, Australia}
\affiliation{School of Physics and Astronomy and Institute for Gravitational Wave Astronomy, University of Birmingham, Edgbaston, Birmingham, B15 2TT, UK}

\author{Patrick Meyers}
\affiliation{School of Physics, University of Melbourne, Parkville, Victoria 3010, Australia}
\affiliation{OzGrav, University of Melbourne, Parkville, Victoria 3010, Australia}
\affiliation{Theoretical Astrophysics Group, California Institute of Technology, Pasadena, CA 91125, United States of America}

\author{Patrick Clearwater}
\affiliation{School of Physics, University of Melbourne, Parkville, Victoria 3010, Australia}
\affiliation{OzGrav, University of Melbourne, Parkville, Victoria 3010, Australia}

\author{Deeksha Beniwal}
\affiliation{OzGrav, University of Adelaide, Adelaide, South Australia 5005, Australia}

\author{Lucy Strang}
\author{Andr\'{e}s Vargas}
\author{Andrew Melatos}
\affiliation{School of Physics, University of Melbourne, Parkville, Victoria 3010, Australia}
\affiliation{OzGrav, University of Melbourne, Parkville, Victoria 3010, Australia}

\date{\today}

\begin{abstract}
Following up large numbers of candidates in continuous gravitational wave searches presents a challenge, particularly in regard to computational power and the time required to manually scrutinize each of the candidates. It is important to design and test good follow-up procedures that are safe (i.e., minimize false dismissals) and computationally efficient across many search configurations. 
We investigate two follow-up procedures, or ``vetoes'', both of which exploit the Doppler modulation predicted in astrophysical signals.
\textcolor{black}{In particular, we introduce the concept of using an effective point spread function as part of our veto criteria.} 
We take advantage of a well-established semicoherent search algorithm based on a hidden Markov model to study \textcolor{black}{various} search configurations and to generalize the veto criteria by considering the overall veto performance in terms of efficiency and safety. 
The results can serve as a guideline for follow-up studies in future continuous gravitational wave searches using a hidden Markov model algorithm. The results also apply qualitatively to other semicoherent search algorithms.
\end{abstract}

\maketitle

\section{Introduction}
\label{sec:intro}

Gravitational waves (GWs), perturbations in spacetime which propagate at the speed of light, were first directly observed in 2015 when the Hanford and Livingston detectors of the Advanced Laser Interferometer Gravitational-Wave Observatory (Advanced LIGO) detected a merging binary black hole system (GW150914) \cite{Abbott2016, LIGO2014}.
In the years since the first detection, the addition of Advanced Virgo~\cite{Virgo2014} and KAGRA~\cite{akutsu2020overview} to the network of observatories, together with improved sensitivity, has produced increasingly frequent detections of compact binary coalescences (CBCs) \cite{Abbott2018,Abbott2021gwtc2,gwtc3}.
Other types of GW sources that are predicted to radiate at frequencies within the observational band of ground-based detectors remain undetected, e.g., the continuous gravitational waves (CWs) produced by spinning neutron stars. CWs, once detected, will provide invaluable information regarding the structure of neutron stars as well as the nuclear equation of state~\cite{lasky2015}. A great deal of work has been carried out to develop methods to search for CWs~\cite{Riles2017,Tenorio2021, Piccinni2022}.

Since the expected strain amplitudes of CWs are orders of magnitude smaller than those produced by CBCs, the computational cost of searching large template banks (including parameters such as the signal frequency and time derivatives thereof) coherently over a long period of time, e.g., $\sim1$~yr, is high~\cite{Riles2017}. In addition, intrinsic, stochastic wandering of the frequency, sometimes called ``timing noise" in the context of radio pulsars, could degrade the sensitivity of a CW search \cite{Hobbs2010,Mukherjee2018,Ashton2015}.
In this paper we focus on a computationally efficient, semicoherent search strategy based on a hidden Markov model (HMM) which is equipped to track a signal frequency that wanders stochastically and spins down secularly~\cite{Sun2018-2,Suvorova2016}. 
The tracking scheme has its origins in engineering applications, and has recently been used in many directed CW searches (e.g.,~\cite{ScoX1ViterbiO1,Sun2019,Abbott2019-2,O2SNR-Viterbi,Sun2020-CygX1,Jones_2021,Beniwal2021,O3aSNR,O3amxp,O3scoX1}).

The HMM-based searches are made up of two main procedures: (1) dividing the total observing time into subintervals and coherently calculating the signal power within each consecutive time interval (e.g., with length of $\sim 1$~d) using a frequency domain matched filter (e.g., the $\mathcal{F}$-statistic)~\cite{Jaranowski1998}, and (2) using the Viterbi algorithm (an HMM tracking scheme) to find the most probable signal evolution over the total observing time (e.g., $\sim1$~yr)~\cite{Sun2018-2,ScoX1ViterbiO1}. This optimal signal evolution, referred to as the Viterbi path, consists of a frequency estimated at each discrete time step. The search over the full frequency band is usually parallelized and carried out in narrow sub-bands (with width of \mbox{$\sim 1$}~Hz). The Viterbi path obtained in each sub-band, corresponding to the most likely CW candidate in that sub-band, is assigned a Viterbi score. This score evaluates the candidate significance, such that a higher score signifies a greater probability that the candidate is inconsistent with random noise fluctuations~\cite{Sun2018-2,ScoX1ViterbiO1}. For the mathematical details behind these procedures, see Secs. II A--D in Ref.~\cite{Jones_2021}.
Some searches rely on an unnormalized log-likelihood instead of the Viterbi score to evaluate the significance of a candidate~\cite{O2SNR-Viterbi,Beniwal2021,O3aSNR,O3amxp,O3scoX1}.

In this study, we focus on directed CW searches where the HMM-based methods have been most widely used. A directed search targets an astrophysical source at a known sky position with unknown ephemerides and hence is conducted over a wide frequency band (e.g., $\sim 20$--1000~Hz). A typical \textcolor{black}{HMM-based} directed search will produce on the order of $10^3$ CW candidates, many of which will require the use of multiple verification techniques to be identified as noise. As such, a computationally efficient follow-up procedure is needed in order to comb through the results and distinguish the candidates caused by noise artifacts from any real astrophysical signal.
\textcolor{black}{This is perhaps even more true when considering an all-sky search, which is a survey done over the whole sky to look for CW signals and which typically produces on the order of \textcolor{black}{at least} $10^4$ candidates that require follow-up \textcolor{black}{(although some searches produce a list of candidates orders of magnitude larger, e.g., Einstein@Home searches~\cite{Steltner2021})}. As outlined in Refs.~\cite{O3all-sky,O3scalar-bosons}, a threshold is defined within these searches to obtain a reasonable total number of candidates in order to keep the follow-up computationally feasible. Defining more efficient candidate follow-up procedures would allow us to lower this threshold and process a greater number of candidates, thereby improving search sensitivity. This is clearly demonstrated in, e.g., Eq. (67) in Ref.~\cite{Astone2014}, where the lowest strain amplitude that the search is sensitive to, denoted by $h_{0,\rm min}$, is directly related to the threshold ${\rm CR}_{\rm thr}$, such that a decrease in threshold would yield an improvement in sensitivity.}

Many CW candidates, which have Viterbi scores above a threshold $S_{\rm th}$ (e.g., corresponding to a 1\% false alarm probability), can be easily eliminated using already well-established procedures. Candidates are initially passed through a known-line veto and a single-interferometer veto, which are defined in Refs.~\cite{ScoX1ViterbiO1, Jones_2021, Beniwal2021, O3aSNR}. The known-line veto involves eliminating every candidate whose frequency evolution overlaps any known instrumental line present in any of the detectors~\cite{Covas2018, Davis2021}. The single interferometer veto involves eliminating unidentified instrumental artifacts by checking if a candidate is significantly louder (i.e., if it has a larger Viterbi score) in one detector than in the other~\cite{Keitel2014, Leaci2015}. After this, most searches use two additional well-defined vetoes to further identify noise artifacts, also described in Refs.~\cite{ScoX1ViterbiO1, Jones_2021, Beniwal2021, O3aSNR}. The total observing time can be split into multiple subintervals, and any candidates that are significantly louder in one particular subinterval are eliminated. Then, if the estimated frequency evolution rate is sufficiently low, the coherent time, $T_{\rm coh}$, over which the data are integrated coherently can be increased and any candidate whose Viterbi score decreases when $T_{\rm coh}$ increases can be vetoed.

Although most candidates will have been eliminated after applying the aforementioned vetoes, some closely resemble astrophysical signals and require more careful inspection. Two additional veto strategies, both based on Doppler modulation (DM), have been developed for the candidate follow-up procedure~\cite{Zhu_2017, O2SNR-Viterbi, Isi2020}. 
They prove to be both useful and complementary to each other, in that one often vetoes the candidates that the other does not.
\textcolor{black}{In fact, the order of applying a series of vetoes is generally interchangeable. More computationally efficient vetoes are recommended to be used earlier. These two DM-based vetoes are more efficient compared to the vetoes that involve rerunning the search in subintervals or increasing the coherent time. Thus they are usually applied directly following the single-interferometer veto.}

The DM-based vetoes are quite dependent on the search configuration used in a particular study. For the first of these two strategies, called the \emph{DM-off veto}, the signal significance is calculated with the DM correction for the Doppler shift due to Earth’s motion switched on and off. The candidates that are louder when the DM correction is switched off are not likely to be of astrophysical origin. For the second strategy, called the \emph{off-target veto}, the search is shifted \textcolor{black}{to one or more sky position(s)} away from the true position of the target, i.e., a shifted DM correction is applied. If the candidates become louder, they are vetoed. \textcolor{black}{Despite the wide applications of these two vetoes in existing searches~\cite{Jones_2021, Beniwal2021, O3aSNR, O3amxp}, the general safety (i.e., no astrophysical signal is falsely eliminated) of these veto procedures still remains to be studied for various search configurations for HMM-based searches.}

In this paper, we carry out an in-depth study of how the DM changes across the sky and investigate the impact on synthetic signals and noise outliers through simulations. \textcolor{black}{In particular, we derive a set of rigorous criteria for when and how to use the off-target veto to discriminate between astrophysical signals and noise artifacts. We introduce a new concept of using an effective point spread function (EPSF) of the detection statistics obtained at various sky locations, which can serve as a particularly useful veto criterion. For the DM-off veto, which has been previously studied in coherent searches~\cite{Zhu_2017}, we carry out additional simulations to verify the criteria and safety in semicoherent searches.}
Although the simulations are designed for signals from isolated sources, the results can be applied to signals from CW sources in binary systems, assuming the Doppler shift caused by the orbital motion is fully accounted for. The impact from the imperfect removal of the binary orbital modulation is out of the scope of this paper.
While this study is carried out using an HMM-based method, in principle the resulting guidelines broadly apply to stack-slide-based semicoherent algorithms~\cite{Brady2000}. Moreover, our results can be generalized to follow up candidates in an all-sky search.

The organization of this paper is as follows. In Section~\ref{sec:doppler_pattern}, an analytic investigation of the Doppler modulation is presented. Section~\ref{sec:methods} outlines the search methods used in this study. \textcolor{black}{Section~\ref{sec:off-target} introduces the EPSF and details the results of the off-target veto study. Section~\ref{sec:DM-off} describes the verification of the DM-off veto in semicoherent searches.} A discussion of the results and concluding remarks are presented in Section~\ref{sec:conclusion}.

\section{Doppler Signature}
\label{sec:doppler_pattern}

In this section, we briefly review the CW signal model and calculate analytically how the DM affects CW signals. This provides a foundation for the empirical studies described in Sections~\ref{sec:off-target} and \ref{sec:DM-off}.

The phase of a CW signal can be modeled as follows \cite{Jaranowski1998}:
\begin{equation}
    \label{eqn:phase}
    \Psi(t) = \Phi_0 + 2\pi \sum_{k=0}^{s} \frac{f_0^{(k)}t^{k+1}}{(k+1)!} +  \frac{2\pi}{c}\mathbf{n}_0\cdot\mathbf{r}_{\rm d}(t) \sum_{k=0}^{s}\frac{f_0^{(k)}t^k}{k!}.
\end{equation}
Here $f_0$ is the signal frequency at reference time $t=0$ with respect to the solar system barycenter (SSB), $f_0^{(k)}$ is the $k$th time derivative of the instantaneous frequency evlatuated at $t=0$ at the SSB, $\mathbf{n}_0$ is the constant unit vector in the source direction within the SSB reference frame, and $\mathbf{r}_{\rm d}(t)$ is the position vector of the detector relative to the SSB origin. For the coordinate system with the SSB reference frame, we take the $x$-axis parallel to the $x$-axis of the celestial sphere coordinate system and the $z$-axis perpendicular to the ecliptic $z$-axis. Then, the unit vector $\mathbf{n}_0$ has the components \cite{Jaranowski1998}
\begin{equation}
    \label{eqn:unit_vector}
    \mathbf{n}_0 = \begin{pmatrix} 1 & 0 & 0 \\ 0 & \cos \epsilon & \sin \epsilon \\ 0 & -\sin \epsilon & \cos \epsilon \end{pmatrix} \begin{pmatrix} \cos \alpha \cos \delta \\ \sin \alpha \sin \delta \\ \sin \delta \end{pmatrix},
\end{equation}
where $\epsilon$ is the angle between the ecliptic plane and Earth's equator (i.e., $\epsilon = 23.5 \degree$), $\alpha$ is the right ascension (RA) of the source, and $\delta$ is the declination (Dec). Meanwhile, the detector's position vector $\mathbf{r}_{\rm d}$ has the components \cite{Jaranowski1998}
\begin{eqnarray}
    \label{eqn:position_vector}
    \nonumber
    \mathbf{r}_{\rm d} &=& R_{ES} \begin{pmatrix} \cos(\phi_0 + \Omega_0 t) \\ \sin(\phi_0 + \Omega_0 t) \\ 0 \end{pmatrix} \\
    && +  R_E \begin{pmatrix} 1 & 0 & 0 \\ 0 & \cos \epsilon & \sin \epsilon \\ 0 & -\sin \epsilon & \cos \epsilon \end{pmatrix} \begin{pmatrix} \cos \lambda \cos(\phi_r + \Omega_r t) \\ \cos \lambda \sin(\phi_r + \Omega_r t) \\ \sin \lambda \end{pmatrix},
\end{eqnarray}
where $R_{ES}$ = 1 AU is the mean distance between Earth's center and the SSB, $R_E$ is the mean radius of Earth, $\Omega_0$ is Earth's mean orbital angular velocity, $\Omega_r$ is Earth's rotational angular velocity, $\lambda$ is the detector latitude, and $\phi_0$ and $\phi_r$ are phases specifying the exact location of Earth in its orbital and diurnal motion, respectively, at $t = 0$. 

Substituting Equations \eqref{eqn:unit_vector} and \eqref{eqn:position_vector} into \eqref{eqn:phase} yields the following expression \cite{Jaranowski1998}:
\begin{widetext}
\begin{eqnarray}
	\nonumber
    \Psi(t) &=& \Phi_0 + 2\pi \sum_{k=0}^{s} \frac{f_0^{(k)}t^{k+1}}{(k+1)!} 
    + \frac{2\pi}{c} \{R_{ES}[\cos \alpha \cos \delta \cos(\phi_0 + \Omega_0 t) 
    + (\cos \epsilon \sin \alpha \cos \delta + \sin \epsilon \sin \delta) \sin(\phi_0 + \Omega_0 t)] \\
    &&+ R_E[\sin \lambda \sin \delta + \cos \lambda \cos \delta \cos(\alpha - \phi_r - \Omega_r t)]\} \sum_{k=0}^{s}\frac{f_0^{(k)}t^k}{k!}.
    \label{eqn:phase2}
\end{eqnarray}
\end{widetext}
We calculate the time derivative of $\Psi(t)$ to obtain the signal frequency:
\begin{eqnarray}
	\label{eqn:frequency}
	f(t) &\approx& (f_0 + f_0^{(1)}t) \left(1 + \frac{\mathbf{v_d} \cdot \mathbf{n_0}}{c} \right) \\
	\nonumber
	&\approx&(f_0 + f_0^{(1)}t) \{1+\frac{R_{ES} \Omega_0}{c} [ -\cos \alpha \cos \delta \sin(\phi_0 + \Omega_0 t) \\ 
	&+& (\cos \epsilon \sin \alpha \cos \delta + \sin \epsilon \sin \delta) \cos(\phi_0 + \Omega_0 t)] \}, \label{eqn:time_derivative}
\end{eqnarray}
where $f_0^{(1)}$ is the first time derivative of the signal frequency at reference time $t=0$ and $\mathbf{v_d}$ is the velocity vector of the detector.
In Equations~(\ref{eqn:frequency})--(\ref{eqn:time_derivative}), we omit the term corresponding to the DM effect due to Earth's rotation, which is negligible compared to the DM effect caused by Earth's orbital motion.
We also omit the higher order time derivatives and other negligible terms in order to simplify calculations. 
(See Appendix~\ref{appendix:signal_freq} for the full derivation.)
For convenience, we define the frequency shift induced by the DM as
\begin{equation}
	\label{eqn:delta}
	\Delta \equiv \frac{\mathbf{v_d} \cdot \mathbf{n_0}}{c} \approx \frac{R_{ES} \Omega_0 \kappa(\alpha,\delta,t)}{c}
\end{equation}
with
\begin{eqnarray}
    \label{eqn:kappa}
    \nonumber
    \kappa &=& -\cos \alpha \cos \delta \sin(\phi_0 + \Omega_0 t) \\
    &&+ (\cos \epsilon \sin \alpha \cos \delta + \sin \epsilon \sin \delta) \cos(\phi_0 + \Omega_0 t).
\end{eqnarray}

\begin{figure*}[hbt!]
	\centering
	\includegraphics[scale=.54]{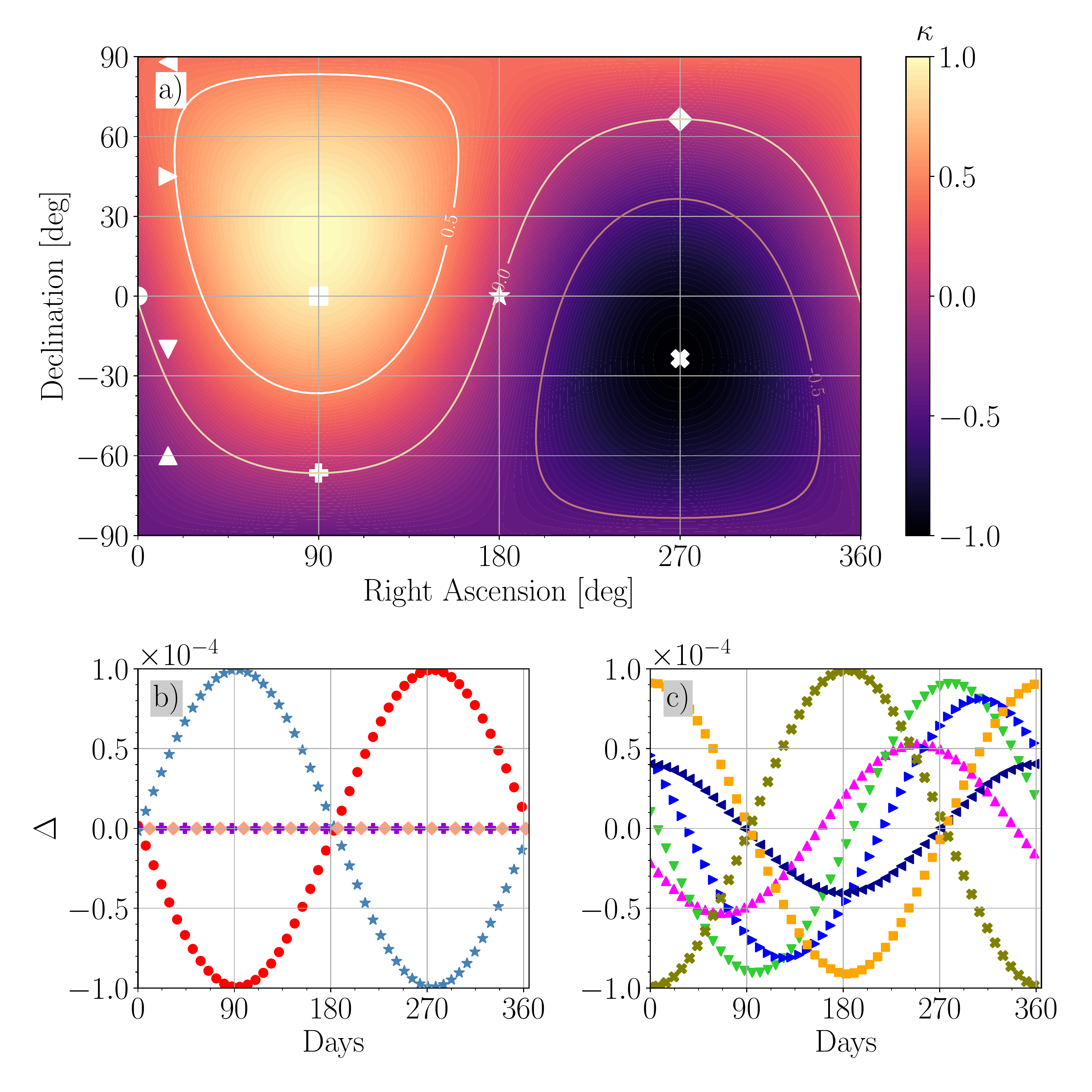}
	\caption{Value of $\kappa$ as a function of RA and Dec (both in degrees) for the GPS time 1167545066 (a). The $\Delta$-values as a function of time over a year (starting from 1167545066) for the sky positions highlighted in (a) are plotted in (b) and (c). Each curve, generated using one of these sky positions, is drawn with the same symbol as the white marker at this specific sky position in panel (a). All curves in panel (b) originate from the same contour in panel (a) with $\kappa=0$. See more detailed discussions in Sec.~\ref{sec:off-target}; sky positions are given in Table~\ref{tab:peaks2}.}
	\label{fig:doppler_pattern}
\end{figure*}

In panel (a) of Figure~\ref{fig:doppler_pattern}, $\kappa$ is plotted with $\phi_0 = 0$, $\Omega_0 = {2\pi}/{31557600}$~rad\,s$^{-1}$, and $t$ set to GPS time 1167545066, for a grid of evenly spaced RA and Dec values spanning the entire sky. Similar contour plots are computed for an equatorial coordinate system in Ref.~\cite{Prix_2005} and an ecliptic coordinate system in Ref.~\cite{Intini_2020}.
Panels (b) and (c) show two sets of curves where $\Delta$ is plotted as a function of time for an entire year (starting from GPS time 1167545066) and where each curve corresponds to a different sky position. 
These sky positions are marked on the contour map in panel (a) with white markers, each with a different shape that can be matched to its corresponding $\Delta$-curve (as a function of time) in panels (b) and (c).

We examine the DM of the signal frequency closely here in preparation for connecting the DM patterns plotted in Figure~\ref{fig:doppler_pattern} to the empirical results discussed in later sections of this paper.
In particular, for any series of markers that lie along the same $\kappa$ contour in panel (a), the $\Delta$-curves that correspond to these markers will all intersect at the particular time used to calculate $\kappa$ in (a). This is because the DM for sources at sky positions along a single contour are the same for this particular time.
For instance, the $\Delta$-curves for four locations along the contour where $\kappa = 0$ are shown in panel (b), marked by red dots, light pink stars, green diamonds, and purple plus signs. 
The $\Delta$ value evolves over time in different ways depending on the sky position, so although the $\Delta$-curves all intersect at the start time, they no longer overlap as time shifts forward. 
While the dot and star markers are near the equator and experience strong DM effects, the diamond and plus sign markers are near the poles and are only weakly impacted by the DM.
Examples of $\Delta$-curves for locations with a range of other $\kappa$ values are displayed in panel (c).

For a source with a known sky position, Figure~\ref{fig:doppler_pattern} helps us to better understand how targeting a sky position shifted away from the true position of the source impacts the search results. That is, for all sky positions along a $\kappa$ contour, despite being off-target from the true position, the DM effect is the same as that seen at the true position for a given instant in time. As the integration time is increased, most of these similarities tend to disappear as the Doppler correction at each sky position evolves differently. However, the DM effects integrated over time for sky positions that lie along certain directions could still mimic each other, especially for sky positions that have similar $\kappa$ values over the integration time.
\textcolor{black}{Therefore, when correcting the DM using an incorrect sky position, which is the basic principle behind the off-target veto}, the recovered signal at the offset position could mimic the signal from the true sky position and produce similar detection statistics in a CW search.
Indeed, if one searches a grid of locations around the true position of the source, the detection statistic for an elliptical area around the true sky position is higher than other sky regions.
\textcolor{black}{The nuances of this behavior are explained in Ref.~\cite{Isi2020} and discussed in detail, along with the different ways that the off-target veto has been implemented in recent searches to take advantage of this behavior, in Section~\ref{sec:off-target}.}

\section{Generating Candidates}
\label{sec:methods}

In order to study the CW validation techniques, we first generate candidates using the HMM-based semicoherent search method. 
The two fundamental procedures that make up the searches are as follows. First, we divide the observation time into subintervals and coherently calculate the signal power within each time segment using the maximum likelihood matched filter, $\mathcal{F}$-statistic. Second, we use an HMM tracking scheme to discover the most probable frequency evolution over the full observing time. These two procedures are briefly reviewed in Sections~\ref{sec:f-stat} and \ref{sec:HMM}, respectively. The detection statistic used in this study, known as the Viterbi score, is outlined in Section~\ref{sec:score}. More details can be found in Refs.~\cite{ScoX1ViterbiO1,Sun2018-2}.

\subsection{$\mathcal{F}$-statistic}
\label{sec:f-stat}

We can write the time-domain data collected in the detector as
\begin{eqnarray}
    x(t) &=& h(t) + n(t) \\
    &=& \mathcal{A}^\mu h_\mu(t) + n(t),
\end{eqnarray}
where $h(t) = \mathcal{A}^\mu h_\mu(t)$ is the signal, and $n(t)$ is stationary, additive noise \cite{Jaranowski1998}. 
The amplitudes $\mathcal{A}^\mu$, depending on the characteristic GW strain amplitude $h_0$, source orientation, and initial phase, are associated with the four linearly independent components $h_\mu(t)$ that depend on the phase in Eq.~\eqref{eqn:phase} and the detector antenna patterns. (For more details, see Refs.~\cite{Jaranowski1998, Sun2018-2}.)

The $\mathcal{F}$-statistic is a frequency-domain matched filter that estimates the likelihood that a signal, parameterized by its frequency and the frequency time derivatives, is present in the data~\cite{Jaranowski1998}. 
We first define a scalar product ($\cdot | \cdot$) as a sum over single-detector inner products:
\begin{eqnarray}
    (x|y) &=& \mathop{\sum} \limits_{X} (x^X|y^X) \\
    &=& \mathop{\sum} \limits_{X} 4\Re \int_{0}^{\infty}df \frac{\tilde{x}^X(f)\tilde{y}^{X*}(f)}{S_h^X(f)}.
\end{eqnarray}
Here $X$ indexes the detector, $x^X$ is the data in detector $X$, $S_h^X(f)$ is the single-sided power spectral density (PSD) of detector $X$, the tilde denotes a Fourier transform, and $\Re$ is the real part of a complex number \cite{Prix2007}. Using this definition, the $\mathcal{F}$-statistic can be expressed as
\begin{equation}
    \mathcal{F} = \frac{1}{2} x_\mu \mathcal{M}^{\mu \nu} x_\nu,
\end{equation}
where we define $x_\mu = (x|h_\mu)$, and $\mathcal{M}^{\mu \nu}$ is the matrix inverse of $\mathcal{M}_{\mu \nu}=(h_\mu|h_\nu)$ \cite{Cutler_2005}.

If we assume that the noise is Gaussian and that the single-sided PSD is the same in all detectors, the probability of having a signal in the data depends on the signal-to-noise ratio (SNR) when analyzing the data coherently for a time $T_{\rm coh}$, given by (cf. Sec.~IIIC in Ref.~\cite{Jaranowski1998})
\begin{equation}
    \label{eqn:snr2}
    \rho_0^2=\frac{K h_0^2T_{\rm coh}}{S_h(f)},
\end{equation}
where the constant $K$ depends on the sky position, orientation of the source, and number of detectors.

\subsection{HMM tracking and Viterbi algorithm}
\label{sec:HMM}
A general description of the HMM method can be found in Refs.~\cite{Suvorova2016,Sun2018-2}. We briefly summarize the method as follows.

A Markov chain is a stochastic process transitioning from one discrete state to another at discrete times \mbox{$\{t_1, \cdots, t_{N_T}\}$}, where $N_T$ is the total number of time steps. An HMM is made up of two variables, the unobservable, hidden state variable \mbox{$q(t) \in \{q_1, \cdots, q_{N_Q}\}$} and the observable, measurement state variable \mbox{$o(t) \in \{o_1, \cdots, o_{N_O}\}$}, where $N_Q$ and $N_O$ are the total number of hidden and measurement states, respectively. The hidden state at time $t_{n+1}$ is solely dependent on the state at time $t_n$ (this is the Markovian assumption) and has a transition probability of
\begin{equation}
	\label{eqn:transition_prob}
	A_{q_j q_i} = P [q(t_{n+1})=q_j|q(t_n)=q_i].
\end{equation}
At time $t_n$, the likelihood that the hidden state $q_i$ is observed in state $o_j$ is described by the emission probability, defined as
\begin{equation}
	L_{o_j q_i} = P [o(t_n)=o_j|q(t_n)=q_i].
\end{equation}
The prior can be written as
\begin{equation}
	\Pi_{q_i} = P [q(t_1)=q_i].
\end{equation}
Then, the probability that an observed sequence \mbox{$O = {o(t_1),...,o(t_{N_T})}$} results from a hidden state path $Q = {q(t_1),...,q(t_{N_T})}$ via a Markov chain can be expressed as
\begin{eqnarray}
	\nonumber
	P(Q|O) &\propto& L_{o(t_{N_T})q(t_{N_T})} A_{q(t_{N_T})q(t_{N_T-1})} \cdots L_{o(t_2)q(t_2)} \\ 
	&&\times A_{q(t_2)q(t_1)} \Pi_{q(t_1)}.
\end{eqnarray}
The most probable path, obtained by maximizing $P(Q|O)$, is simply \cite{Sun2018-2}
\begin{equation}
	Q^*(O)= \arg\max P(Q|O),
\end{equation}
where $\arg\max(\cdots)$ returns the argument that maximizes~$(\cdots)$.

In this study, we track  $q(t) = f(t)$, where $f(t)$ is the signal frequency at time $t$.
We then map the discrete hidden states one-to-one with the frequency bins in the output of $\mathcal{F}$-statistic calculated over a span of length $T_{\rm coh}$ (see Section~\ref{sec:f-stat}), with each frequency bin size being $\Delta f = 1/(2T_{\rm coh})$.
Thus we choose $T_{\rm coh}$ to satisfy 
\begin{equation}
    \left|\int_t^{t+T_{\rm coh}}dt' \dot{f}(t')\right| \leq \Delta f,
    \label{eqn:Tcoh}
\end{equation}
where $\dot{f}(t)$ is the first time derivative of the signal frequency. This relationship must remain valid throughout the total observing time $ T_{\rm obs}$.

The choice of $A_{q_j q_i}$ does not greatly impact the sensitivity of an HMM as long as it captures the general behavior of the signal, and as such, the particular transition matrix that is chosen should not impact the guidelines presented in this paper~\cite{Quinn2001,Suvorova2016}.
Without loss of generality, we assume that, in the CW signal we are searching for, the effect of timing noise on the frequency evolution is orders of magnitude smaller than that of the secular spin down of the star, and that $|\dot{f} (t)|$ is uniformly distributed in the range from zero to the maximum estimated frequency derivative $|\dot{f}|_{\rm max}$. \textcolor{black}{It should be noted that even if we are dealing with a situation where the timing noise is large and has a comparable effect on the signal's evolution as the star's spin down, as long as an appropriate transition matrix is chosen based on our belief of the expected timescale on which the signal frequency evolves, the HMM remains capable of tracking the signal~\cite{Sun2018-2}.}
By substituting $|\dot{f}|_{\rm max}$ into \eqref{eqn:Tcoh}, we are able to simplify \eqref{eqn:transition_prob} to
\begin{equation}
    A_{q_{i-1} q_i} = A_{q_i q_i} = \frac{1}{2},
\end{equation}
with all other $A_{q_j q_i}$ entries vanishing. (In searches where the signal frequency is assumed to walk randomly, such as in Ref.~\cite{Suvorova2016}, we would instead have $A_{q_{i+1} q_i} = A_{q_{i-1} q_i} = A_{q_i q_i} = 1/3$.) 
\textcolor{black}{The observed state $o(t)$ at time $t$ is represented by the data observed by the detectors over $[t, t + T_{\rm coh}]$.}
Using the definition of the $\mathcal{F}$-statistic, we express the emission probability as
\begin{eqnarray}
    L_{o(t) q_i} &=& P [o(t)|{f}_i \leq f(t') \leq {f}_i+\Delta f] \\ 
    &\propto& \exp[\mathcal{F}({f}_i)],
\end{eqnarray}
for $t \leq t' \leq t + T_{\rm coh}$, where $f_i$ denotes the central frequency in the $i$th bin. A uniform prior of $\Pi_{q_i} = N_Q^{-1}$ is chosen because there is no independent knowledge of the signal frequency at $t_1$~\cite{Sun2018-2}.

We use the classic Viterbi algorithm~\cite{Viterbi1967} to efficiently solve the HMM, which outputs the most likely frequency evolution path $Q^*(O)$ over the entire $T_{\rm obs}$, i.e., the Viterbi path.

\subsection{Detection statistic}
\label{sec:score}
We use the Viterbi score, denoted by $S$, to evaluate the significance of a candidate in this study~\cite{ScoX1ViterbiO1,Sun2018-2}.
In each sub-band searched (with width of 1 Hz in this paper), $S$ is defined such that the log likelihood of the optimal Viterbi path is equal to the mean log likelihood of all paths ending in different bins of the sub-band plus $S$ standard deviations at final step $N_T$. This is written as
\begin{equation}
\label{eqn:viterbi_score}
S = \frac{\ln \delta_{q^*}{(t_{N_T})} -\mu_{\ln \delta}(t_{N_T})}{\sigma_{\ln \delta}(t_{N_T})},
\end{equation}
with mean
\begin{equation}
\mu_{\ln \delta}(t_{N_T}) = N_Q^{-1} \sum_{i=1}^{N_Q} \ln \delta_{q_i}(t_{N_T}),
\end{equation}
and variance
\begin{equation}
\sigma_{\ln \delta}(t_{N_T})^2 = N_Q^{-1} \sum_{i=1}^{N_Q} [\ln \delta_{q_i}(t_{N_T}) - \mu_{\ln \delta}(t_{N_T}) ]^2.
\end{equation}
Here, $\delta_{q_i}(t_{N_T})$ is the likelihood of the path with the maximum probability ending in state $q_i$ ($1\leq i \leq N_Q$) at step $N_T$ and $\delta_{q^*}{(t_{N_T})}$ is the likelihood of the optimal Viterbi path. 
In some applications of the HMM algorithm, e.g., \cite{O2SNR-Viterbi,O3amxp,O3aSNR,O3scoX1,Beniwal2021}, the total log-likelihood along the Viterbi path, $\ln \delta_{q^*}(t_{N_T})$, is directly used as the detection statistic. Transforming to the Viterbi score in Eq.~\eqref{eqn:viterbi_score} does not impact the results in this paper.

To conclude this section, it is worth noting that the $\mathcal{F}$-statistic is well-equipped to deal with the DM correction; it calculates the expected Doppler shift for a given signal and thus corrects for the Doppler modulation when it computes the signal likelihood. The Viterbi score, which is calculated using the resulting likelihoods, is then also impacted by the DM correction. 
Then, the DM effect acts as an interesting basis for vetoes; it provides a useful way to study the difference between signals, whose intrinsic frequencies can only be recovered with the DM correction, and noise artifacts, whose frequencies are not impacted by the DM (since they originate on Earth).

\section{Searching off target}
\label{sec:off-target}

When an incorrect sky position is used to perform the DM correction on an astrophysical signal, the detection statistic decreases. The off-target veto makes use of this knowledge and of the fact that noise artifacts originate on Earth and are not impacted by DM due to Earth's motion---that is, performing a DM correction on a noise artifact at a sky position that is slightly off target from the true sky position of the source should not cause the Viterbi score to drop significantly.
\textcolor{black}{In particular, we introduce a new concept of using an effective point spread function (EPSF) of the detection statistics obtained at various sky locations around the true position of the source to serve as a veto criterion in this section.}

\subsection{Implementation in other searches}
\label{sec:off-target_other_searches}

In existing searches, the off-target veto is usually done as follows. The detection statistic is computed for one or more offsets from the source's true sky position. For an astrophysical source, one would expect the detection statistic to drop and remain below a threshold once a certain offset from the true sky position is reached. On the other hand, the detection statistic of candidates arising from noise artifacts should remain consistently above this threshold regardless of the offset~\cite{O2SNR-Viterbi}.
\textcolor{black}{Implementations of the off-target veto which are similar to the one considered in this study---namely, one or a handful of offsets is taken along one or two spatial directions---can be found in the following references: for a signal offset along one direction, see Refs.~\cite{O2SNR-Viterbi,Middleton2020,O3amxp}; for two offsets, one along a line of constant RA and the other along a line of constant Dec, see Ref.~\cite{O3aSNR}; and for multiple offsets along both a line of constant RA and a line of constant Dec, see Ref.~\cite{Jones_2021}.}

\textcolor{black}{However, this is not the only way the off-target veto is used in CW searches.}
In a recent search for CWs from accreting millisecond X-ray pulsars, for each candidate that remains after the initial veto procedures, the log-likelihood is computed for a grid of off-target positions around the source's true position~\cite{O3amxp}. Every candidate whose log likelihood contours do not match simulations in Gaussian noise gets eliminated.

\textcolor{black}{In Ref.~\cite{Isi2020}, the authors use what they call the ``sky-shifting" method where they run the search at many off-target sky positions across an entire hemisphere of the sky in order to build up an empirical background distribution of noise to be used as a detection statistic. Because an incorrect Doppler correction is being used for each of the off-target positions, the noise background is effectively blinded to the presence of any astrophysical signals.
In addition to being more robust than simply choosing a handful of off-target sky positions, building up a background distribution is extremely useful in that it is better able to account for the non-Gaussianities found in real detector noise. However, the disadvantage here lies in the computational expense of generating the map for an entire hemisphere. Moreover, while this method would certainly be useful when targeting individual sources, it becomes computationally expensive when candidates are obtained for many sky positions, as in an all-sky search.}

\textcolor{black}{Another recent implementation of the off-target veto is presented in Ref.~\cite{Intini_2020}. For an astrophysical signal, one can predict the relation between a mismatch in sky position and an error in estimated signal frequency and spin down, using the DM relation indicated by Eq.~\eqref{eqn:frequency}. That is, for certain off-target sky positions, the incorrect Doppler correction still yields a significant detection statistic, but with some small quantifiable shift in the estimated frequency and frequency time derivative(s). Thus one can look for these expected relations, or ``patterns,'' that would be present were there a real signal. This method is robust in its ability to safely eliminate candidates caused by noise artifacts. However, it is more suitable for an all-sky search, where a group of candidates is usually found above the detection threshold across a patch in the sky and for a narrow band of estimated signal frequency and time derivative(s), than for a directed search. In addition, this method has only been specifically applied in a Hough-transform-based pipeline (FrequencyHough~\cite{Astone2014}).}

\textcolor{black}{Building upon these studies, robust and effective criteria are needed in HMM-based searches so that the off-target veto can be applied to a wide range of search configurations.} In this section we discuss an in-depth empirical study of the off-target veto and define veto criteria based on this investigation for two different scenarios depending on the number of candidates that require processing. 



\begin{figure*}[hbt!]
	\centering
	\includegraphics[scale=.62]{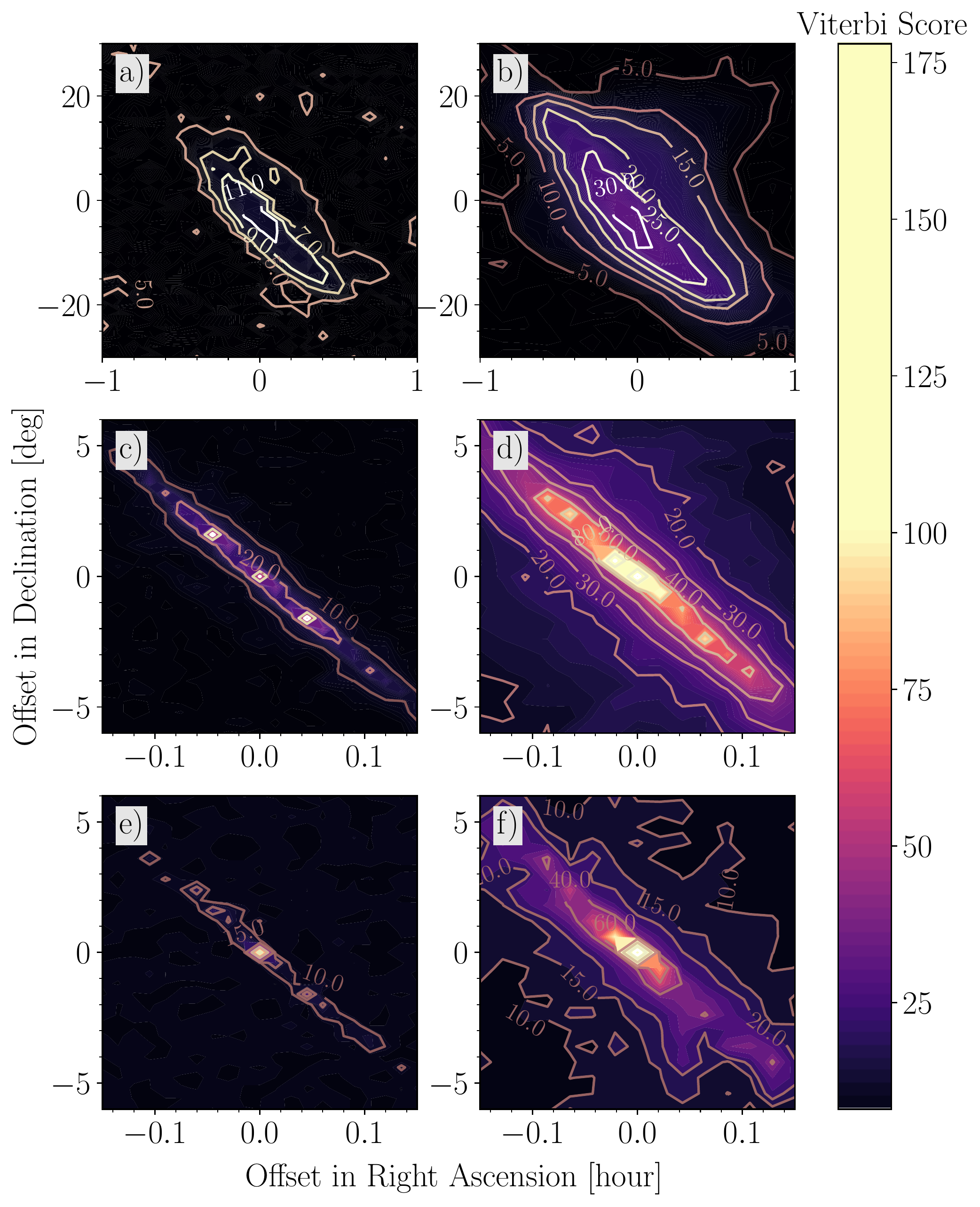}
	\caption{Contour of Viterbi scores as a function of the offset in RA and Dec for an injection at RA = 00~h 00~m 00~s and Dec = $00\degree$ $00'$ $00''$ with signal strength $h_0 = 4.0 \times 10^{-26}$ (left) and $h_0 = 8.0 \times 10^{-26}$ (right). From top to bottom, coherent lengths $T_{\rm coh}=12$~hr, 5~d, and 30~d are used with $T_{\rm obs}=180$~d. (Other simulation parameters: \mbox{$S_h^{1/2} = 4 \times 10^{-24}$~Hz$^{-1/2}$}, $\cos\iota =1$.) See Table~\ref{tab:peaks1} for the Viterbi score obtained at the injection location ($S_{\rm target}$) in each panel.}
	\label{fig:off-target_Viterbi}
\end{figure*}

\begin{figure*}[hbt!]
	\centering
	\includegraphics[scale=.45]{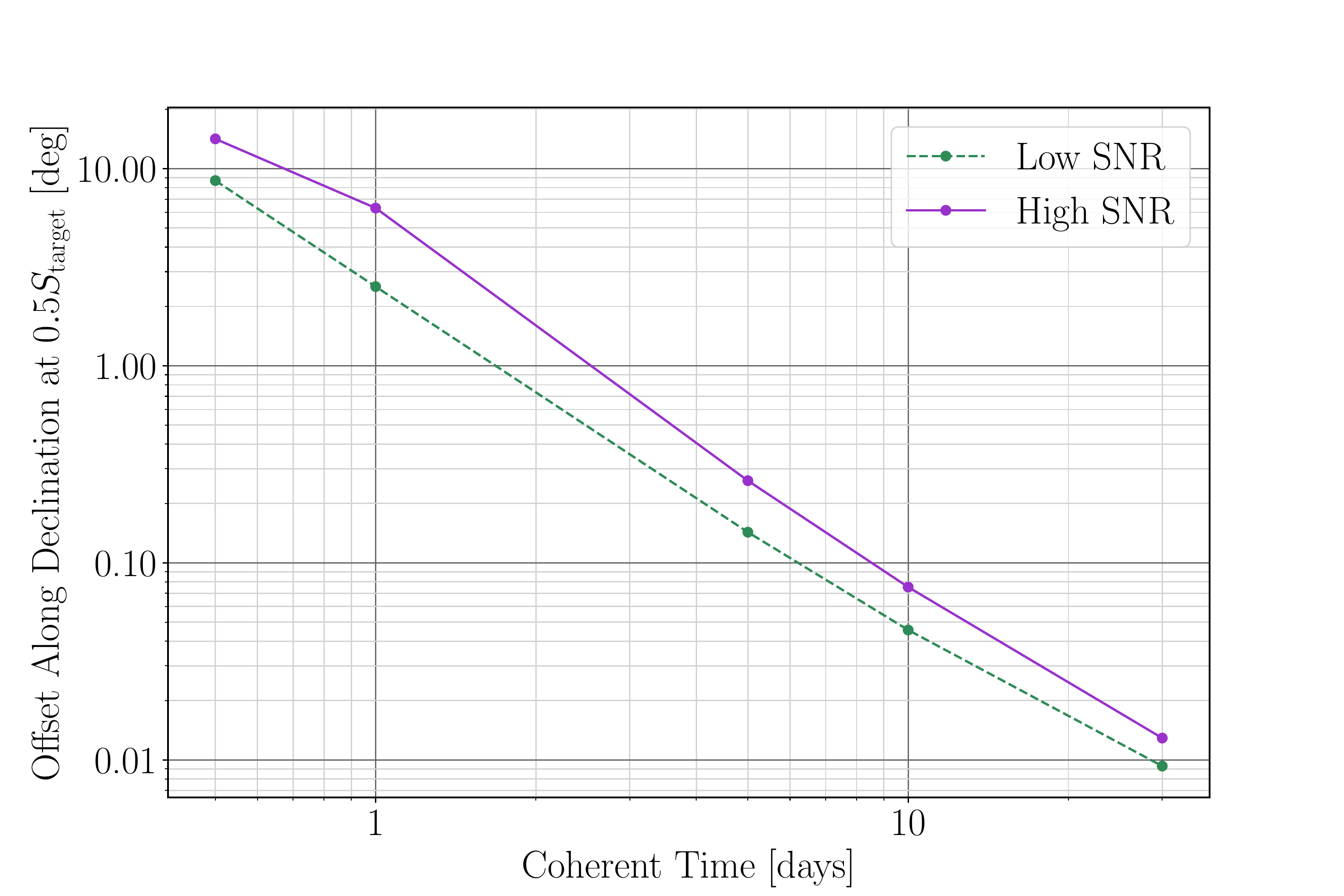}
	\caption{Dec offset at which the Viterbi score drops below $0.5 S_{\rm target}$ as a function of $T_{\rm coh}$. The coherent lengths tested are $T_{\rm coh}$ = 12~hr, 1~d, 5~d, 10~d, and 30~d. The lines that join the sample points give a rough idea of where this offset would occur for other choices of $T_{\rm coh}$ within this range. For the low (green) and high (purple) SNRs, the signal strength of the injection is $h_0 = 4.0 \times 10^{-26}$ and $h_0 = 8.0 \times 10^{-26}$, respectively. (Other simulation parameters: \mbox{$S_h^{1/2} = 4 \times 10^{-24}$~Hz$^{-1/2}$}, $\cos\iota =1$). \textcolor{black}{See Table~\ref{tab:errors} for the offset grid spacing used for each configuration and the error on each point.}}
	\label{fig:offset_vs_Tcoh}
\end{figure*}

\begin{figure*}[hbt!]
	\centering
	\includegraphics[scale=.5]{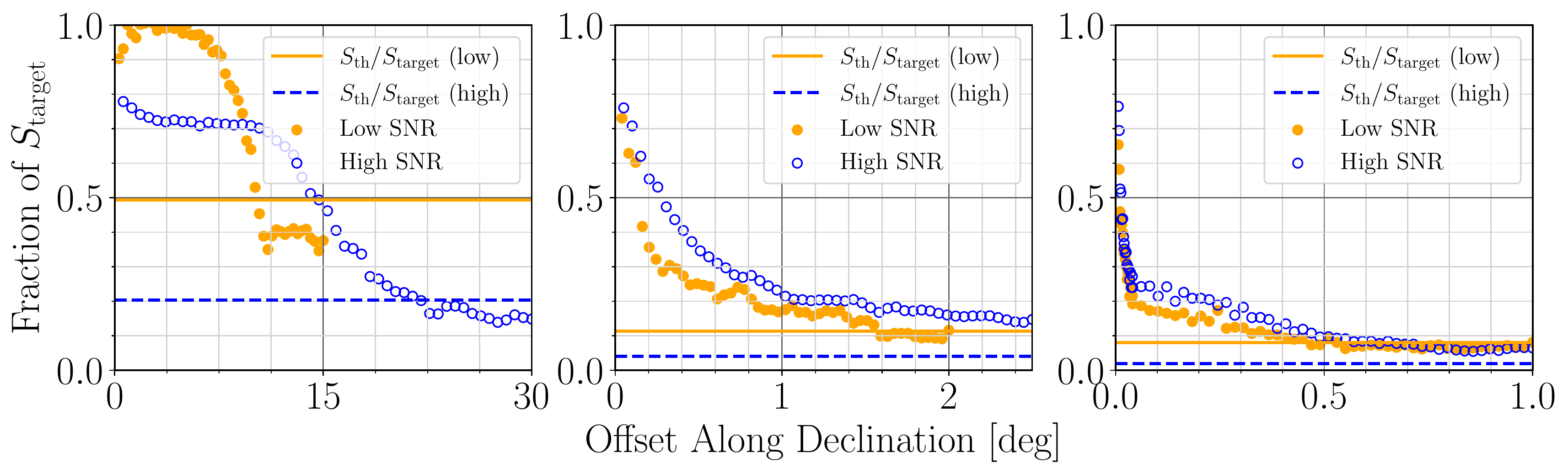}
	\caption{Viterbi score (as a fraction of $S_{\rm target}$) obtained at positions away from the center along Dec (from left to right: $T_{\rm coh}=12$~hr, 5~d, 30~d). In each panel, orange dots and blue circles correspond to the low SNR case with $h_0 = 4.0 \times 10^{-26}$ and high SNR case with $h_0 = 8.0 \times 10^{-26}$, respectively. The solid orange and dashed blue lines mark the threshold, each one displayed as a fraction of $S_{\rm target}$ for its corresponding SNR. (Although $S_{\rm th}$ remains the same for a particular $T_{\rm coh}$ regardless of SNR, $S_{\rm target}$ obtained is larger for a higher SNR, and thus the orange and blue lines do not overlap.)}
	\label{fig:percent_loss_vs_offset}
\end{figure*}

\subsection{$T_{\rm coh}$}
\label{sec:off-target_Tcoh}

Recalling Eq.~\eqref{eqn:Tcoh}, we can select a $T_{\rm coh}$ value that satisfies
\begin{equation}
T_{\rm coh} \leq (2|\dot{f}_{\rm max}|)^{-1/2},
\label{eqn:Tcoh2}
\end{equation}
in a directed search, depending on a rough estimate of $|\dot{f}_{\rm max}|$ for the source.
Because $T_{\rm coh}$ directly depends on the rate of frequency evolution, it varies for searches targeting different sources. For example, for younger sources, a shorter $T_{\rm coh}$ is required~\cite{Sun2018-2}.
In the first stage of the investigation, we test three different coherent lengths in order to help generalize the study to other stack-slide-based searches on various timescales as well as to allow for a broader application to other stack-slide-based semicoherent CW searches. We carry out the tests in both low and high SNR scenarios. As we increase $T_{\rm coh}$, the recovered candidate is more significant at the position where the signal was injected, and the score drops below threshold more steeply as we move off target. This is exactly what we would expect to see; moreover, it holds true in both low and high SNR scenarios. The methods and results are detailed as follows.

The detection threshold $S_{\rm th}$ is estimated for each $T_{\rm coh}$ by running a series of 600 Monte-Carlo simulations in pure Gaussian noise, with the detector's amplitude spectral density (ASD) set to \mbox{$S_h^{1/2} = 4 \times 10^{-24}$~Hz$^{-1/2}$} here and throughout the rest of this study, in the sub-bands 200--201~Hz and 500--501~Hz~\footnote{\textcolor{black}{We use the software package LALSuite for all the simulations in this paper~\cite{lalsuite}.}}. The resulting Viterbi scores are sorted and the score at the 99th percentile is set as the threshold (corresponding to a 1\% false alarm probability per 1~Hz sub-band). For $T_{\rm coh}=12$~hr and 5~d, we find $S_{\rm th} = 5.56$ and 7.14, respectively. (These values are cross-checked with previous studies using similar search configurations, e.g., Refs.~\cite{Jones_2021, O3aSNR}.)

Next we run a series of simulations in which a synthetic signal is injected into Gaussian noise in the 1~Hz sub-band starting at 200~Hz and the Viterbi score is computed for a grid of offsets around the center (i.e., the sky position of the injection). 
Two signal strengths are studied: a weak signal with $h_0 = 4.0 \times 10^{-26}$ and a loud one with $h_0 = 8.0 \times 10^{-26}$.
We assume the signals are circularly polarized with $\cos \iota$, where $\iota$ is the source inclination angle, fixed at unity such that $h_0^{\rm eff}/h_0$ is held fixed, \textcolor{black}{where $h_0^{\rm eff}$ is defined as~\cite{Sun2018-2}:
\begin{equation}
h_0^{\mathrm{eff}}=h_0 2^{-1 / 2}\left\{\left[\left(1+\cos^2 \iota\right) / 2\right]^2+\cos^2 \iota\right\}^{1/2}.
\end{equation}}
When $\cos \iota$ is equal to 1 or $-1$, the signal is circularly polarized and its strength is maximized.
The simulation is run over 180~d \textcolor{black}{(chosen to keep the computational cost of simulations to a reasonable level and given the fact that several recent CW searches use half-year observational data, e.g.,~\cite{O3aINS, O3aBNS, O3aCassVela, O3aSNR})} starting from the arbitrarily chosen GPS time 1167545066 (same as the one used in Figure~\ref{fig:doppler_pattern} (a)).
Three search configurations are tested in order to get a more complete picture of how a signal behaves around its center position in the sky: $T_{\rm coh}=12$~hr ($N_T=360$), $T_{\rm coh}=5$~d ($N_T=36$), and $T_{\rm coh}=30$~d ($N_T=6$).
\textcolor{black}{The fraction of the sky spanned by the grid of off-target positions varies depending on the signal SNR, where louder signals require a larger span in order for the extent of their features to be fully captured. However, to minimize the computational time needed for these simulations, a fixed grid of $21 \times 15$ data points is used regardless of the fraction of the sky spanned.}

Figure~\ref{fig:off-target_Viterbi} shows the Viterbi scores (plotted as the color) of the recovered signal at its injected sky position (RA = 00~h 00~m 00~s, Dec = $00\degree$ $00'$ $00''$) and for a grid of offsets around that position.
The left and right columns correspond to the low and high SNR scenarios, respectively. 
The three rows, from top to bottom, correspond to $T_{\rm coh}=12$~hr, 5~d, and 30~d. 
Table~\ref{tab:peaks1} lists the Viterbi scores at the source's true position, denoted as $S_{\rm target}$, for each panel.

\begin{table}[tbh]
	\centering
	\setlength{\tabcolsep}{15pt}
	\renewcommand\arraystretch{1.2}
	\begin{tabular}{clll}
		\hline
		Label & SNR & $T_{\rm coh}$ & $S_{\rm target}$ \\
		\hline
		a & low & 12~hr & 11.28 \\
		b & high & 12~hr & 27.39 \\
		c & low & 5~d & 62.70 \\
		d & high & 5~d & 176.59 \\
		e & how & 30~d & 116.06 \\
		f & high & 30~d & 478.04 \\
		\hline
	\end{tabular}
	\caption{Viterbi score obtained at the injection site for each panel shown in Fig.~\ref{fig:off-target_Viterbi}.}
	\label{tab:peaks1}
\end{table}

The bright ellipse (with the brightest spot in the center) shown in these plots, which we refer to as the EPSF, is exactly what one would expect for an astrophysical signal in any stack-slide-based semicoherent search algorithm. \textcolor{black}{(See Appendix~\ref{sec:off-target_cross-corr} for an example using another stack-slide-based semicoherent search algorithm.)}
Moreover, such patterns are consistent with those in existing literature: e.g., see Fig. 1 in Ref.~\cite{Mastrogiovanni_2018}, Fig. 1 in Ref.~\cite{Prix2007}, and Fig. 14 in Ref.~\cite{Isi_2019}. \textcolor{black}{The ellipse likely results from the different dependencies in RA versus Dec, outlined in Ref.~\cite{Prix2007}, as well as the symmetries present in the Doppler modulation, as shown in Eq.~\eqref{eqn:kappa} (discussed in Ref.~\cite{Isi2020}).} The faint periodic features seen in some plots but not in others may also be related to the periodic functions within the Doppler modulation~\cite{Intini_2020}. However, we rely only on the broader elliptical patterns in this study, so the minor features do not impact the results. A more detailed investigation of these minor features lies outside the scope of this paper. It should be noted that the ellipses shown in panels (a) and (b), both with $T_{\rm coh}=12$~hr, appear to be slightly off-center such that the brightest point is not exactly at the injection site. This is most likely due to the relatively poor sky resolution when a short $T_{\rm coh} < 1$~d is used. Indeed, the peak is recovered almost exactly at the injection location for $T_{\rm coh}=5$~d and 30~d.

In comparing the images shown in Fig.~\ref{fig:off-target_Viterbi}, it is clear that as $T_{\rm coh}$ increases, the EPSF becomes brighter in the center and narrower. In fact, it has already been established that the offset at which the detection statistic drops below threshold is related to $T_{\rm coh}$; as $T_{\rm coh}$ increases, this offset decreases~\cite{Mastrogiovanni_2018}. 
This relationship is shown in Figure~\ref{fig:offset_vs_Tcoh}; the offset in Dec (holding RA fixed at the injection coordinate) at which the Viterbi score drops below $0.5 S_{\rm target}$ is tracked as a function of $T_{\rm coh}$. (To quantify this relation with better resolution, we conduct additional searches with two more coherent lengths $T_{\rm coh}=1$~d and 10~d in Dec.)
\textcolor{black}{
The error bars for the offsets in Dec (too small to be shown in Figure~\ref{fig:offset_vs_Tcoh}) are set by the grid spacing used to estimate these offsets, which is listed in  Table~\ref{tab:errors} for each case, along with their errors ($\lesssim 1\%$).}
Indeed, for both SNRs tested, the shortest coherent time, 12~hr, has the greatest Dec offset. As we increase $T_{\rm coh}$, this offset decreases approximately linearly in log-log scale. We choose to take our offsets only in Dec because the behavior of the detection statistics is more dependent on the source position when moving only in RA, so taking offsets in Dec allows us to set more generally applicable guidelines for the veto that are independent of sky position. (More details are discussed in Section~\ref{sec:off-target_loc}.)

\begin{table}[tbh]
	\centering
	\setlength{\tabcolsep}{15pt}
	\renewcommand\arraystretch{1.2}
	\begin{tabular}{clll}
		\hline
		SNR & $T_{\rm coh}$ & Grid spacing & Error \\
		\hline
		low & 12~hr & 0.01~deg & 0.11\% \\
		high & 12~hr & 0.01~deg & 0.07\% \\
		low & 1~d & 0.01~deg & 0.40\% \\
		high & 1~d & 0.01~deg & 0.16\% \\
		low & 5~d & 0.001~deg & 0.70\% \\
		high & 5~d & 0.001~deg & 0.38\% \\
		low & 10~d & 0.0001~deg & 0.22\% \\
		high & 10~d & 0.0001~deg & 0.13\% \\
		low & 30~d & 0.0001~deg & 1.08\% \\
		high & 30~d & 0.0001~deg & 0.78\% \\
		\hline
	\end{tabular}
	\caption{\textcolor{black}{Offset grid spacing used for each configuration and relative error on each point shown in Fig.~\ref{fig:offset_vs_Tcoh}}}
	\label{tab:errors}
\end{table}

Figure~\ref{fig:percent_loss_vs_offset} shows further details on how the Viterbi score decreases as the Dec offset increases for $T_{\rm coh}=12$~hr, 5~d, and 30~d, from left to right. 
In general, in the first few steps away from the center, the scores drop steeply, especially for longer coherent times, then level out such that further increasing the offset leads to only small decreases in the score. 
Once a large enough offset is reached, the score drops below threshold and the signal is no longer detectable. This offset at which the signal becomes undetectable heavily depends on $T_{\rm coh}$ and SNR. 
For extremely loud signals, the detection statistic may never drop below threshold, regardless of the sky position searched. However, current detectors do not operate in the high-SNR regime for CW sources---in fact, such a loud signal would be astrophysically implausible because the source would likely fall outside the range of extreme ellipticities and would be spinning down so rapidly that its $\dot{f}$-value would not lie within the ranges used in this study. Thus we do not take into consideration the extremely high-SNR scenario.

\begin{figure*}[hbt!]
	\centering
	\includegraphics[scale=.6385]{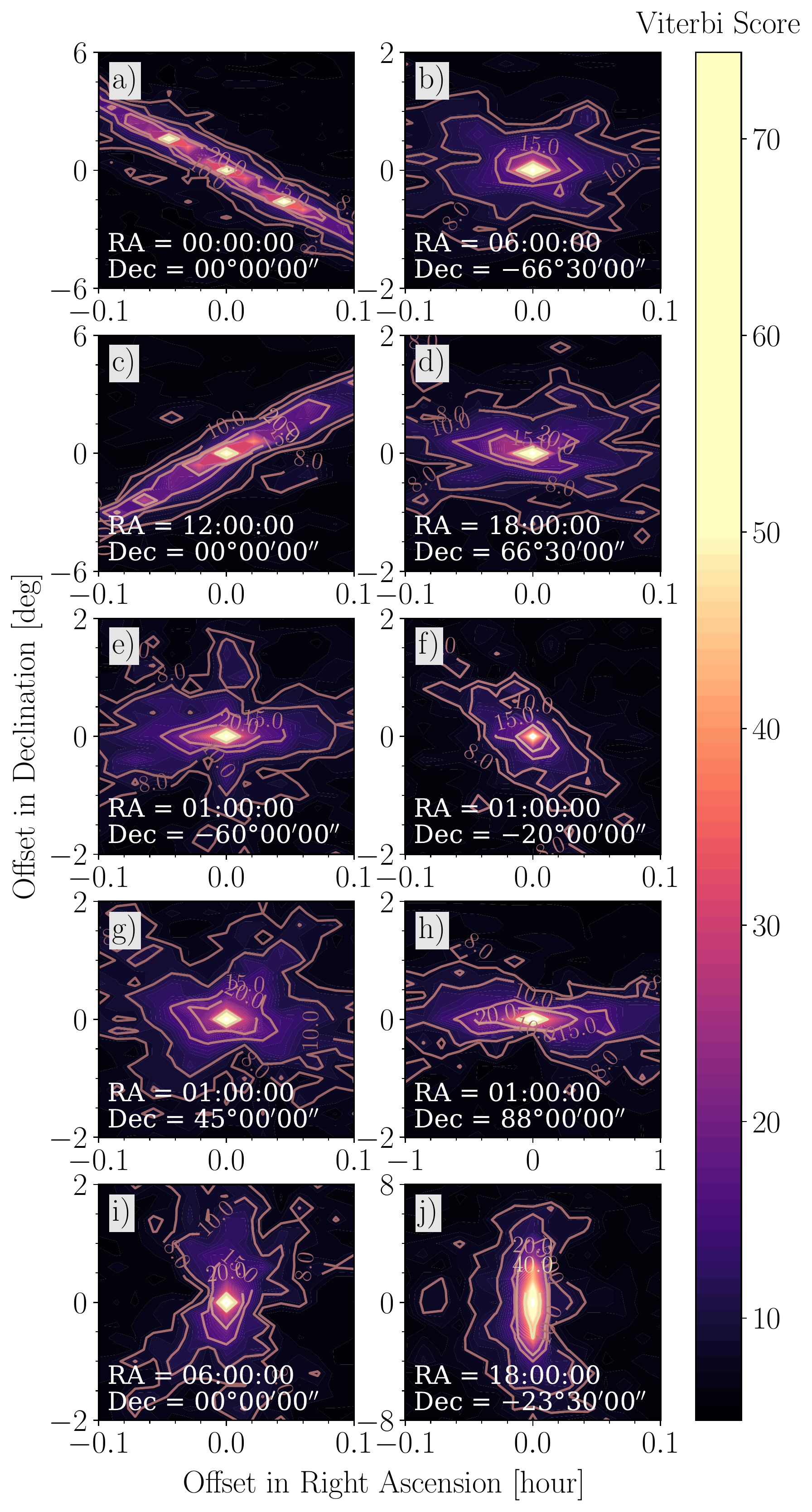}
	\caption{Contour of Viterbi scores as a function of the offset in RA and Dec for injections at a variety of different sky positions with signal strength $h_0 = 4.0 \times 10^{-26}$ (Gaussian noise). A coherent length $T_{\rm coh}=5$~d is used with $T_{\rm obs}=180$~d. (Other simulation parameters: \mbox{$S_h^{1/2} = 4 \times 10^{-24}$~Hz$^{-1/2}$}, $\cos\iota =1$.) See Table~\ref{tab:peaks2} for the Viterbi score obtained at each injection location ($S_{\rm target}$).}
	\label{fig:off-target_Viterbi_loc}
\end{figure*}

\begin{figure*}[hbt!]
	\centering
	\includegraphics[scale=.5]{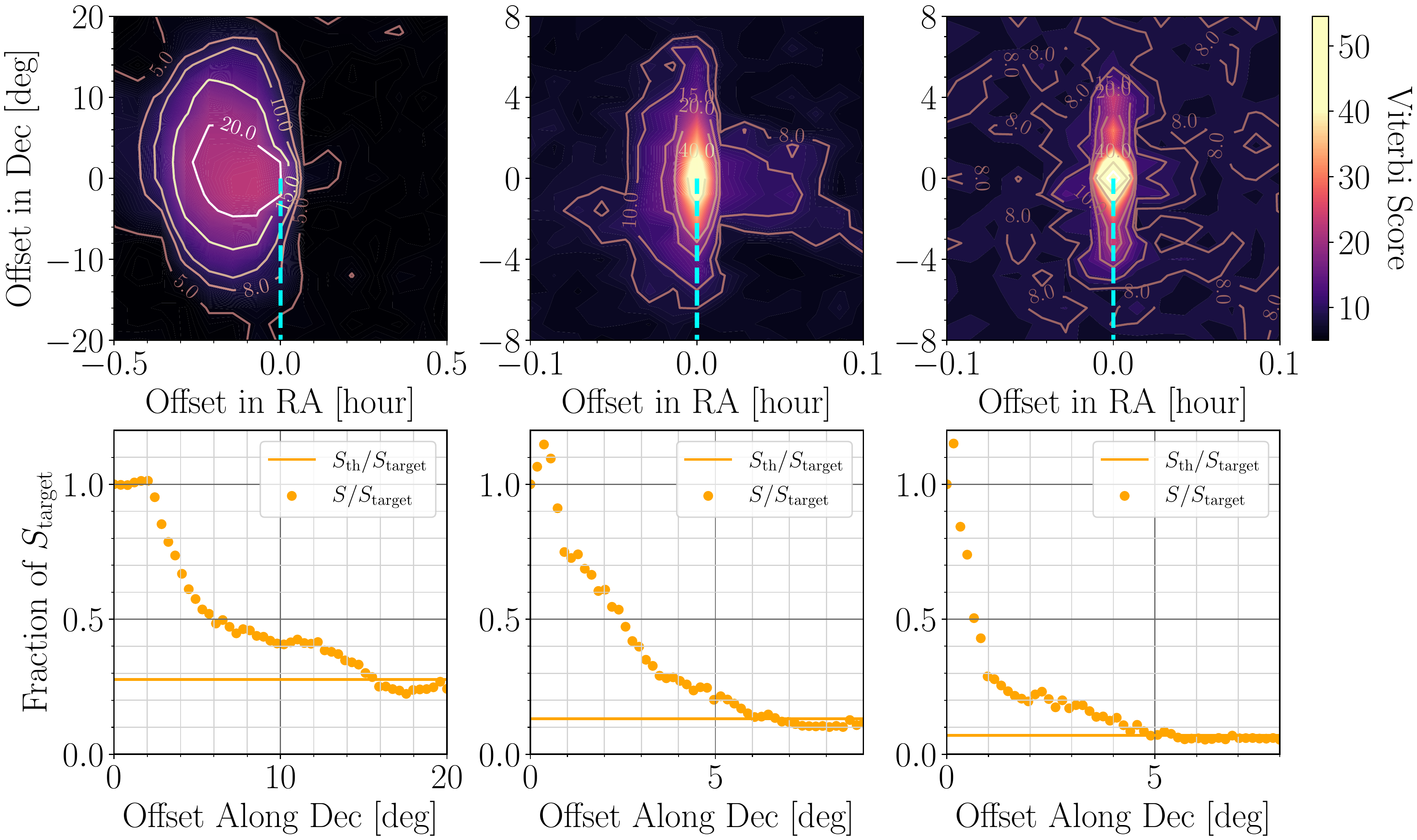}
	\caption{Top row: Contour of Viterbi scores as a function of the offset in RA and Dec for an injection at RA = 18~h 00~m 00~s and Dec = $-23\degree$ $30'$ $00''$ with signal strength $h_0 = 5.0 \times 10^{-26}$ (left plot) and $h_0 = 4.0 \times 10^{-26}$ (middle and right plots) (Gaussian noise), using three different choices of $T_{\rm coh}$. From left to right, we have $T_{\rm coh}=12$~hr, 5~d, and 30~d, with $S_{\rm target}$ = 22.55, 54.44, and 131.16, respectively. The total observing time is 180~d. (Other simulation parameters: \mbox{$S_h^{1/2} = 4 \times 10^{-24}$~Hz$^{-1/2}$}, $\cos\iota =1$.)
		Bottom row: Viterbi score $S$ (as a fraction of $S_{\rm target}$) obtained at positions away from the center along Dec. In each panel, the search configuration is the same as that used in the contour image above it (from left to right: $T_{\rm coh}=12$~hr, 5~d, 30~d). The solid lines mark the threshold for each search configuration displayed as a fraction of $S_{\rm target}$. 
		The dashed cyan line in each contour image in the top row marks the direction along which the series of offsets shown in the bottom row are taken.}
	\label{fig:off-target_Viterbi_percent_loss}
\end{figure*}

\subsection{Sky position}
\label{sec:off-target_loc}

The elliptical EPSF shown in Figure~\ref{fig:off-target_Viterbi} is further explored in this section by varying the location of the injection. A series of simulations are run with different combinations of RA and Dec.
We find that sky position is the dominant factor in determining the shape (i.e., the ratio of the major and minor axes) and orientation of the EPSF. The EPSF can act as a precise marker of an astrophysical signal, so we use it in the off-target veto criteria.
The details are discussed below.

In this set of simulations, we fix the injection SNR using the signal strength $h_0 = 4.0 \times 10^{-26}$ and conduct the search with $T_{\rm coh} = 5$~d.
The other simulation parameters are the same as those in Section~\ref{sec:off-target_Tcoh}.
Figure~\ref{fig:off-target_Viterbi_loc} shows the resulting contour plots, where the Viterbi score (color) is plotted as a function of RA and Dec for a grid of offsets centered on the injection site; see Table~\ref{tab:peaks2} for the ten injection locations and their corresponding scores.
Random variations aside, the shape and orientation of the EPSF change systematically as functions of RA and Dec. Matching up each panel in Figure~\ref{fig:off-target_Viterbi_loc} to its position in Figure~\ref{fig:doppler_pattern}~(a), we observe that the inclination of the ellipse at a particular sky position roughly follows the slope of the $\kappa$ contour in Figure~\ref{fig:doppler_pattern}~(a) passing through that same position.

\begin{table}[tbh]
	\centering
	\setlength{\tabcolsep}{10pt}
	\renewcommand\arraystretch{1.2}
	\begin{tabular}{clll}
		\hline
		Label & RA & Dec & $S_{\rm target}$ \\
		\hline
		a & 00:00:00 & $00\degree 00' 00''$ & 62.70 \\
		b & 06:00:00 & $-66\degree 30' 00''$ & 73.20 \\
		c & 12:00:00 & $00\degree 00' 00''$ & 60.35 \\
		d & 18:00:00 & $66\degree 30' 00''$ & 73.71 \\
		e & 01:00:00 & $-60\degree 00' 00''$ & 70.93 \\
		f & 01:00:00 & $-20\degree 00' 00''$ & 55.82 \\
		g & 01:00:00 & $45\degree 00' 00''$ & 74.35 \\
		h & 01:00:00 & $88\degree 00' 00''$ & 69.42 \\
		i & 06:00:00 & $00\degree 00' 00''$ & 68.89 \\
		j & 18:00:00 & $-23\degree 30' 00''$ & 61.64 \\
		\hline
	\end{tabular}
	\caption{Viterbi scores at the ten injection sites shown in Fig.~\ref{fig:off-target_Viterbi_loc}.}
	\label{tab:peaks2}
\end{table}

Two notable examples are discussed below. The first is presented in Figure~\ref{fig:off-target_Viterbi_loc}~(h); this image is centered on RA = 01~h 00~m 00~s and Dec = $88\degree$ $00'$ $00''$, and it shows an ellipse that spans more than two hours in RA---a significant fraction of the sky. Referring to where this position is found within the contour map in Figure~\ref{fig:doppler_pattern}~(a)---the triangle marker pointing towards the left---we immediately notice that the contour passing through this point moves along RA at a roughly fixed Dec. The same holds true for a source found at, for example, Dec = $-88\degree$ $00'$ $00''$. Indeed, as the position approaches either pole, the EPSF extends broadly along RA because the DM effect is weak near the poles. This fact could lead to astrophysical signals being falsely vetoed as noise if the off-target veto is applied to sources near the poles using an offset only in RA.
Certainly, choosing our criteria so that the off-target veto can be safely applied to candidates at the poles is important since it is unlikely that the DM-off veto will eliminate such candidates.

The second example worth discussing is shown in Figure~\ref{fig:off-target_Viterbi_loc}~(j). This image is centered on RA = 18~h 00~m 00~s and Dec = $-23\degree$ $30'$ $00''$ (the latter corresponding to the tilt of Earth's axis) and shows an ellipse spanning roughly $15\degree$ in Dec. Once again, referring to the corresponding sky position in Figure~\ref{fig:doppler_pattern}~(a), we can see that the injection is located in the center of the dark region---the ``dark spot" where $\kappa = -1$ (the cross marker). 
Based on the simulation results obtained at various positions and on how the EPSFs tend to vary with injection location, we find that this sky position at the dark spot---along with the ``bright spot" with $\kappa = 1$---in general produces the most extended EPSF in Dec. We use this to set veto guidelines in Sec.~\ref{sec:off-target_guide}.

Since the extension of the ellipse also largely depends on $T_{\rm coh}$ (for a given SNR), we further conduct a set of tests for injections at the dark spot, using another two choices of coherent length: $T_{\rm coh} = 12$~hr and 30~d.
In Figure~\ref{fig:off-target_Viterbi_percent_loss}, the top row shows the EPSFs at the dark spot for $T_{\rm coh}=12$~hr (left), 5~d (middle), and 30~d (right). We inject a louder signal, $h_0 = 5.0 \times 10^{-26}$, for the search which uses $T_{\rm coh}=12$~hr (as opposed to $h_0 = 4.0 \times 10^{-26}$ for the other two coherent times) so that the EPSF is better resolved. Of all the search configurations tested throughout this study, the ellipse at this sky position using the shortest coherent length ($T_{\rm coh}=12$~hr) spans the largest angular distance in Dec---approximately $30 \degree$.
The bottom row of Fig.~\ref{fig:off-target_Viterbi_percent_loss} shows how the Viterbi score decreases as the Dec offset increases along the dashed cyan line plotted in the top panels, for $T_{\rm coh}=12$~hr, 5~d, and 30~d, from left to right (similar to Figure~\ref{fig:percent_loss_vs_offset}).  

\subsection{\textcolor{black}{Real LIGO data in the second observing run}}
\label{sec:off-target_real_data}

\begin{figure*}[hbt!]
	\centering
	\includegraphics[scale=.5]{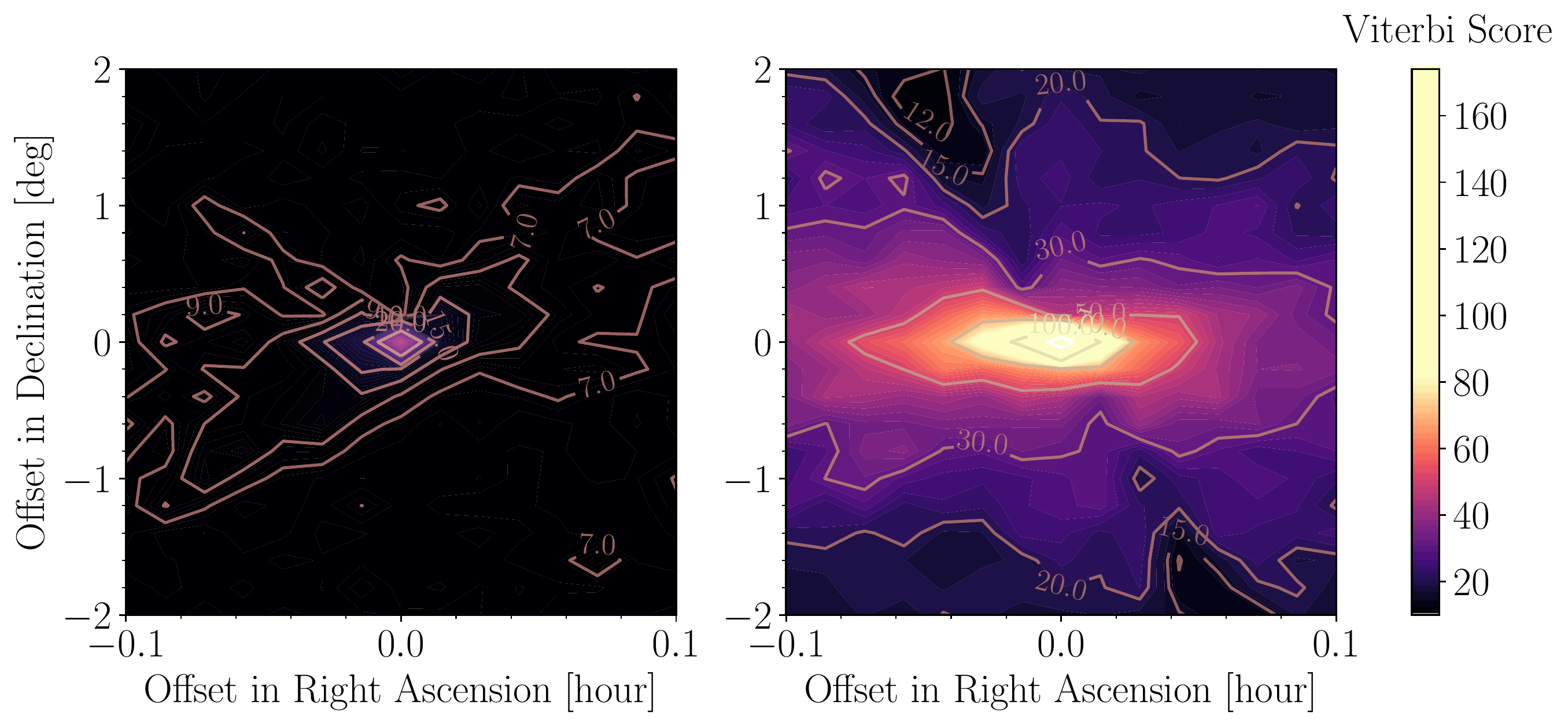}
	\caption{\textcolor{black}{Contour of Viterbi scores as a function of the offset in RA and Dec for a synthetic signal ($h_0 = 8.0 \times 10^{-26}$, $\cos\iota =1$) injected in LIGO O2 data (left, \textcolor{black}{$S_h^{1/2} = 7 \times 10^{-24}$~Hz$^{-1/2}$}) and Gaussian noise (right, $S_h^{1/2} = 4 \times 10^{-24}$~Hz$^{-1/2}$) \textcolor{black}{in the 200--201~Hz band}. The images are centered at the sky position of the injection (RA = 01~h 00~m 00~s and Dec = $-60\degree$ $00'$ $00''$). A coherent length $T_{\rm coh}=5$~d is used with $T_{\rm obs}=180$~d.}}
	\label{fig:off-target_real_noise}
\end{figure*}

\textcolor{black}{In addition to investigating the behavior of the EPSF of a signal in Gaussian noise, we look at some examples with a real noise background. The left panel in Figure~\ref{fig:off-target_real_noise} shows the EPSF of a synthetic signal ($h_0 = 8.0 \times 10^{-26}$, $\cos\iota =1$) injected \textcolor{black}{into the 200--201~Hz band} in the Advanced LIGO data collected in the second observing run (O2)~\cite{Abbott-O2}, recovered using the HMM pipeline with $T_{\rm coh}=5$~d. \textcolor{black}{The effective averaged ASD in that band is $S_h^{1/2} = 7 \times 10^{-24}$~Hz$^{-1/2}$.}
For consistency, a GPS start time of 1167545066 is used with a total observing time of 180~d. The right panel shows a signal with the same parameters injected into Gaussian noise ($S_h^{1/2} = 4 \times 10^{-24}$~Hz$^{-1/2}$). Despite the effects of non-Gaussianities in real detector noise and the higher noise ASD, the EPSF in LIGO O2 data is still clearly resolvable and, despite some additional features that seem to have caused a slight shift in the EPSF inclination, the overall shapes of the EPSF are still comparable in these two sets of simulations.}

\begin{figure*}[hbt!]
	\centering
	\includegraphics[scale=.5]{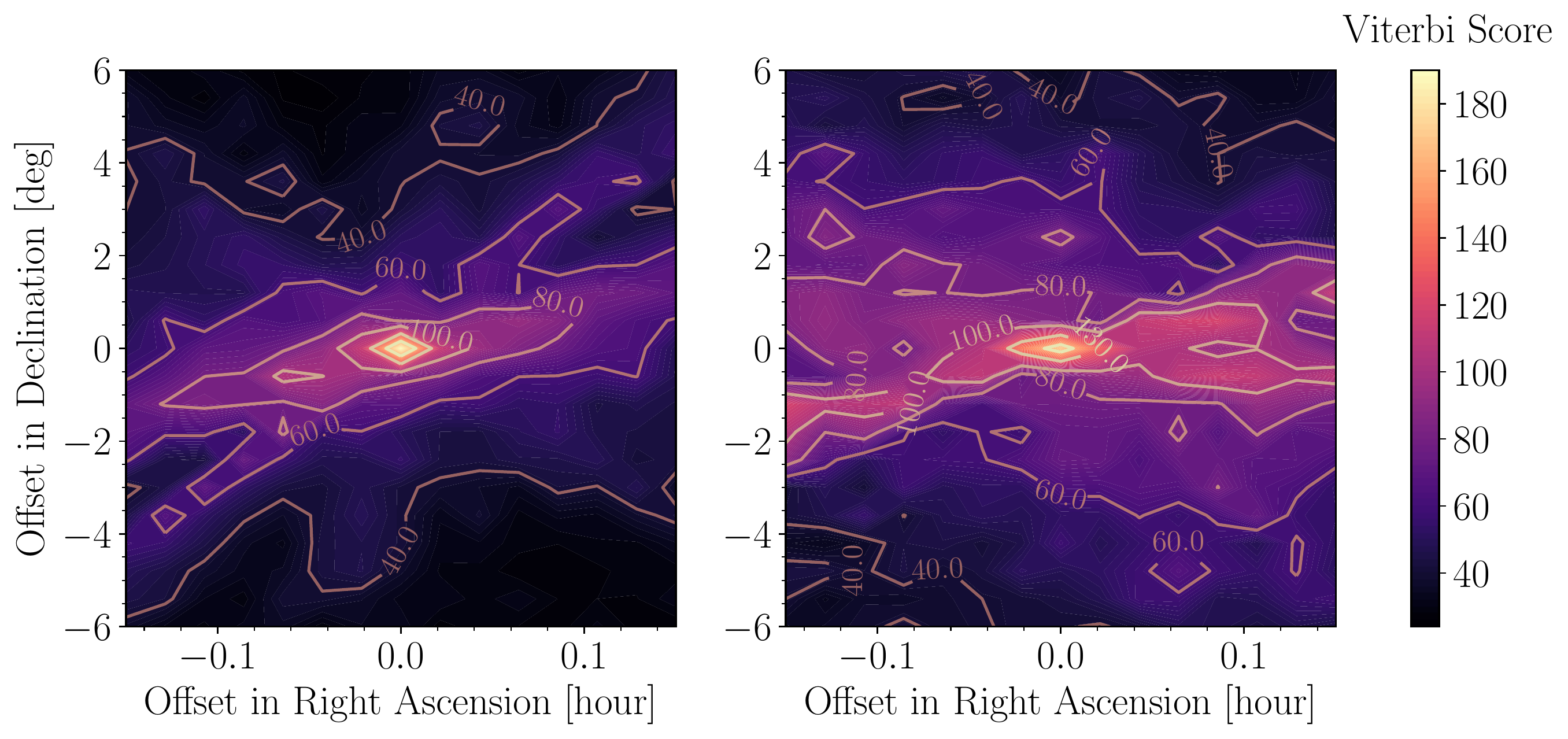}
	\caption{\textcolor{black}{Contour of Viterbi scores as a function of the offset in RA and Dec for two hardware injections in LIGO O2 data. Left: The hardware injection P3 ($h_0 = 8.23 \times 10^{-25}$, $\cos\iota = -0.08$) in the 108--109~Hz band and centered at RA = 11~h 53~m 29.4~s and Dec = $-33\degree$ $26'$ $11.8''$. Right: The hardware injection P9 ($h_0 = 3.01 \times 10^{-24}$, $\cos\iota = -0.62$) in the 763--764~Hz band and centered at RA = 13~h 15~m 32.5~s and Dec = $75\degree$ $41'$ $22.5''$. We use $T_{\rm coh}=5$~d and $T_{\rm obs}=180$~d for both injections.}}
	\label{fig:off-target_hardware}
\end{figure*}

\textcolor{black}{Fifteen hardware injections were included in the LIGO O2 observing run for testing purposes~\textcolor{black}{\cite{Biwer2017}}. These injections mimic real CW signals by moving the detector mirrors as though CWs are passing through, and they are an important complement to the software injections we have looked at exclusively up until now. Of the fifteen injections in O2, we look at two, P3 and P9, as examples. 
We have chosen these two injections because they have signal parameters that fall within the parameter space covered by a typical search configuration studied in this paper ($T_{\rm coh} = 5$ d), enabling a consistent comparison and the use of statistics we have already derived here. 
We still search for $T_{\rm obs}=180$~d, starting from the GPS time 1167545066. 	
The EPSFs of these injections are shown in Figure~\ref{fig:off-target_hardware}, with P3 on the left and P9 on the right. P3 is located in the 108--109~Hz band and is centered at RA = 11~h 53~m 29.4~s and Dec = $-33\degree$ $26'$ $11.8''$. P9 is located in the 763--764~Hz band and is centered at RA = 13~h 15~m 32.5~s and Dec = $75\degree$ $41'$ $22.5''$. See the Gravitational Wave Open Science Center O2 data release~\cite{Abbott-O2} for a more detailed parameter list. Overall, these EPSFs are in line with what we have predicted for astrophysical signals using software injections; if these injections were candidates in a real search, the off-target veto would not eliminate them (as demonstrated in Section~\ref{sec:off-target_guide}).}

\begin{figure*}[hbt!]
	\centering
	\includegraphics[scale=.5]{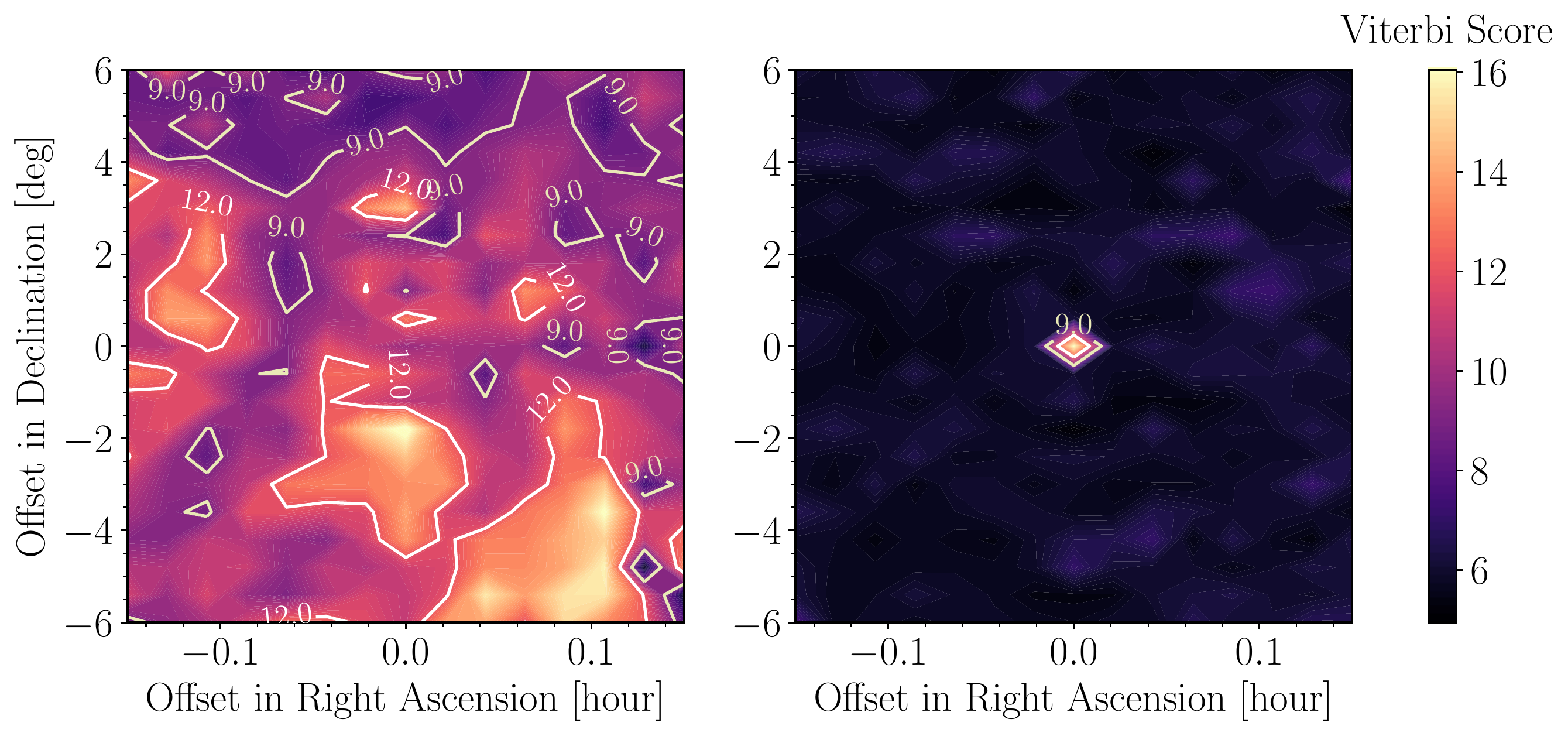}
	\caption{Contour of Viterbi scores as a function of the offset in RA and Dec in the 462--463~Hz band in LIGO O2 data where an instrumental line lies (left) and Gaussian noise where a synthetic signal ($h_0 = 2.0 \times 10^{-26}$, $\cos\iota =1$) is injected (right). Both images are centered at RA = 22~h 57~m 39.1~s and Dec = $-29\degree$ $37'$ $20.0''$. A coherent length $T_{\rm coh}=5$~d is used with a total observing time $T_{\rm obs}=180$~d.}
	\label{fig:off-target_noise}
\end{figure*}

We also look at the Viterbi scores for a grid of sky positions centered at RA = 22~h 57~m 39.1~s and Dec = $-29\degree$ $37'$ $20.0''$ in the 462--463~Hz sub-band in LIGO O2 data, where an unidentified noise artifact originating in the Hanford detector is located (identified in Sec. IV of Ref.~\cite{Jones_2021}).
The same duration of $180$~d starting from 1167545066 is searched.
Then we repeat this procedure in Gaussian noise with a synthetic signal ($h_0 = 2.0 \times 10^{-26}$, $\cos\iota =1$) injected at RA = 22~h 57~m 39.1~s and Dec = $-29\degree$ $37'$ $20.0''$.
The EPSFs from LIGO O2 data and the simulated data containing a synthetic signal are shown in the left and right panels in Figure~\ref{fig:off-target_noise}, respectively. The left panel show a spread of above-threshold Viterbi scores which fluctuate randomly across the sky. There is no trace of the bright peak in the center seen in the right panel resulting from a synthetic signal. This is the expected behavior of a candidate caused by a noise artifact, which would not be impacted by DM because it originates on Earth.

\subsection{\textcolor{black}{Physical insights from the EPSF}}
\label{sec:off-target_EPSF}

\begin{figure*}[hbt!]
	\centering
	\includegraphics[scale=.5]{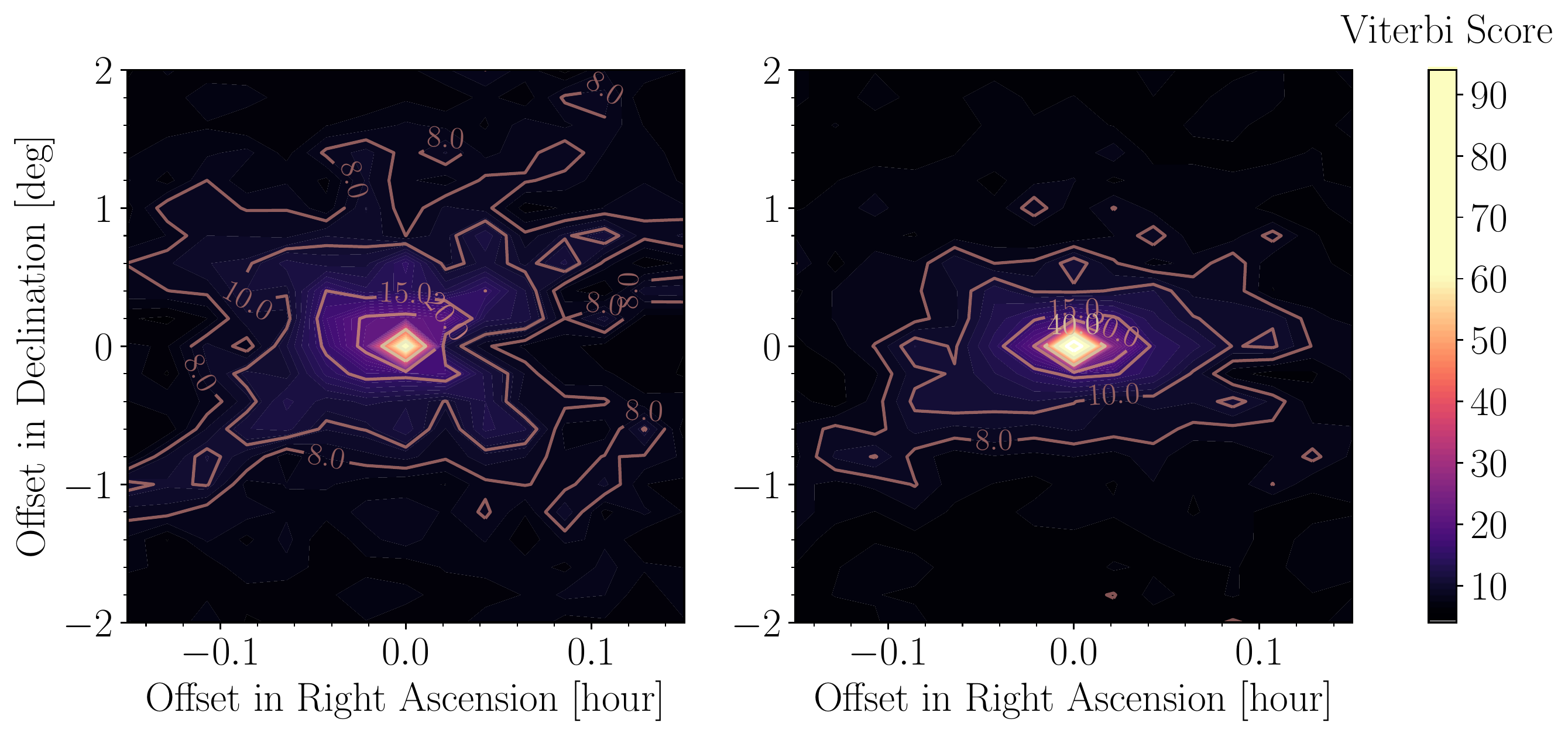}
	\caption{\textcolor{black}{Contour of Viterbi scores as a function of the offset in RA and Dec for an injection at RA = 18~h 00~m 00~s and Dec = $66\degree$ $30'$ $00''$ with signal strength $h_0 = 4.0 \times 10^{-26}$ ($\cos\iota =1$). Results are obtained for $T_{\rm obs}=180$~d (left) and $T_{\rm obs}=365$~d (right), with $T_{\rm coh}=5$~d for both. We find $S_{\rm target}=68.84$ (left) and $S_{\rm target}=93.94$ (right).}}
	\label{fig:off-target_diffTobs}
\end{figure*}

In practice, determining the exact shape and orientation of the EPSF of a CW signal is more complicated than simply following the direction of a Doppler pattern contour, e.g., in Figure~\ref{fig:doppler_pattern}~(a) (also see Ref.~\cite{Intini_2020}). In fact, the shape and orientation of the EPSF mainly depend on a combination of three factors: the source position, the start time of the observation, and the amount of time the signal is integrated over.
\textcolor{black}{First, the sky position of the source is important, as this determines where it lies along the Doppler pattern shown in Figure~\ref{fig:doppler_pattern}~(a) and what Doppler correction should be made.}
Then, as the start time is shifted forward, this Doppler pattern also shifts across the sky.
The EPSF becomes less extended as the observing time increases, because the bright and dark regions from different instants in time are included in the integration and cancel out.
\textcolor{black}{This is demonstrated in Figure~\ref{fig:off-target_diffTobs}, where the EPSF integrated over the full year (right) is brighter and more concentrated in the center than the EPSF integrated over half a year (left).}
Despite the impact from terrestrial noise, knowing these three factors allows one to roughly predict the shape and orientation of the EPSF around the true location in a real CW search, which can be used as a veto criterion. 

\subsection{Veto guidelines}
\label{sec:off-target_guide}

To conclude this discussion of the off-target veto, we outline concrete guidelines for how to best apply this veto in two different scenarios: (1) when following up a large number of candidates from various sky positions, and (2) when following up a handful of candidates from a limited number of positions. 
First, in the follow-up procedure of either a directed search which is targeting many sources at different sky positions (e.g.,~\cite{O3aSNR, O3amxp}) or an all-sky CW search, if the earlier-stage vetoes are not effective enough, we need to use the off-target veto to process a large number of CW candidates from many sky positions. Thus, both safety and efficiency are important.
Creating and inspecting detailed images of the EPSF for every candidate would be computationally expensive.
This study allows us to use the detection statistic at a single Dec offset (as offsets in Dec have been shown in Sec.~\ref{sec:off-target_loc} to be safer than offsets in RA) to decide whether or not to veto a candidate. Moreover, we choose a conservative offset such that it is valid for all sky positions; however, the choice of offset does depend on the $T_{\rm coh}$ and $T_{\rm obs}$ used in a particular search.


\subsubsection{Veto a large number of candidates}
We propose the following broad criteria for applying the off-target veto to many candidates at once: a candidate can be safely vetoed \emph{only if} the score at a certain conservative Dec offset remains above $0.5S_{\rm target}$ or $S_{\rm th}$ (whichever is larger), with the added condition that the recovered Viterbi path at this offset position must overlap the original path. 
In practice, one can choose a different fraction other than 0.5 as needed, based on empirical studies.
The reason we stipulate a fraction of the original score rather than a particular value of the score is so that, for a relatively loud signal with original score $S > 2 S_{\rm th}$, the dependence on the SNR is generally removed (as demonstrated in Figure~\ref{fig:offset_vs_Tcoh}). If such an astrophysical signal exists, the score should drop at least 50\% by the chosen Dec offset.
We keep any candidate whose score is below $S_{\rm th}$ at this offset position because a weak signal  becomes undetectable at very small offsets from the true sky position.
One could even consider setting the offset based on a more sophisticated parameter-space-based distance, as presented in Ref.~\cite{Tenorio2021_2}, but such a detailed study is beyond the scope of this paper. 

Figure~\ref{fig:off-target_Viterbi_percent_loss} shows an example of how one might find this optimal offset position for the three $T_{\rm coh}$ choices used most frequently in this paper. 
When processing many candidates at once, a plot like those along the bottom row of Fig.~\ref{fig:off-target_Viterbi_percent_loss} can be created using a synthetic signal injected into Gaussian noise. 
To obtain a safe offset that can be applied to candidates from various sky positions, we carry out the simulations at the sky position where the detection statistic varies the slowest in Dec (i.e., the position at which the EPSF will be the most extended in Dec).
The offset for applying the off-target veto can then be determined by looking at where the score drops below $0.5 S_{\rm target}$ for a relatively loud signal such that $0.5 S_{\rm target}>S_{\rm th}$.
For our particular search configuration, using the data presented in Fig.~\ref{fig:off-target_Viterbi_percent_loss}, we find this offset for $T_{\rm coh}=12$~hr, 5~d, and 30~d (for $T_{\rm obs}=180$~d) to be $7\degree$, $3\degree$, and $1\degree$, respectively. It should be noted for $T_{\rm coh}=12$~hr that, despite the EPSF being significantly off-center, as often occurs for short coherent times which have poorer resolution, the veto can still be safely applied as long as a conservative offset is chosen using the results from simulations in Gaussian noise.

\textcolor{black}{The safety of this veto guideline is demonstrated by running a search for the two O2 hardware injections discussed in Sec.~\ref{sec:off-target_real_data} (Fig.~\ref{fig:off-target_hardware}). Because both injections are very loud, we have $0.5 S_{\rm target} > S_{\rm th}$. (In fact, the score at every location shown in both panels of Fig.~\ref{fig:off-target_hardware} is above the detection threshold.) We veto the candidate if the score fails to drop below $0.5 S_{\rm target}$ at our chosen offset. Choosing to look at a single offset of $3\degree$ along Dec (as defined in the previous paragraph for this search configuration with $T_{\rm coh}=5$~d), we find that for both injections the score has dropped well below 50\% of $S_{\rm target}$. Thus, neither hardware injection is vetoed using the criteria we have defined.}


\subsubsection{Veto a handful of candidates}
When we inspect a handful of candidates individually, a more comprehensive procedure is preferred, similar to what was done in Ref.~\cite{O3amxp}.
That is, an image should be produced showing the Viterbi scores for a grid of sky positions centered at the candidate position. 
A second image also showing the EPSF, this time from a synthetic signal injected into Gaussian noise at the candidate sky position, should also be produced. 
The search configuration should remain unchanged when creating both images (i.e., observation start time, $T_{\rm coh}$, and $T_{\rm obs}$ should all be the same), and the signal strength of the injection should be chosen such that the recovered signal at the center has roughly the same $S_{\rm target}$ as the CW candidate.
Then, these two images can be compared. Although we would not expect the EPSF of a real CW signal to perfectly match the simulations, the elliptical pattern should, in general, be centered around the same position and have roughly the same shape and orientation. 
If it does not, this candidate may be safely vetoed.

Indeed, in Ref.~\cite{O3amxp}, for the single candidate which remained after passing all the candidates through a hierarchy of vetoes, the candidate's EPSF in real data was not centered on the true sky position of the target and did not match the EPSF produced in Gaussian noise, so the candidate was vetoed.
We verify this procedure in Figure~\ref{fig:off-target_noise}, where the Viterbi scores for a grid of sky positions centered at a candidate position in LIGO O2 data (vetoed as noise in Ref.~\cite{Jones_2021}) and the EPSF for a synthetic signal injected into Gaussian noise at the same sky position are compared. No trace of the EPSF shown in the right panel can be seen in the real data in the left panel. Thus, as expected, we would veto this candidate.

\textcolor{black}{It should be noted that in a real search, data gaps are usually present (due to maintenance, upgrading, etc.). These gaps are generally dealt with by filling the emission probability in each frequency bin with equal probabilities for the periods during which no observational data are available, which will cause a slight decrease in overall sensitivity---akin to using a slightly shorter total observation time---but should not have any major impact on veto procedures such as the EPSFs. (See the sensitivity scaling due to gaps in the real interferometer data in, e.g., Fig. 2 of Ref.~\cite{ScoX1ViterbiO1}.) Nevertheless, for a more rigorous comparison, one can include data gaps in the Gaussian noise simulations.}

\section{Switching off the DM Correction}
\label{sec:DM-off}

\textcolor{black}{Vetoes based on switching on and off the DM correction (i.e., the DM-off veto) have been studied in fully-coherent searches~\cite{Zhu_2017} and used in existing studies~\cite{Jones_2021, Beniwal2021, O3aSNR, O3amxp}.
In this section, we revisit the method and establish veto criteria in the context of semicoherent HMM-based searches. We verify that the veto is safe using Monte Carlo simulations.}

The DM-off veto, as first presented in Ref.~\cite{Zhu_2017}, is a technique in which the DM correction, which accounts for the Doppler shift due to Earth’s orbital and rotational motion, is switched off and the detection statistic is reevaluated and compared to the one obtained with the DM correction applied. 
In a Viterbi search, the DM correction is switched off within the $\mathcal{F}$-statistic calculation. When the DM correction is switched off, the Viterbi score of a signal of astrophysical origin should drop below the threshold $S_{\rm th}$ and a different Viterbi path should be returned.
A candidate resulting from a noise artifact, on the other hand, should yield a higher score (except in the rare case that a noise line wanders in a way that mimics the DM) and a Viterbi path that overlaps with the original.
As such, a candidate is vetoed only if the Viterbi score increases when the DM correction is switched off and an overlapping Viterbi path is returned. In Ref.~\cite{Zhu_2017}, the DM-off veto is described for a coherent $\mathcal{F}$-statistic search.

\subsection{Search configurations}

In order to safely use the DM-off veto in a semicoherent Viterbi search, we need to ensure that a single set of criteria can be applied to various search configurations.
\textcolor{black}{Thus, we test two different $T_{\rm coh}$ values in our search, a longer one of 5~d and a shorter one of 12~hr}, and we use a total observation time of 180~d for the analysis (starting at the same arbitrarily chosen GPS time as in the off-target veto study).

\subsection{Test synthetic signals}

\begin{figure*}[hbt!]
	\centering
	\includegraphics[scale=.6]{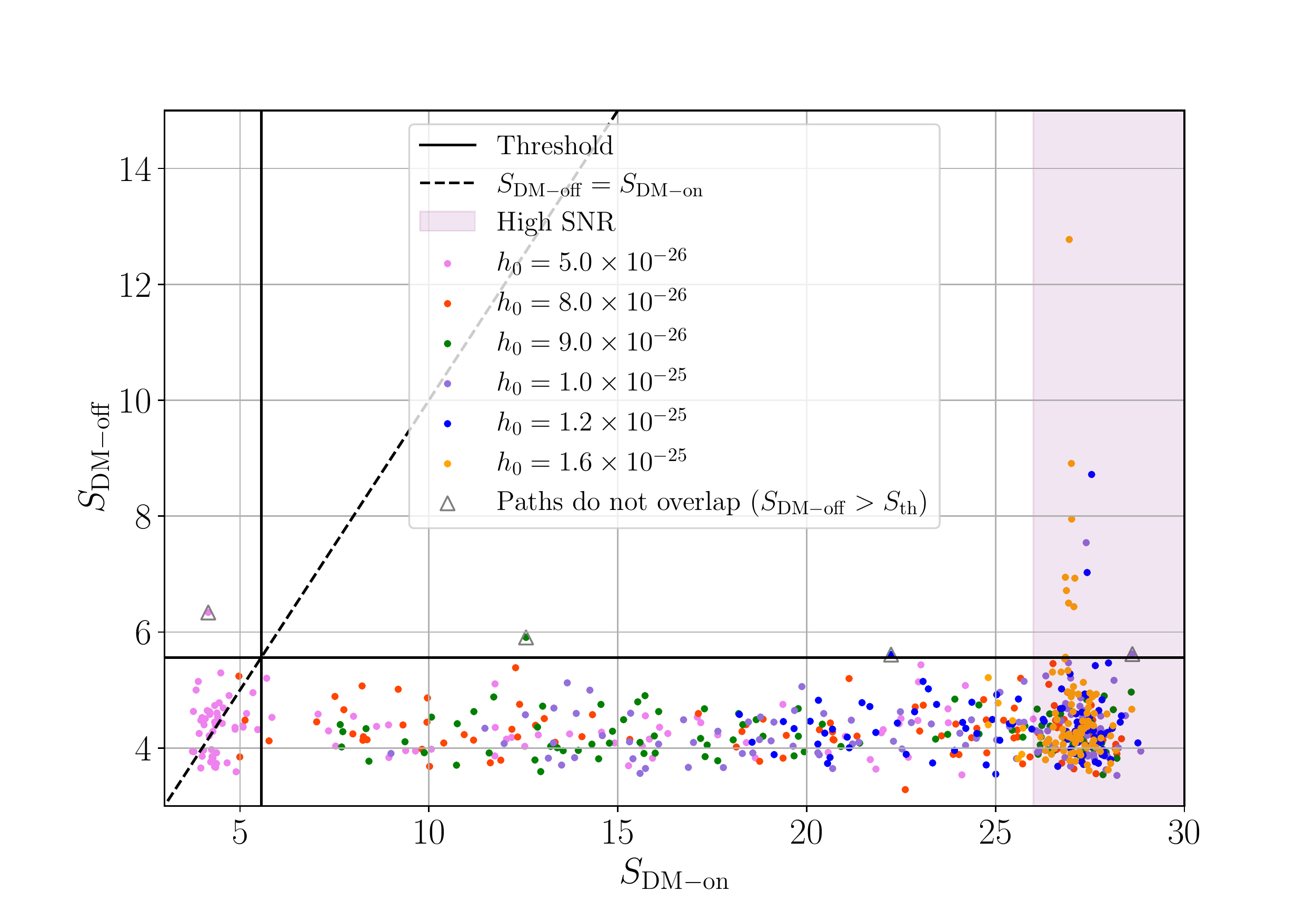}
	\caption{Comparison of the Viterbi scores $S_{\rm DM-on}$ and $S_{\rm DM-off}$ for synthetic signals with $T_{\rm coh}=12$~hr. Each color corresponds to a different signal strength, as indicated in the legend. The solid black lines mark $S_{\rm th}$, the dashed black line marks the diagonal $S_{\rm DM-off} = S_{\rm DM-on}$, and the shaded purple region marks where the signal is so strong that it is detected in the search even without the DM correction applied. Any candidates marked with black triangles are not recovered in the DM-off search (the frequency paths in the DM-on and DM-off runs do not overlap) and thus can be ignored.}
	\label{fig:DM-off12h}
\end{figure*}

\begin{figure*}[hbt!]
	\centering
	\includegraphics[scale=.6]{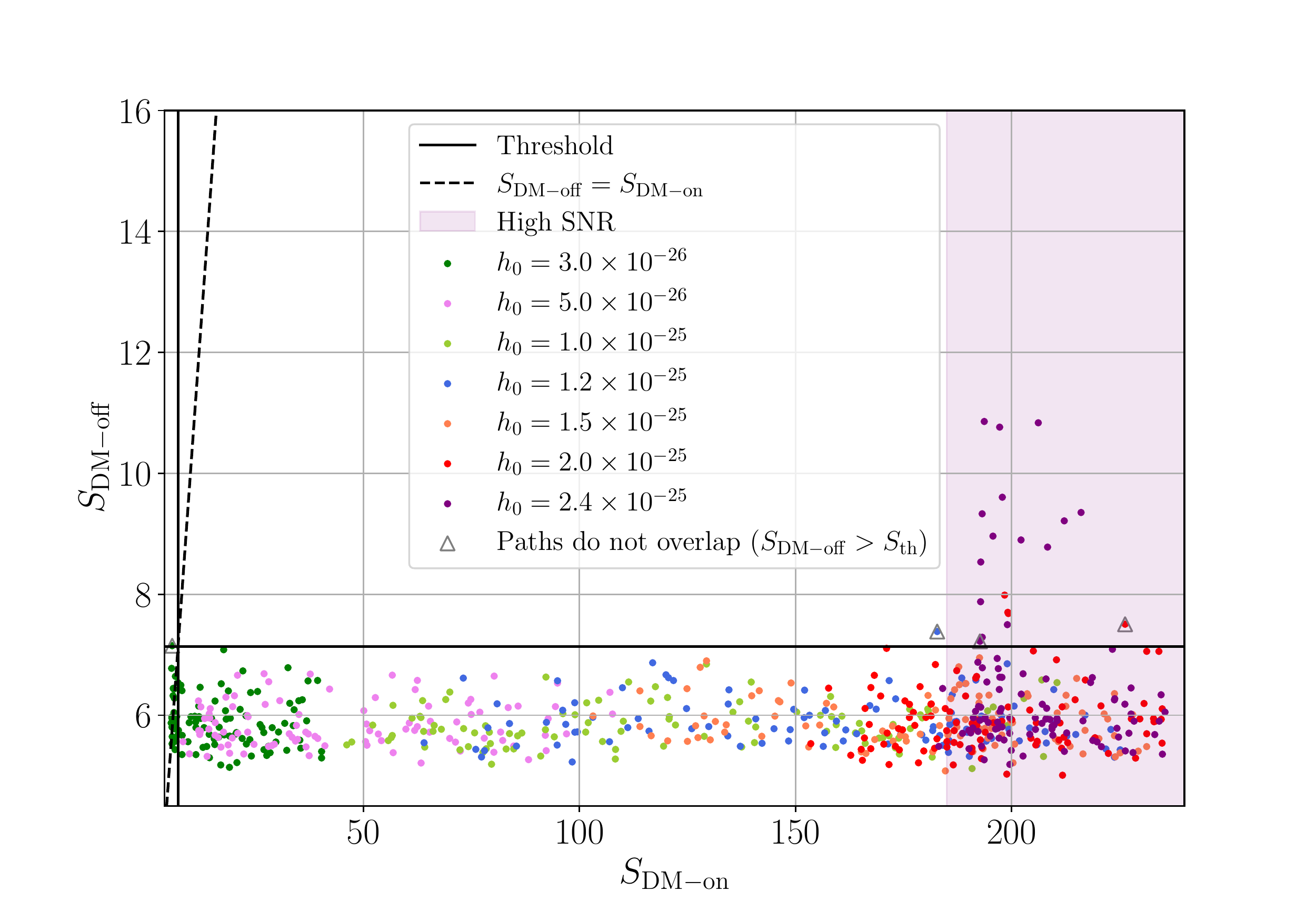}
	\caption{Comparison of the Viterbi scores $S_{\rm DM-on}$ and $S_{\rm DM-off}$ for synthetic signals with $T_{\rm coh}=5$~d. Each color corresponds to a different signal strength, as indicated in the legend. The solid black lines mark $S_{\rm th}$, the dashed black line marks the diagonal $S_{\rm DM-off} = S_{\rm DM-on}$, and the shaded purple region marks where the signal is so strong that it is detected in the search even without the DM correction applied. Any candidates marked with black triangles are not recovered in the DM-off search (the frequency paths in the DM-on and DM-off runs do not overlap) and thus can be ignored.}
	\label{fig:DM-off5d}
\end{figure*}

To establish criteria for the DM-off veto, we run pairs of simulations in which we inject synthetic signals into Gaussian noise and first search with the DM correction applied (DM-on), then with DM-off. We compare the Viterbi scores and paths between the pairs of simulations to see if the candidate present in the DM-on case disappears in the DM-off case, as would be expected for a real astrophysical signal. Figures~\ref{fig:DM-off12h} and~\ref{fig:DM-off5d} show the DM-off score ($S_{\rm DM-off}$) plotted against the DM-on score ($S_{\rm DM-on}$) for $T_{\rm coh}=12$~hr and $T_{\rm coh}=5$~d, respectively, for randomly chosen sky positions distributed uniformly across the sky. A variety of different signal strengths are tested, ranging from $h_0 = 3.0 \times 10^{-26}$ to $h_0 = 2.4 \times 10^{-25}$, as shown in the legends. For each injected signal, we randomly draw $\cos \iota$ from a uniform distribution over the range $[-1, 1]$. In each figure, the horizontal and vertical black lines indicate the same detection threshold $S_{\rm th}$ for that particular $T_{\rm coh}$.

The expected behavior of a synthetic signal is as follows: although $S_{\rm DM-on} > S_{\rm th}$ could lie anywhere along the horizontal axis to the right of the threshold depending on the SNR, $S_{\rm DM-off}$ should fall below the dashed diagonal line (i.e.,  $S_{\rm DM-off} <  S_{\rm DM-on}$). This behavior is confirmed in all of the injections. 
In fact, almost all of the markers fall below the horizontal black line marking $S_{\rm th}$, with several exceptional cases.
Among these exceptional candidates with $S_{\rm DM-on} > S_{\rm DM-off} > S_{\rm th}$, a black triangle marks each candidate whose Viterbi path recovered by the DM-off search does not overlap with the path returned by the DM-on search. When we describe the two paths as ``overlapping" we mean that the DM-off path, widened by $\pm 1 \times 10^{-4} f_0$~Hz to account for the maximum Doppler shift, intersects the DM-on path at some point along its frequency evolution.

We now further discuss the two scenarios that result in the candidates produced in this study falling above the threshold in the DM-off search. 
The first is for those without triangle markers, all found at $S_{\rm DM-on} \gtrsim 27$ for $T_{\rm coh}=12$~hr ($h_0 \geq 1.0 \times 10^{-25}$) and $S_{\rm DM-on} \gtrsim 200$ for $T_{\rm coh}=5$~d ($h_0 \geq 2.0 \times 10^{-25}$), shown within the shaded purple region which marks where the signals are so strong that we do not miss them even without the DM correction applied~\cite{O3aSNR}.
The signal is so loud in this case that it bleeds into nearby frequency bins within the $\mathcal{F}$-statistic, so the likelihood values for the bins around the true frequency bin are all high. When this occurs, even if we do not apply the DM correction in the $\mathcal{F}$-statistic, the total log likelihood is still significant enough to produce a score above the threshold (but lower than the score with the DM correction switched on), and the Viterbi path still roughly tracks the true signal frequency.
Nevertheless, we set the criteria to only veto candidates with $S_{\rm DM-off}$ increased, which we do not see here, so the veto is still reliable for such loud signals.

\begin{figure*}[hbt!]
	\centering
	\includegraphics[scale=.6]{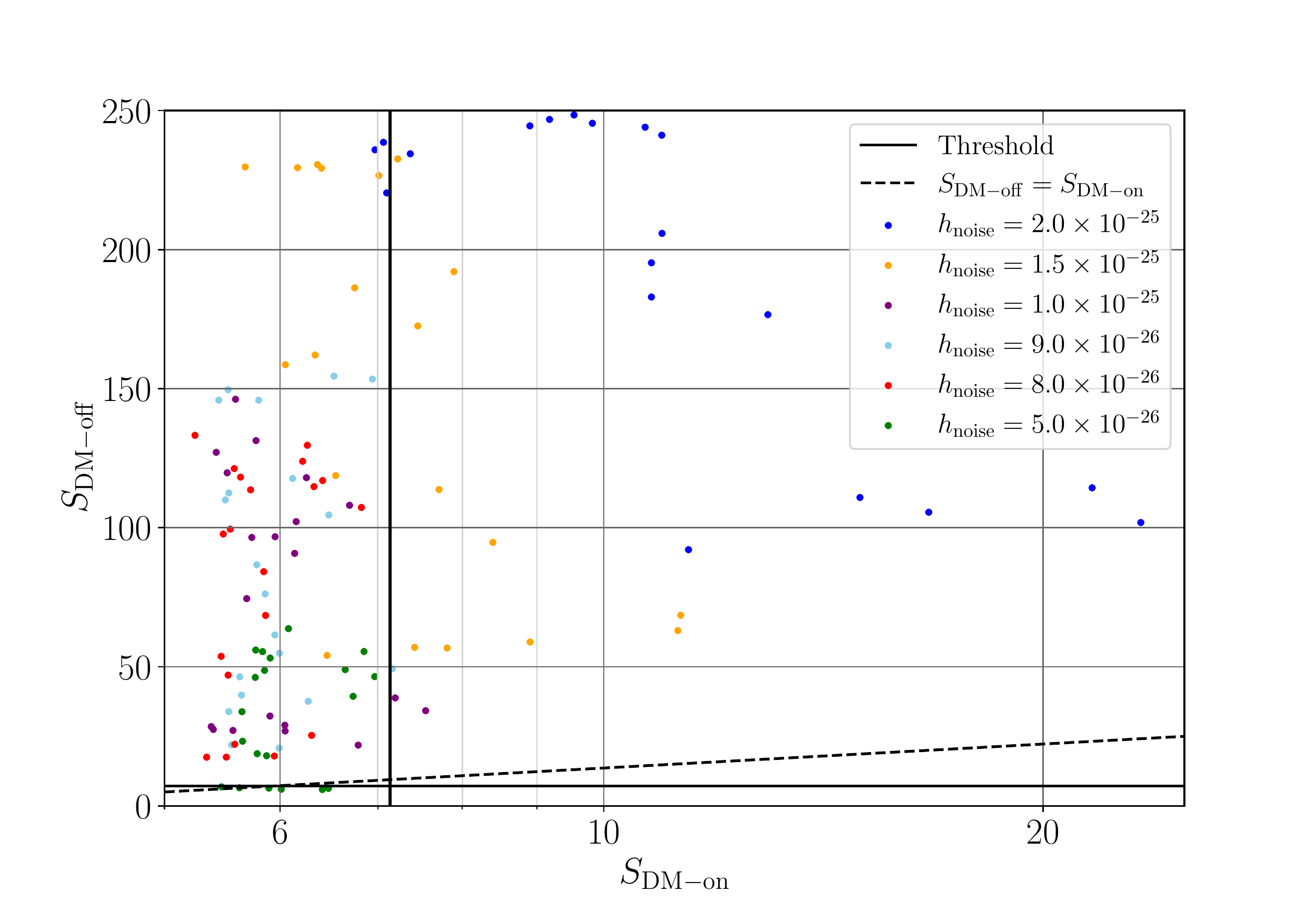}
	\caption{Comparison of the Viterbi scores $S_{\rm DM-on}$ and $S_{\rm DM-off}$ for synthetic noise lines with $T_{\rm coh}=5$~d. Each color corresponds to a different injection strength, as indicated in the legend. The solid black lines mark $S_{\rm th}$, and the dashed black line marks the diagonal $S_{\rm DM-off} = S_{\rm DM-on}$, above which we would expect the outliers caused by noise lines to lie.}
	\label{fig:fake_noise_lines}
\end{figure*}

\textcolor{black}{The second scenario is for those weaker signals to the left of the shaded region but with $S_{\rm DM-off} > S_{\rm th}$. They are all marked by triangles, meaning that the DM-on and DM-off paths do not match each other.
Given that we choose a threshold corresponding to a 1\% false alarm probability per 1-Hz sub-band, these DM-off candidates are consistent with false alarms.}
We note that in both Figures~\ref{fig:DM-off12h} and~\ref{fig:DM-off5d}, one false alarm in the DM-off run (with $S_{\rm DM-off} \approx 6.5$ and 7.0, respectively) does have an increased score relative to $S_{\rm DM-on}$ that is above threshold (with a different Viterbi path), but in fact the injected signal is too weak to be identified as a candidate in the first place (i.e., $S_{\rm DM-on} < S_{\rm th}$).
According to these simulation results, one can even safely veto a candidate if the score remains above $S_{\rm th}$ and the same Viterbi path is returned in the DM-off search, as long as the candidate is not in the high SNR shaded region. 
(In practice, with the current detector sensitivity, candidates found in a real search with Viterbi scores $S > 27$ ($T_{\rm coh}=12$~hr) and $S > 200$ ($T_{\rm coh}=5$~d) are caused by noise artifacts and would most likely be eliminated by another CW veto.)
However, comparing $S_{\rm DM-off}$ to $S_{\rm DM-on}$ rather than the threshold is generally a more conservative option, i.e., to be cautious, we keep the candidates with $S_{\rm DM-on} > S_{\rm DM-off} > S_{\rm th}$ for further scrutiny.
In particular, setting veto criteria by comparing $S_{\rm DM-off}$ to $S_{\rm DM-on}$ rather than $S_{\rm th}$ ensures the veto safety for sources with sky positions close to the ecliptic poles, since the DM correction at these positions would be minimal and so switching off the DM correction would not have much impact on the significance of the Viterbi score. Still, even for these sky positions, we would not expect the Viterbi score to increase when switching the DM correction off. Thus, based on the criteria we have defined, \textcolor{black}{for the 1300 injections tested here}, none are falsely eliminated and the DM-off veto remains safe.

\subsection{Compare with noise}

To provide a baseline for comparison, a series of synthetic monochromatic noise lines at fixed frequency (no DM is added in the simulation code) with different strain amplitudes $h_{\rm noise}$ are injected into the Gaussian noise background in the 200--201~Hz sub-band, and the results of the search are plotted in a similar fashion to the synthetic signals, shown in Figure~\ref{fig:fake_noise_lines}. Noise lines, unlike astrophysical signals, should increase in significance when the DM correction is switched off and lie above the diagonal (dashed line) in the figure. Indeed, all outliers caused by synthetic noise lines show this behavior other than the weakest few with $h_{\rm noise} < 5.0 \times 10^{-26}$ that are below threshold for both the $S_{\rm DM-on}$ and $S_{\rm DM-off}$ searches (these exceptional ones are not identified as candidates in the first place).

\begin{figure*}[hbt!]
	\centering
	\includegraphics[scale=.6]{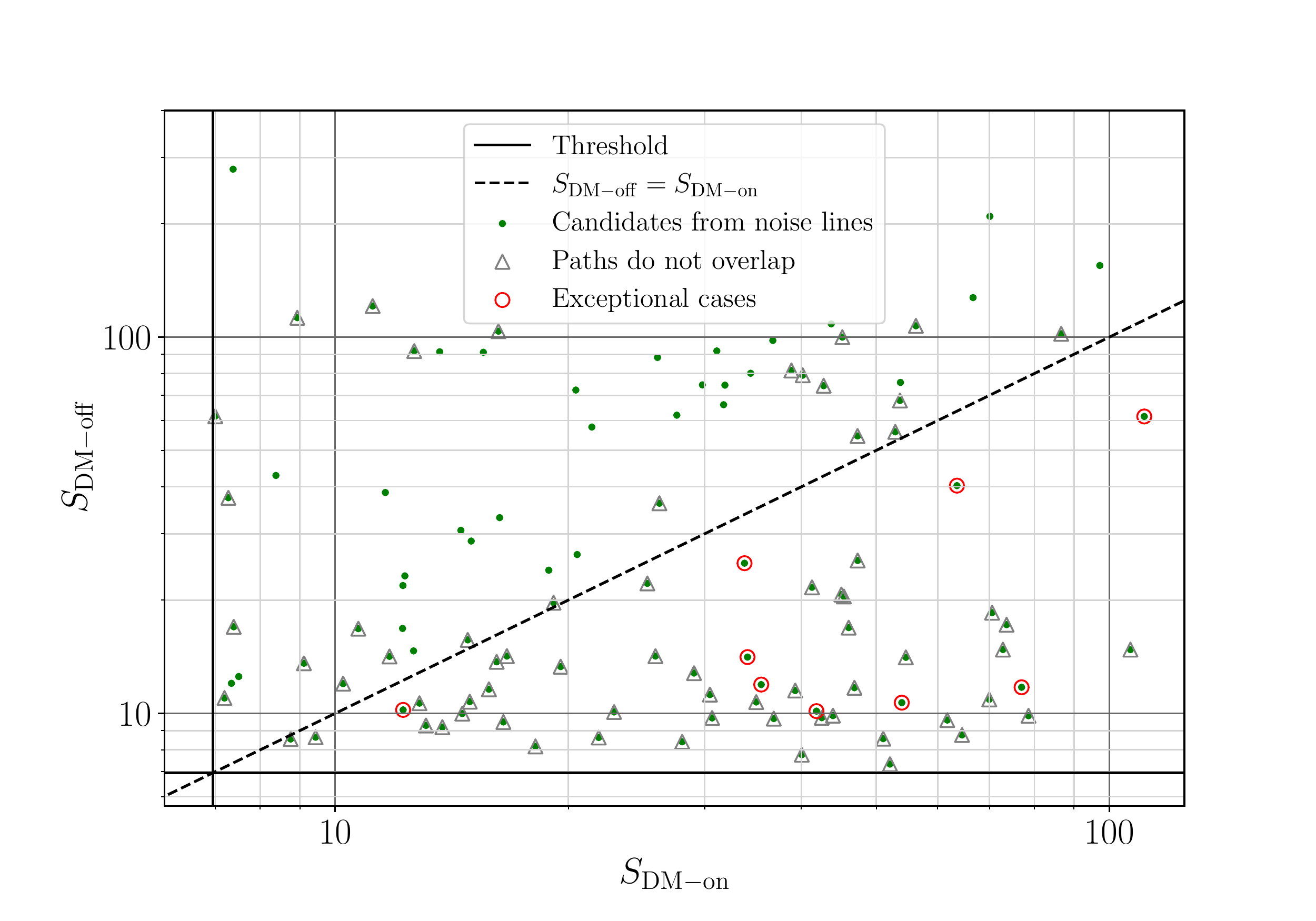}
	\caption{Comparison of the Viterbi scores $S_{\rm DM-on}$ and $S_{\rm DM-off}$ for real noise lines with $T_{\rm coh}=5$~d. The solid black lines mark $S_{\rm th}$, and the dashed black line marks the diagonal $S_{\rm DM-off} = S_{\rm DM-on}$, above which we would expect the outliers caused by noise lines to lie. Candidates marked by black triangles are found at different frequencies in the DM-on and DM-off searches and thus cannot be regarded as a pair of results with and without the DM correction applied. Candidates marked with red circles have overlapping frequency paths with decreased $S_{\rm DM-off}$ and are thus considered exceptional cases (see details in text).}
	\label{fig:noise_lines}
\end{figure*}

\begin{table}[tbh]
	\centering
	\setlength{\tabcolsep}{6pt}
	\renewcommand\arraystretch{1.2}
	\begin{tabular}{cll}
		\hline
		Starting frequency & Detector & Veto stage \\
		\hline
		331.89~Hz & Hanford & Known line \\
		444.55~Hz & Livingston & Known line \\
		468.48~Hz & Livingston & Known line \\
		471.50~Hz & Livingston & Known line \\
		487.40~Hz & Livingston & Known line \\
		492.06~Hz & Hanford & Single interferometer \\
		504.80~Hz & Hanford & Known line \\
		541.68~Hz & Hanford & Single interferometer \\
		543.08~Hz & Hanford & Single interferometer \\
		\hline
	\end{tabular}
	\caption{\textcolor{black}{Starting frequency and detector of origin for the nine noise artifacts in LIGO O2 data that the DM-off veto failed to veto (marked with red circles in Fig.~\ref{fig:noise_lines}). The last column lists the veto stage at which they are eliminated in the search presented in Ref.~\cite{Jones_2021}.}}
	\label{tab:DM-off_real_noise}
\end{table}

The behavior caused by noise lines is also confirmed by studying the real noise lines from Advanced LIGO O2 (selected from the candidates identified as noise artifacts in Ref.~\cite{Jones_2021}). The search results are shown in Figure~\ref{fig:noise_lines}.
Although quite a few candidates caused by noise lines lie below the dashed line, many do not have overlapping frequency paths between the DM-on and DM-off runs (signified with black triangles), meaning that the original candidates are not recovered by the DM-off runs, so they cannot be evaluated in the DM-off veto procedure. There are nine candidates caused by noise lines that fall below the dashed line that do have overlapping frequency paths, marked with red circles. These artifacts are inspected individually; \textcolor{black}{Table~\ref{tab:DM-off_real_noise} presents a summary of their main attributes, namely the starting frequency for each artifact, which detector the artifact originates in, and at what stage in the veto hierarchy (as detailed in Ref.~\cite{Jones_2021}) it is eliminated.}
\textcolor{black}{There are a couple of theories as to why these noise artifacts are not eliminated by the DM-off veto:} (i) the noise lines wander in a way that mimics the Doppler modulation in an astrophysical signal; or (ii) when we integrate data from both detectors and switch off the DM correction, the noise artifact at the candidate frequency remains the most significant, but the likelihoods in other frequency bins happen to increase, decreasing the relative significance of the original artifact. In a real search, these candidates would not be eliminated by the DM-off veto because they do not satisfy the veto criteria, so they would be followed up using other methods. \textcolor{black}{In fact, as shown in Table~\ref{tab:DM-off_real_noise}, all nine candidates are eliminated by one of the two initial CW vetoes---the known-line veto, which identifies candidates as known instrumental lines, and the single-interferometer veto, which identifies candidates as previously unknown noise artifacts present in only one detector---so in all likelihood artifacts such as these would not even reach the stage wherein they would be analyzed using DM-off. In summation, regardless of the small number of false positives (nine out of 108) when dealing with noise artifacts, the DM-off veto remains safe in that it does not falsely reject any of the 1300 synthetic signals we have tested above.}

\section{Conclusion}
\label{sec:conclusion}

In this study, we investigate the off-target veto and the DM-off veto in order to establish veto criteria for a semicoherent CW search.
\textcolor{black}{To aid the design of generalized and rigorous veto criteria for the off-target veto, we introduce an EPSF, discuss its physical insights, and demonstrate the applications.}
We draw conclusions about the safety of the DM-based vetoes through Monte Carlo simulations in which synthetic signals are injected into Gaussian noise. We use a combination of the coherent $\mathcal{F}$-statistic and an HMM scheme to track the signal, and we take the Viterbi score as our primary detection statistic.

We show that the off-target veto can be used in two different ways to follow up CW candidates depending on how many remain to be investigated. If many candidates from various sky positions need to be processed at once, a single Dec offset which is safe for all sky positions can be chosen such that if the Viterbi score at this offset remains above 50\% of the original score and above $S_{\rm th}$, and the recovered Viterbi path overlaps the original candidate path, the candidate is vetoed.
On the other hand, if only a few candidates require follow-up, a more detailed investigation can be carried out by calculating and comparing the EPSFs around the candidate location using both injections in Gaussian noise and the real data.
A candidate is vetoed if these two patterns do not generally agree.
Thus, the EPSF is used as a precise marker of an astrophysical signal, expanding the applicability of the off-target veto.

Furthermore, we demonstrate that the DM-off veto is safe  \textcolor{black}{in the HMM-based semicoherent searches} for all configurations tested (i.e., for all sky positions and for both a short coherent length of 12 hr and a longer coherent length of 5 d). Although the veto is not able to eliminate every noise line, the majority are successfully eliminated. More importantly,  \textcolor{black}{no synthetic signals out of 1300 are falsely eliminated in any configuration tested.}

With the conclusion of the third observing run of Advanced LIGO and Virgo and the impending fourth observing run, we will see continued improvements in detector sensitivity over the next few years. As a result, we expect to see searches produce more CW candidates (because the data are more likely to be polluted by weak noise artifacts), and these candidates will become more time-consuming to verify or veto. 
This study verifies the safety of the off-target and DM-off vetoes and proposes a refined veto procedure in preparation for future CW searches and the first CW detection.

\section{Acknowledgments}
We thank Evan Goetz for the helpful review and suggestions during the LIGO Publications and Presentations (P\&P) review procedure.
This material is based upon work supported by the United States National Science Foundation (NSF)'s LIGO Laboratory, which is a major facility fully funded by NSF.
The authors are grateful for computational resources provided by the LIGO Laboratory and supported by the NSF Grants PHY--0757058 and PHY--0823459.
The authors acknowledge the support of the Australian Research Council Centre of Excellence for Gravitational Wave Discovery (OzGrav), Project No. CE170100004.
The authors would like to thank all of the essential workers who put their health at risk during the COVID--19 pandemic, without whom we would not have been able to complete this work.
This paper carries LIGO Document Number LIGO--P2200057.

\onecolumngrid
\appendix
\section{Signal frequency}
\label{appendix:signal_freq}

We simplify the CW phase shown in Eq.~\eqref{eqn:phase2} by omitting the term that accounts for the DM effect due to Earth's rotation, which is two orders of magnitude smaller than the effect of Earth's orbital motion (i.e., $\Omega_0 R_{ES} / c \sim 1.0 \times 10^{-4}$, compared to $\Omega_R R_E / c \sim 1.5 \times 10^{-6}$). The approximate phase becomes
\begin{eqnarray}
	\nonumber
    \Psi(t) &\approx& \Phi_0 + 2\pi \sum_{k=0}^{s} \frac{f_0^{(k)}t^{k+1}}{(k+1)!} \\
    &&+ \frac{2\pi}{c} \{R_{ES}[\cos \alpha \cos \delta \cos(\phi_0 + \Omega_0 t) 
    + (\cos \epsilon \sin \alpha \cos \delta + \sin \epsilon \sin \delta) \sin(\phi_0 + \Omega_0 t)]\} \sum_{k=0}^{s}\frac{f_0^{(k)}t^k}{k!}.
\end{eqnarray}
We can then write the signal frequency as
\begin{eqnarray}
	\nonumber
	f(t) &\approx& (f_0 + f_0^{(1)}t) + \frac{R_{ES} \Omega_0}{c}[ -\cos \alpha \cos \delta \sin(\phi_0 + \Omega_0 t) 
	+ (\cos \epsilon \sin \alpha \cos \delta + \sin \epsilon \sin \delta) \cos(\phi_0 + \Omega_0 t)](f_0 + f_0^{(1)}t) \\
	&& + \frac{R_{ES}}{c}[\cos \alpha \cos \delta \cos(\phi_0 + \Omega_0 t) 
    + (\cos \epsilon \sin \alpha \cos \delta + \sin \epsilon \sin \delta) \sin(\phi_0 + \Omega_0 t)]f_0^{(1)},
    \label{eq:approx_freq}
\end{eqnarray}
where we have omitted higher order frequency derivative terms $f_0^{(k)}$ with $k \geq 2$. Considering the fact that $\Omega_0 f_0 \gg f_0^{(1)}$ in the parameter space searched, the last term in Eq.~\eqref{eq:approx_freq} can be omitted, and we obtain
\begin{eqnarray}
	f(t) \approx (f_0 + f_0^{(1)}t) \{1+\frac{R_{ES} \Omega_0}{c} [ -\cos \alpha \cos \delta \sin(\phi_0 + \Omega_0 t) 
	+ (\cos \epsilon \sin \alpha \cos \delta + \sin \epsilon \sin \delta) \cos(\phi_0 + \Omega_0 t)] \},
\end{eqnarray}
i.e., Eq.~\eqref{eqn:frequency} in Section~\ref{sec:doppler_pattern}.

\section{\textcolor{black}{EPSF in a semicoherent cross-correlation search}}
\label{sec:off-target_cross-corr}

\textcolor{black}{To demonstrate that the EPSF and the off-target veto procedure can be generalized to other stack-slide-based semicoherent search methods, we run an analysis using another method, the cross-correlation method (detailed in Refs.~\cite{Dhurandhar2008,Chung2011,Crosscorr}). A similar EPSF image is produced to those output in the HMM-based analyses.}

\textcolor{black}{In a typical semicoherent search, detector data is divided into 30-minute time segments and a short Fourier transform (SFT) is computed for each segment. When using the cross-correlation search pipeline, these SFTs are multiplied pairwise (where these pairs are chosen according to criteria that may involve incorporating data from multiple detectors and establishing a maximum time lag), yielding a cross-correlation variable for each SFT pair. The detection statistic $\rho$ output by the search represents a weighted sum of these cross-correlation variables over all SFT pairs~\cite{Crosscorr}.}

\begin{figure*}[hbt!]
	\centering
	\includegraphics[scale=.5]{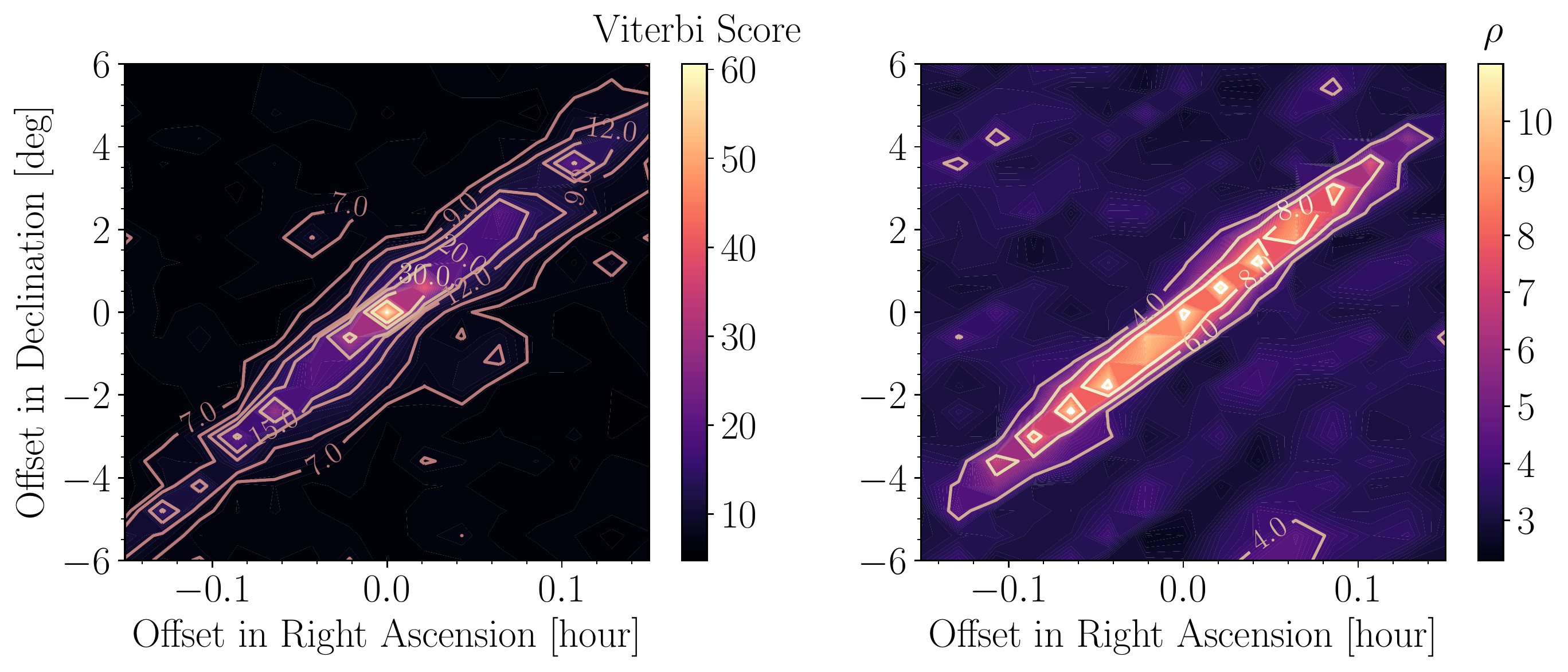}
	\caption{\textcolor{black}{Comparison of the EPSFs generated using the HMM method and the cross-correlation method. Left: Contour of Viterbi scores as a function of the offset in RA and Dec for an injection in Gaussian noise at RA = 12~h 00~m 00~s and Dec = $00\degree$ $00'$ $00''$ with signal strength $h_0 = 2.0 \times 10^{-26}$ ($\cos\iota =1$) ($T_{\rm coh}=5$~d). Right: Contour of $\rho$ (used as the detection statistic in the cross-correlation method) as a function of the offset in RA and Dec for the same injection. (The maximum time lag is set to 1 hr.) Both are for $T_{\rm obs}=180$~d.}}
	\label{fig:off-target_cc}
\end{figure*}

\textcolor{black}{Figure~\ref{fig:off-target_cc} shows a side-by-side comparison of a synthetic signal with the same parameters (see figure caption) injected into Gaussian noise and recovered using the HMM method (left) and the cross-correlation method (right). Although the two search methods use different detection statistics ($S$ and $\rho$, respectively), they produce very similar EPSFs, demonstrating that the off-target veto methods detailed here are indeed generalizable to other semicoherent searches.}

\twocolumngrid

\bibliography{CW}

\begin{thebibliography}{58}%
\makeatletter
\providecommand \@ifxundefined [1]{%
 \@ifx{#1\undefined}
}%
\providecommand \@ifnum [1]{%
 \ifnum #1\expandafter \@firstoftwo
 \else \expandafter \@secondoftwo
 \fi
}%
\providecommand \@ifx [1]{%
 \ifx #1\expandafter \@firstoftwo
 \else \expandafter \@secondoftwo
 \fi
}%
\providecommand \natexlab [1]{#1}%
\providecommand \enquote  [1]{``#1''}%
\providecommand \bibnamefont  [1]{#1}%
\providecommand \bibfnamefont [1]{#1}%
\providecommand \citenamefont [1]{#1}%
\providecommand \href@noop [0]{\@secondoftwo}%
\providecommand \href [0]{\begingroup \@sanitize@url \@href}%
\providecommand \@href[1]{\@@startlink{#1}\@@href}%
\providecommand \@@href[1]{\endgroup#1\@@endlink}%
\providecommand \@sanitize@url [0]{\catcode `\\12\catcode `\$12\catcode
  `\&12\catcode `\#12\catcode `\^12\catcode `\_12\catcode `\%12\relax}%
\providecommand \@@startlink[1]{}%
\providecommand \@@endlink[0]{}%
\providecommand \url  [0]{\begingroup\@sanitize@url \@url }%
\providecommand \@url [1]{\endgroup\@href {#1}{\urlprefix }}%
\providecommand \urlprefix  [0]{URL }%
\providecommand \Eprint [0]{\href }%
\providecommand \doibase [0]{http://dx.doi.org/}%
\providecommand \selectlanguage [0]{\@gobble}%
\providecommand \bibinfo  [0]{\@secondoftwo}%
\providecommand \bibfield  [0]{\@secondoftwo}%
\providecommand \translation [1]{[#1]}%
\providecommand \BibitemOpen [0]{}%
\providecommand \bibitemStop [0]{}%
\providecommand \bibitemNoStop [0]{.\EOS\space}%
\providecommand \EOS [0]{\spacefactor3000\relax}%
\providecommand \BibitemShut  [1]{\csname bibitem#1\endcsname}%
\let\auto@bib@innerbib\@empty
\bibitem [{\citenamefont {Abbott}\ \emph {et~al.}(2016)\citenamefont {Abbott}
  \emph {et~al.}}]{Abbott2016}%
  \BibitemOpen
  \bibfield  {author} {\bibinfo {author} {\bibfnamefont {B.~P.}\ \bibnamefont
  {Abbott}} \emph {et~al.} (\bibinfo {collaboration} {LIGO Scientific
  Collaboration and Virgo Collaboration}),\ }\bibfield  {title} {\enquote
  {\bibinfo {title} {Observation of gravitational waves from a binary black
  hole merger},}\ }\href {\doibase 10.1103/PhysRevLett.116.061102} {\bibfield
  {journal} {\bibinfo  {journal} {Phys. Rev. Lett.}\ }\textbf {\bibinfo
  {volume} {116}},\ \bibinfo {pages} {061102} (\bibinfo {year}
  {2016})}\BibitemShut {NoStop}%
\bibitem [{\citenamefont {Aasi}\ \emph {et~al.}(2015)\citenamefont {Aasi} \emph
  {et~al.}}]{LIGO2014}%
  \BibitemOpen
  \bibfield  {author} {\bibinfo {author} {\bibfnamefont {J.}~\bibnamefont
  {Aasi}} \emph {et~al.} (\bibinfo {collaboration} {LSC}),\ }\bibfield  {title}
  {\enquote {\bibinfo {title} {{Advanced LIGO}},}\ }\href {\doibase
  10.1088/0264-9381/32/7/074001} {\bibfield  {journal} {\bibinfo  {journal}
  {Classical and Quantum Gravity}\ }\textbf {\bibinfo {volume} {32}},\ \bibinfo
  {pages} {074001} (\bibinfo {year} {2015})}\BibitemShut {NoStop}%
\bibitem [{\citenamefont {Acernese}\ \emph {et~al.}(2015)\citenamefont
  {Acernese} \emph {et~al.}}]{Virgo2014}%
  \BibitemOpen
  \bibfield  {author} {\bibinfo {author} {\bibfnamefont {F.}~\bibnamefont
  {Acernese}} \emph {et~al.} (\bibinfo {collaboration} {Virgo}),\ }\bibfield
  {title} {\enquote {\bibinfo {title} {{Advanced Virgo: A second-generation
  interferometric gravitational wave detector}},}\ }\href {\doibase
  10.1088/0264-9381/32/2/024001} {\bibfield  {journal} {\bibinfo  {journal}
  {Classical and Quantum Gravity}\ }\textbf {\bibinfo {volume} {32}},\ \bibinfo
  {pages} {024001} (\bibinfo {year} {2015})}\BibitemShut {NoStop}%
\bibitem [{\citenamefont {Akutsu}\ \emph {et~al.}(2020)\citenamefont {Akutsu}
  \emph {et~al.}}]{akutsu2020overview}%
  \BibitemOpen
  \bibfield  {author} {\bibinfo {author} {\bibfnamefont {T.}~\bibnamefont
  {Akutsu}} \emph {et~al.},\ }\bibfield  {title} {\enquote {\bibinfo {title}
  {{Overview of KAGRA: Detector design and construction history}},}\
  }\href@noop {} {\  (\bibinfo {year} {2020})},\ \Eprint
  {http://arxiv.org/abs/2005.05574} {arXiv:2005.05574 [physics.ins-det]}
  \BibitemShut {NoStop}%
\bibitem [{\citenamefont {Abbott}\ \emph
  {et~al.}(2019{\natexlab{a}})\citenamefont {Abbott} \emph
  {et~al.}}]{Abbott2018}%
  \BibitemOpen
  \bibfield  {author} {\bibinfo {author} {\bibfnamefont {B.~P.}\ \bibnamefont
  {Abbott}} \emph {et~al.} (\bibinfo {collaboration} {LIGO Scientific
  Collaboration and Virgo Collaboration}),\ }\bibfield  {title} {\enquote
  {\bibinfo {title} {{GWTC-1: A gravitational-wave transient catalog of compact
  binary mergers observed by LIGO and Virgo during the first and second
  observing runs}},}\ }\href {\doibase 10.1103/PhysRevX.9.031040} {\bibfield
  {journal} {\bibinfo  {journal} {Phys. Rev. X}\ }\textbf {\bibinfo {volume}
  {9}},\ \bibinfo {pages} {031040} (\bibinfo {year}
  {2019}{\natexlab{a}})}\BibitemShut {NoStop}%
\bibitem [{\citenamefont {Abbott}\ \emph
  {et~al.}(2021{\natexlab{a}})\citenamefont {Abbott} \emph
  {et~al.}}]{Abbott2021gwtc2}%
  \BibitemOpen
  \bibfield  {author} {\bibinfo {author} {\bibfnamefont {R.}~\bibnamefont
  {Abbott}} \emph {et~al.} (\bibinfo {collaboration} {LIGO Scientific
  Collaboration and Virgo Collaboration}),\ }\bibfield  {title} {\enquote
  {\bibinfo {title} {{GWTC-2: Compact binary coalescences observed by LIGO and
  Virgo during the first half of the third observing run}},}\ }\href {\doibase
  10.1103/PhysRevX.11.021053} {\bibfield  {journal} {\bibinfo  {journal} {Phys.
  Rev. X}\ }\textbf {\bibinfo {volume} {11}},\ \bibinfo {pages} {021053}
  (\bibinfo {year} {2021}{\natexlab{a}})}\BibitemShut {NoStop}%
\bibitem [{\citenamefont {Abbott}\ \emph
  {et~al.}(2021{\natexlab{b}})\citenamefont {Abbott} \emph {et~al.}}]{gwtc3}%
  \BibitemOpen
  \bibfield  {author} {\bibinfo {author} {\bibfnamefont {R.}~\bibnamefont
  {Abbott}} \emph {et~al.},\ }\bibfield  {title} {\enquote {\bibinfo {title}
  {{GWTC-3: Compact binary coalescences observed by LIGO and Virgo during the
  second part of the third observing run}},}\ }\href@noop {} {\  (\bibinfo
  {year} {2021}{\natexlab{b}})},\ \Eprint {http://arxiv.org/abs/2111.03606}
  {arXiv:2111.03606 [gr-qc]} \BibitemShut {NoStop}%
\bibitem [{\citenamefont {Lasky}(2015)}]{lasky2015}%
  \BibitemOpen
  \bibfield  {author} {\bibinfo {author} {\bibfnamefont {P.~D.}\ \bibnamefont
  {Lasky}},\ }\bibfield  {title} {\enquote {\bibinfo {title} {Gravitational
  waves from neutron stars: A review},}\ }\href {\doibase 10.1017/pasa.2015.35}
  {\bibfield  {journal} {\bibinfo  {journal} {Publications of the Astronomical
  Society of Australia}\ }\textbf {\bibinfo {volume} {32}},\ \bibinfo {pages}
  {e034} (\bibinfo {year} {2015})}\BibitemShut {NoStop}%
\bibitem [{\citenamefont {Riles}(2017)}]{Riles2017}%
  \BibitemOpen
  \bibfield  {author} {\bibinfo {author} {\bibfnamefont {K.}~\bibnamefont
  {Riles}},\ }\bibfield  {title} {\enquote {\bibinfo {title} {Recent searches
  for continuous gravitational waves},}\ }\href {\doibase
  10.1142/S021773231730035X} {\bibfield  {journal} {\bibinfo  {journal} {Mod.
  Phys. Lett. A}\ }\textbf {\bibinfo {volume} {32}},\ \bibinfo {pages}
  {1730035} (\bibinfo {year} {2017})}\BibitemShut {NoStop}%
\bibitem [{\citenamefont {Tenorio}\ \emph
  {et~al.}(2021{\natexlab{a}})\citenamefont {Tenorio}, \citenamefont {Keitel},\
  and\ \citenamefont {Sintes}}]{Tenorio2021}%
  \BibitemOpen
  \bibfield  {author} {\bibinfo {author} {\bibfnamefont {R.}~\bibnamefont
  {Tenorio}}, \bibinfo {author} {\bibfnamefont {D.}~\bibnamefont {Keitel}}, \
  and\ \bibinfo {author} {\bibfnamefont {A.~M.}\ \bibnamefont {Sintes}},\
  }\bibfield  {title} {\enquote {\bibinfo {title} {{Search methods for
  continuous gravitational-wave signals from unknown sources in the
  advanced-detector era}},}\ }\href {\doibase 10.3390/universe7120474}
  {\bibfield  {journal} {\bibinfo  {journal} {Universe}\ }\textbf {\bibinfo
  {volume} {7}},\ \bibinfo {pages} {474} (\bibinfo {year}
  {2021}{\natexlab{a}})},\ \Eprint {http://arxiv.org/abs/2111.12575}
  {arXiv:2111.12575 [gr-qc]} \BibitemShut {NoStop}%
\bibitem [{\citenamefont {Piccinni}(2022)}]{Piccinni2022}%
  \BibitemOpen
  \bibfield  {author} {\bibinfo {author} {\bibfnamefont {O.~J.}\ \bibnamefont
  {Piccinni}},\ }\bibfield  {title} {\enquote {\bibinfo {title} {Status and
  perspectives of continuous gravitational wave searches},}\ }\href@noop {} {\
  (\bibinfo {year} {2022})},\ \Eprint {http://arxiv.org/abs/2202.01088}
  {arXiv:2202.01088 [gr-qc]} \BibitemShut {NoStop}%
\bibitem [{\citenamefont {Hobbs}\ \emph {et~al.}(2010)\citenamefont {Hobbs},
  \citenamefont {Lyne},\ and\ \citenamefont {Kramer}}]{Hobbs2010}%
  \BibitemOpen
  \bibfield  {author} {\bibinfo {author} {\bibfnamefont {G.}~\bibnamefont
  {Hobbs}}, \bibinfo {author} {\bibfnamefont {A.~G.}\ \bibnamefont {Lyne}}, \
  and\ \bibinfo {author} {\bibfnamefont {M.}~\bibnamefont {Kramer}},\
  }\bibfield  {title} {\enquote {\bibinfo {title} {An analysis of the timing
  irregularities for 366 pulsars},}\ }\href {\doibase
  10.1111/j.1365-2966.2009.15938.x} {\bibfield  {journal} {\bibinfo  {journal}
  {Monthly Notices of the Royal Astronomical Society}\ }\textbf {\bibinfo
  {volume} {402}},\ \bibinfo {pages} {1027--1048} (\bibinfo {year}
  {2010})}\BibitemShut {NoStop}%
\bibitem [{\citenamefont {Mukherjee}\ \emph {et~al.}(2018)\citenamefont
  {Mukherjee}, \citenamefont {Messenger},\ and\ \citenamefont
  {Riles}}]{Mukherjee2018}%
  \BibitemOpen
  \bibfield  {author} {\bibinfo {author} {\bibfnamefont {A.}~\bibnamefont
  {Mukherjee}}, \bibinfo {author} {\bibfnamefont {C.}~\bibnamefont
  {Messenger}}, \ and\ \bibinfo {author} {\bibfnamefont {K.}~\bibnamefont
  {Riles}},\ }\bibfield  {title} {\enquote {\bibinfo {title}
  {{Accretion-induced spin-wandering effects on the neutron star in Scorpius
  X-1: Implications for continuous gravitational wave searches}},}\ }\href
  {\doibase 10.1103/PhysRevD.97.043016} {\bibfield  {journal} {\bibinfo
  {journal} {Phys. Rev. D}\ }\textbf {\bibinfo {volume} {97}},\ \bibinfo
  {pages} {043016} (\bibinfo {year} {2018})}\BibitemShut {NoStop}%
\bibitem [{\citenamefont {Ashton}\ \emph {et~al.}(2015)\citenamefont {Ashton},
  \citenamefont {Jones},\ and\ \citenamefont {Prix}}]{Ashton2015}%
  \BibitemOpen
  \bibfield  {author} {\bibinfo {author} {\bibfnamefont {G.}~\bibnamefont
  {Ashton}}, \bibinfo {author} {\bibfnamefont {D.~I.}\ \bibnamefont {Jones}}, \
  and\ \bibinfo {author} {\bibfnamefont {R.}~\bibnamefont {Prix}},\ }\bibfield
  {title} {\enquote {\bibinfo {title} {Effect of timing noise on targeted and
  narrow-band coherent searches for continuous gravitational waves from
  pulsars},}\ }\href {\doibase 10.1103/PhysRevD.91.062009} {\bibfield
  {journal} {\bibinfo  {journal} {Phys. Rev. D}\ }\textbf {\bibinfo {volume}
  {91}},\ \bibinfo {pages} {062009} (\bibinfo {year} {2015})}\BibitemShut
  {NoStop}%
\bibitem [{\citenamefont {Sun}\ \emph {et~al.}(2018)\citenamefont {Sun},
  \citenamefont {Melatos}, \citenamefont {Suvorova}, \citenamefont {Moran},\
  and\ \citenamefont {Evans}}]{Sun2018-2}%
  \BibitemOpen
  \bibfield  {author} {\bibinfo {author} {\bibfnamefont {L.}~\bibnamefont
  {Sun}}, \bibinfo {author} {\bibfnamefont {A.}~\bibnamefont {Melatos}},
  \bibinfo {author} {\bibfnamefont {S.}~\bibnamefont {Suvorova}}, \bibinfo
  {author} {\bibfnamefont {W.}~\bibnamefont {Moran}}, \ and\ \bibinfo {author}
  {\bibfnamefont {R.~J.}\ \bibnamefont {Evans}},\ }\bibfield  {title} {\enquote
  {\bibinfo {title} {{Hidden Markov model tracking of continuous gravitational
  waves from young supernova remnants}},}\ }\href {\doibase
  10.1103/PhysRevD.97.043013} {\bibfield  {journal} {\bibinfo  {journal} {Phys.
  Rev. D}\ }\textbf {\bibinfo {volume} {97}},\ \bibinfo {pages} {043013}
  (\bibinfo {year} {2018})}\BibitemShut {NoStop}%
\bibitem [{\citenamefont {Suvorova}\ \emph {et~al.}(2016)\citenamefont
  {Suvorova}, \citenamefont {Sun}, \citenamefont {Melatos}, \citenamefont
  {Moran},\ and\ \citenamefont {Evans}}]{Suvorova2016}%
  \BibitemOpen
  \bibfield  {author} {\bibinfo {author} {\bibfnamefont {S.}~\bibnamefont
  {Suvorova}}, \bibinfo {author} {\bibfnamefont {L.}~\bibnamefont {Sun}},
  \bibinfo {author} {\bibfnamefont {A.}~\bibnamefont {Melatos}}, \bibinfo
  {author} {\bibfnamefont {W.}~\bibnamefont {Moran}}, \ and\ \bibinfo {author}
  {\bibfnamefont {Robin~J.}\ \bibnamefont {Evans}},\ }\bibfield  {title}
  {\enquote {\bibinfo {title} {{Hidden Markov model tracking of continuous
  gravitational waves from a neutron star with wandering spin}},}\ }\href
  {\doibase 10.1103/PhysRevD.93.123009} {\bibfield  {journal} {\bibinfo
  {journal} {Physical Review D}\ }\textbf {\bibinfo {volume} {93}},\ \bibinfo
  {pages} {123009} (\bibinfo {year} {2016})}\BibitemShut {NoStop}%
\bibitem [{\citenamefont {Abbott}\ \emph {et~al.}(2017)\citenamefont {Abbott}
  \emph {et~al.}}]{ScoX1ViterbiO1}%
  \BibitemOpen
  \bibfield  {author} {\bibinfo {author} {\bibfnamefont {B.~P.}\ \bibnamefont
  {Abbott}} \emph {et~al.} (\bibinfo {collaboration} {LIGO Scientific
  Collaboration and Virgo Collaboration}),\ }\bibfield  {title} {\enquote
  {\bibinfo {title} {{Search for gravitational waves from Scorpius X-1 in the
  first Advanced LIGO observing run with a hidden Markov model}},}\ }\href@noop
  {} {\bibfield  {journal} {\bibinfo  {journal} {Phys. Rev. D}\ }\textbf
  {\bibinfo {volume} {95}},\ \bibinfo {pages} {122003} (\bibinfo {year}
  {2017})}\BibitemShut {NoStop}%
\bibitem [{\citenamefont {Sun}\ and\ \citenamefont {Melatos}(2019)}]{Sun2019}%
  \BibitemOpen
  \bibfield  {author} {\bibinfo {author} {\bibfnamefont {L.}~\bibnamefont
  {Sun}}\ and\ \bibinfo {author} {\bibfnamefont {A.}~\bibnamefont {Melatos}},\
  }\bibfield  {title} {\enquote {\bibinfo {title} {{Application of hidden
  Markov model tracking to the search for long-duration transient gravitational
  waves from the remnant of the binary neutron star merger GW170817}},}\ }\href
  {\doibase 10.1103/PhysRevD.99.123003} {\bibfield  {journal} {\bibinfo
  {journal} {Phys. Rev. D}\ }\textbf {\bibinfo {volume} {99}},\ \bibinfo
  {pages} {123003} (\bibinfo {year} {2019})}\BibitemShut {NoStop}%
\bibitem [{\citenamefont {Abbott}\ \emph
  {et~al.}(2019{\natexlab{b}})\citenamefont {Abbott} \emph
  {et~al.}}]{Abbott2019-2}%
  \BibitemOpen
  \bibfield  {author} {\bibinfo {author} {\bibfnamefont {B.~P.}\ \bibnamefont
  {Abbott}} \emph {et~al.} (\bibinfo {collaboration} {LIGO Scientific
  Collaboration and Virgo Collaboration}),\ }\bibfield  {title} {\enquote
  {\bibinfo {title} {{Search for gravitational waves from Scorpius X-1 in the
  second Advanced LIGO observing run with an improved hidden Markov model}},}\
  }\href {\doibase 10.1103/PhysRevD.100.122002} {\bibfield  {journal} {\bibinfo
   {journal} {Phys. Rev. D}\ }\textbf {\bibinfo {volume} {100}},\ \bibinfo
  {pages} {122002} (\bibinfo {year} {2019}{\natexlab{b}})}\BibitemShut
  {NoStop}%
\bibitem [{\citenamefont {Millhouse}\ \emph {et~al.}(2020)\citenamefont
  {Millhouse}, \citenamefont {Strang},\ and\ \citenamefont
  {Melatos}}]{O2SNR-Viterbi}%
  \BibitemOpen
  \bibfield  {author} {\bibinfo {author} {\bibfnamefont {M.}~\bibnamefont
  {Millhouse}}, \bibinfo {author} {\bibfnamefont {L.}~\bibnamefont {Strang}}, \
  and\ \bibinfo {author} {\bibfnamefont {A.}~\bibnamefont {Melatos}},\
  }\bibfield  {title} {\enquote {\bibinfo {title} {{Search for gravitational
  waves from 12 young supernova remnants with a hidden Markov model in Advanced
  LIGO's second observing run}},}\ }\href {\doibase
  10.1103/PhysRevD.102.083025} {\bibfield  {journal} {\bibinfo  {journal}
  {Phys. Rev. D}\ }\textbf {\bibinfo {volume} {102}},\ \bibinfo {pages}
  {083025} (\bibinfo {year} {2020})}\BibitemShut {NoStop}%
\bibitem [{\citenamefont {Sun}\ \emph {et~al.}(2020)\citenamefont {Sun},
  \citenamefont {Brito},\ and\ \citenamefont {Isi}}]{Sun2020-CygX1}%
  \BibitemOpen
  \bibfield  {author} {\bibinfo {author} {\bibfnamefont {L.}~\bibnamefont
  {Sun}}, \bibinfo {author} {\bibfnamefont {R.}~\bibnamefont {Brito}}, \ and\
  \bibinfo {author} {\bibfnamefont {M.}~\bibnamefont {Isi}},\ }\bibfield
  {title} {\enquote {\bibinfo {title} {{Search for ultralight bosons in Cygnus
  X-1 with Advanced LIGO}},}\ }\href {\doibase 10.1103/PhysRevD.101.063020}
  {\bibfield  {journal} {\bibinfo  {journal} {Phys. Rev. D}\ }\textbf {\bibinfo
  {volume} {101}},\ \bibinfo {pages} {063020} (\bibinfo {year}
  {2020})}\BibitemShut {NoStop}%
\bibitem [{\citenamefont {Jones}\ and\ \citenamefont {Sun}(2021)}]{Jones_2021}%
  \BibitemOpen
  \bibfield  {author} {\bibinfo {author} {\bibfnamefont {D.}~\bibnamefont
  {Jones}}\ and\ \bibinfo {author} {\bibfnamefont {L.}~\bibnamefont {Sun}},\
  }\bibfield  {title} {\enquote {\bibinfo {title} {{Search for continuous
  gravitational waves from Fomalhaut b in the second Advanced LIGO observing
  run with a hidden Markov model}},}\ }\href
  {http://dx.doi.org/10.1103/PhysRevD.103.023020} {\bibfield  {journal}
  {\bibinfo  {journal} {Physical Review D}\ }\textbf {\bibinfo {volume} {103}}
  (\bibinfo {year} {2021})}\BibitemShut {NoStop}%
\bibitem [{\citenamefont {Beniwal}\ \emph {et~al.}(2021)\citenamefont
  {Beniwal}, \citenamefont {Clearwater}, \citenamefont {Dunn}, \citenamefont
  {Melatos},\ and\ \citenamefont {Ottaway}}]{Beniwal2021}%
  \BibitemOpen
  \bibfield  {author} {\bibinfo {author} {\bibfnamefont {D.}~\bibnamefont
  {Beniwal}}, \bibinfo {author} {\bibfnamefont {P.}~\bibnamefont {Clearwater}},
  \bibinfo {author} {\bibfnamefont {L.}~\bibnamefont {Dunn}}, \bibinfo {author}
  {\bibfnamefont {A.}~\bibnamefont {Melatos}}, \ and\ \bibinfo {author}
  {\bibfnamefont {D.}~\bibnamefont {Ottaway}},\ }\bibfield  {title} {\enquote
  {\bibinfo {title} {{Search for continuous gravitational waves from ten
  H.E.S.S. sources using a hidden Markov model}},}\ }\href {\doibase
  10.1103/PhysRevD.103.083009} {\bibfield  {journal} {\bibinfo  {journal}
  {Phys. Rev. D}\ }\textbf {\bibinfo {volume} {103}},\ \bibinfo {pages}
  {083009} (\bibinfo {year} {2021})}\BibitemShut {NoStop}%
\bibitem [{\citenamefont {Abbott}\ \emph
  {et~al.}(2021{\natexlab{c}})\citenamefont {Abbott} \emph {et~al.}}]{O3aSNR}%
  \BibitemOpen
  \bibfield  {author} {\bibinfo {author} {\bibfnamefont {R.}~\bibnamefont
  {Abbott}} \emph {et~al.},\ }\bibfield  {title} {\enquote {\bibinfo {title}
  {{Searches for continuous gravitational waves from young supernova remnants
  in the early third observing run of Advanced LIGO and Virgo}},}\ }\href
  {\doibase 10.3847/1538-4357/ac17ea} {\bibfield  {journal} {\bibinfo
  {journal} {The Astrophysical Journal}\ }\textbf {\bibinfo {volume} {921}},\
  \bibinfo {pages} {80} (\bibinfo {year} {2021}{\natexlab{c}})}\BibitemShut
  {NoStop}%
\bibitem [{\citenamefont {Abbott}\ \emph
  {et~al.}(2022{\natexlab{a}})\citenamefont {Abbott} \emph {et~al.}}]{O3amxp}%
  \BibitemOpen
  \bibfield  {author} {\bibinfo {author} {\bibfnamefont {R.}~\bibnamefont
  {Abbott}} \emph {et~al.} (\bibinfo {collaboration} {LIGO Scientific
  Collaboration, Virgo Collaboration, and KAGRA Collaboration}),\ }\bibfield
  {title} {\enquote {\bibinfo {title} {{Search for continuous gravitational
  waves from 20 accreting millisecond X-ray pulsars in O3 LIGO data}},}\ }\href
  {\doibase 10.1103/PhysRevD.105.022002} {\bibfield  {journal} {\bibinfo
  {journal} {Phys. Rev. D}\ }\textbf {\bibinfo {volume} {105}},\ \bibinfo
  {pages} {022002} (\bibinfo {year} {2022}{\natexlab{a}})}\BibitemShut
  {NoStop}%
\bibitem [{\citenamefont {Abbott}\ \emph
  {et~al.}(2022{\natexlab{b}})\citenamefont {Abbott} \emph {et~al.}}]{O3scoX1}%
  \BibitemOpen
  \bibfield  {author} {\bibinfo {author} {\bibfnamefont {R.}~\bibnamefont
  {Abbott}} \emph {et~al.},\ }\bibfield  {title} {\enquote {\bibinfo {title}
  {{Search for gravitational waves from Scorpius X-1 with a hidden Markov model
  in O3 LIGO data}},}\ }\href@noop {} {\  (\bibinfo {year}
  {2022}{\natexlab{b}})},\ \Eprint {http://arxiv.org/abs/2201.10104}
  {arXiv:2201.10104 [gr-qc]} \BibitemShut {NoStop}%
\bibitem [{\citenamefont {Jaranowski}\ \emph {et~al.}(1998)\citenamefont
  {Jaranowski}, \citenamefont {Kr{\'{o}}lak},\ and\ \citenamefont
  {Schutz}}]{Jaranowski1998}%
  \BibitemOpen
  \bibfield  {author} {\bibinfo {author} {\bibfnamefont {P.}~\bibnamefont
  {Jaranowski}}, \bibinfo {author} {\bibfnamefont {A.}~\bibnamefont
  {Kr{\'{o}}lak}}, \ and\ \bibinfo {author} {\bibfnamefont {B.~F.}\
  \bibnamefont {Schutz}},\ }\bibfield  {title} {\enquote {\bibinfo {title}
  {Data analysis of gravitational-wave signals from spinning neutron stars: The
  signal and its detection},}\ }\href {\doibase 10.1103/PhysRevD.58.063001}
  {\bibfield  {journal} {\bibinfo  {journal} {Physical Review D}\ }\textbf
  {\bibinfo {volume} {58}},\ \bibinfo {pages} {063001} (\bibinfo {year}
  {1998})}\BibitemShut {NoStop}%
\bibitem [{\citenamefont {Steltner}\ \emph {et~al.}(2021)\citenamefont
  {Steltner}, \citenamefont {Papa}, \citenamefont {Eggenstein}, \citenamefont
  {Allen}, \citenamefont {Dergachev}, \citenamefont {Prix}, \citenamefont
  {Machenschalk}, \citenamefont {Walsh}, \citenamefont {Zhu}, \citenamefont
  {Behnke},\ and\ \citenamefont {Kwang}}]{Steltner2021}%
  \BibitemOpen
  \bibfield  {author} {\bibinfo {author} {\bibfnamefont {B.}~\bibnamefont
  {Steltner}}, \bibinfo {author} {\bibfnamefont {M.~A.}\ \bibnamefont {Papa}},
  \bibinfo {author} {\bibfnamefont {H.-B.}\ \bibnamefont {Eggenstein}},
  \bibinfo {author} {\bibfnamefont {B.}~\bibnamefont {Allen}}, \bibinfo
  {author} {\bibfnamefont {V.}~\bibnamefont {Dergachev}}, \bibinfo {author}
  {\bibfnamefont {R.}~\bibnamefont {Prix}}, \bibinfo {author} {\bibfnamefont
  {B.}~\bibnamefont {Machenschalk}}, \bibinfo {author} {\bibfnamefont
  {S.}~\bibnamefont {Walsh}}, \bibinfo {author} {\bibfnamefont {S.~J.}\
  \bibnamefont {Zhu}}, \bibinfo {author} {\bibfnamefont {O.}~\bibnamefont
  {Behnke}}, \ and\ \bibinfo {author} {\bibfnamefont {S.}~\bibnamefont
  {Kwang}},\ }\bibfield  {title} {\enquote {\bibinfo {title} {{Einstein@Home
  all-sky search for continuous gravitational waves in LIGO O2 public data}},}\
  }\href {\doibase 10.3847/1538-4357/abc7c9} {\bibfield  {journal} {\bibinfo
  {journal} {The Astrophysical Journal}\ }\textbf {\bibinfo {volume} {909}},\
  \bibinfo {pages} {79} (\bibinfo {year} {2021})}\BibitemShut {NoStop}%
\bibitem [{\citenamefont {{The LIGO Scientific Collaboration}}\ \emph
  {et~al.}(2022)\citenamefont {{The LIGO Scientific Collaboration}},
  \citenamefont {{The Virgo Collaboration}}, \citenamefont {{The KAGRA
  Collaboration}}, \citenamefont {Abbott} \emph {et~al.}}]{O3all-sky}%
  \BibitemOpen
  \bibfield  {author} {\bibinfo {author} {\bibnamefont {{The LIGO Scientific
  Collaboration}}}, \bibinfo {author} {\bibnamefont {{The Virgo
  Collaboration}}}, \bibinfo {author} {\bibnamefont {{The KAGRA
  Collaboration}}}, \bibinfo {author} {\bibfnamefont {R.}~\bibnamefont
  {Abbott}},  \emph {et~al.},\ }\href {\doibase 10.48550/ARXIV.2201.00697}
  {\enquote {\bibinfo {title} {{All-sky search for continuous gravitational
  waves from isolated neutron stars using Advanced LIGO and Advanced Virgo O3
  data}},}\ } (\bibinfo {year} {2022})\BibitemShut {NoStop}%
\bibitem [{\citenamefont {Abbott}\ \emph
  {et~al.}(2022{\natexlab{c}})\citenamefont {Abbott} \emph
  {et~al.}}]{O3scalar-bosons}%
  \BibitemOpen
  \bibfield  {author} {\bibinfo {author} {\bibfnamefont {R.}~\bibnamefont
  {Abbott}} \emph {et~al.} (\bibinfo {collaboration} {The LIGO Scientific
  Collaboration, the Virgo Collaboration, and the KAGRA Collaboration}),\
  }\bibfield  {title} {\enquote {\bibinfo {title} {{All-sky search for
  gravitational wave emission from scalar boson clouds around spinning black
  holes in LIGO O3 data}},}\ }\href {\doibase 10.1103/PhysRevD.105.102001}
  {\bibfield  {journal} {\bibinfo  {journal} {Phys. Rev. D}\ }\textbf {\bibinfo
  {volume} {105}},\ \bibinfo {pages} {102001} (\bibinfo {year}
  {2022}{\natexlab{c}})}\BibitemShut {NoStop}%
\bibitem [{\citenamefont {Astone}\ \emph {et~al.}(2014)\citenamefont {Astone},
  \citenamefont {Colla}, \citenamefont {D'Antonio}, \citenamefont {Frasca},\
  and\ \citenamefont {Palomba}}]{Astone2014}%
  \BibitemOpen
  \bibfield  {author} {\bibinfo {author} {\bibfnamefont {Pia}\ \bibnamefont
  {Astone}}, \bibinfo {author} {\bibfnamefont {Alberto}\ \bibnamefont {Colla}},
  \bibinfo {author} {\bibfnamefont {Sabrina}\ \bibnamefont {D'Antonio}},
  \bibinfo {author} {\bibfnamefont {Sergio}\ \bibnamefont {Frasca}}, \ and\
  \bibinfo {author} {\bibfnamefont {Cristiano}\ \bibnamefont {Palomba}},\
  }\bibfield  {title} {\enquote {\bibinfo {title} {{Method for all-sky searches
  of continuous gravitational wave signals using the frequency-Hough
  transform}},}\ }\href {\doibase 10.1103/PhysRevD.90.042002} {\bibfield
  {journal} {\bibinfo  {journal} {Phys. Rev. D}\ }\textbf {\bibinfo {volume}
  {90}},\ \bibinfo {pages} {042002} (\bibinfo {year} {2014})}\BibitemShut
  {NoStop}%
\bibitem [{\citenamefont {Covas}\ \emph {et~al.}(2018)\citenamefont {Covas}
  \emph {et~al.}}]{Covas2018}%
  \BibitemOpen
  \bibfield  {author} {\bibinfo {author} {\bibfnamefont {P.~B.}\ \bibnamefont
  {Covas}} \emph {et~al.} (\bibinfo {collaboration} {LSC Instrument Authors}),\
  }\bibfield  {title} {\enquote {\bibinfo {title} {{Identification and
  mitigation of narrow spectral artifacts that degrade searches for persistent
  gravitational waves in the first two observing runs of Advanced LIGO}},}\
  }\href {\doibase 10.1103/PhysRevD.97.082002} {\bibfield  {journal} {\bibinfo
  {journal} {Phys. Rev. D}\ }\textbf {\bibinfo {volume} {97}},\ \bibinfo
  {pages} {082002} (\bibinfo {year} {2018})}\BibitemShut {NoStop}%
\bibitem [{\citenamefont {Davis}\ \emph {et~al.}(2021)\citenamefont {Davis}
  \emph {et~al.}}]{Davis2021}%
  \BibitemOpen
  \bibfield  {author} {\bibinfo {author} {\bibfnamefont {D.}~\bibnamefont
  {Davis}} \emph {et~al.},\ }\bibfield  {title} {\enquote {\bibinfo {title}
  {{LIGO detector characterization in the second and third observing runs}},}\
  }\href {\doibase 10.1088/1361-6382/abfd85} {\bibfield  {journal} {\bibinfo
  {journal} {Classical and Quantum Gravity}\ }\textbf {\bibinfo {volume}
  {38}},\ \bibinfo {pages} {135014} (\bibinfo {year} {2021})}\BibitemShut
  {NoStop}%
\bibitem [{\citenamefont {Keitel}\ \emph {et~al.}(2014)\citenamefont {Keitel},
  \citenamefont {Prix}, \citenamefont {Papa}, \citenamefont {Leaci},\ and\
  \citenamefont {Siddiqi}}]{Keitel2014}%
  \BibitemOpen
  \bibfield  {author} {\bibinfo {author} {\bibfnamefont {David}\ \bibnamefont
  {Keitel}}, \bibinfo {author} {\bibfnamefont {Reinhard}\ \bibnamefont {Prix}},
  \bibinfo {author} {\bibfnamefont {Maria~Alessandra}\ \bibnamefont {Papa}},
  \bibinfo {author} {\bibfnamefont {Paola}\ \bibnamefont {Leaci}}, \ and\
  \bibinfo {author} {\bibfnamefont {Maham}\ \bibnamefont {Siddiqi}},\
  }\bibfield  {title} {\enquote {\bibinfo {title} {Search for continuous
  gravitational waves: Improving robustness versus instrumental artifacts},}\
  }\href {\doibase 10.1103/PhysRevD.89.064023} {\bibfield  {journal} {\bibinfo
  {journal} {Phys. Rev. D}\ }\textbf {\bibinfo {volume} {89}},\ \bibinfo
  {pages} {064023} (\bibinfo {year} {2014})}\BibitemShut {NoStop}%
\bibitem [{\citenamefont {Leaci}(2015)}]{Leaci2015}%
  \BibitemOpen
  \bibfield  {author} {\bibinfo {author} {\bibfnamefont {Paola}\ \bibnamefont
  {Leaci}},\ }\bibfield  {title} {\enquote {\bibinfo {title} {Methods to filter
  out spurious disturbances in continuous-wave searches from gravitational-wave
  detectors},}\ }\href {\doibase 10.1088/0031-8949/90/12/125001} {\bibfield
  {journal} {\bibinfo  {journal} {Physica Scripta}\ }\textbf {\bibinfo {volume}
  {90}},\ \bibinfo {pages} {125001} (\bibinfo {year} {2015})}\BibitemShut
  {NoStop}%
\bibitem [{\citenamefont {Zhu}\ \emph {et~al.}(2017)\citenamefont {Zhu},
  \citenamefont {Papa},\ and\ \citenamefont {Walsh}}]{Zhu_2017}%
  \BibitemOpen
  \bibfield  {author} {\bibinfo {author} {\bibfnamefont {S.~J.}\ \bibnamefont
  {Zhu}}, \bibinfo {author} {\bibfnamefont {M.~A.}\ \bibnamefont {Papa}}, \
  and\ \bibinfo {author} {\bibfnamefont {S.}~\bibnamefont {Walsh}},\ }\bibfield
   {title} {\enquote {\bibinfo {title} {New veto for continuous gravitational
  wave searches},}\ }\href {\doibase 10.1103/physrevd.96.124007} {\bibfield
  {journal} {\bibinfo  {journal} {Physical Review D}\ }\textbf {\bibinfo
  {volume} {96}} (\bibinfo {year} {2017}),\
  10.1103/physrevd.96.124007}\BibitemShut {NoStop}%
\bibitem [{\citenamefont {Isi}\ \emph {et~al.}(2020)\citenamefont {Isi},
  \citenamefont {Mastrogiovanni}, \citenamefont {Pitkin},\ and\ \citenamefont
  {Piccinni}}]{Isi2020}%
  \BibitemOpen
  \bibfield  {author} {\bibinfo {author} {\bibfnamefont {Maximiliano}\
  \bibnamefont {Isi}}, \bibinfo {author} {\bibfnamefont {Simone}\ \bibnamefont
  {Mastrogiovanni}}, \bibinfo {author} {\bibfnamefont {Matthew}\ \bibnamefont
  {Pitkin}}, \ and\ \bibinfo {author} {\bibfnamefont {Ornella~Juliana}\
  \bibnamefont {Piccinni}},\ }\bibfield  {title} {\enquote {\bibinfo {title}
  {Establishing the significance of continuous gravitational-wave detections
  from known pulsars},}\ }\href {\doibase 10.1103/PhysRevD.102.123027}
  {\bibfield  {journal} {\bibinfo  {journal} {Phys. Rev. D}\ }\textbf {\bibinfo
  {volume} {102}},\ \bibinfo {pages} {123027} (\bibinfo {year}
  {2020})}\BibitemShut {NoStop}%
\bibitem [{\citenamefont {Brady}\ and\ \citenamefont
  {Creighton}(2000)}]{Brady2000}%
  \BibitemOpen
  \bibfield  {author} {\bibinfo {author} {\bibfnamefont {P.~R.}\ \bibnamefont
  {Brady}}\ and\ \bibinfo {author} {\bibfnamefont {T.}~\bibnamefont
  {Creighton}},\ }\bibfield  {title} {\enquote {\bibinfo {title} {{Searching
  for periodic sources with LIGO. II. Hierarchical searches}},}\ }\href
  {\doibase 10.1103/PhysRevD.61.082001} {\bibfield  {journal} {\bibinfo
  {journal} {Phys. Rev. D}\ }\textbf {\bibinfo {volume} {61}},\ \bibinfo
  {pages} {082001} (\bibinfo {year} {2000})}\BibitemShut {NoStop}%
\bibitem [{\citenamefont {Prix}\ and\ \citenamefont {Itoh}(2005)}]{Prix_2005}%
  \BibitemOpen
  \bibfield  {author} {\bibinfo {author} {\bibfnamefont {R.}~\bibnamefont
  {Prix}}\ and\ \bibinfo {author} {\bibfnamefont {Y.}~\bibnamefont {Itoh}},\
  }\bibfield  {title} {\enquote {\bibinfo {title} {Global parameter-space
  correlations of coherent searches for continuous gravitational waves},}\
  }\href {\doibase 10.1088/0264-9381/22/18/s14} {\bibfield  {journal} {\bibinfo
   {journal} {Classical and Quantum Gravity}\ }\textbf {\bibinfo {volume}
  {22}},\ \bibinfo {pages} {S1003–S1012} (\bibinfo {year}
  {2005})}\BibitemShut {NoStop}%
\bibitem [{\citenamefont {Intini}\ \emph {et~al.}(2020)\citenamefont {Intini},
  \citenamefont {Leaci}, \citenamefont {Astone}, \citenamefont {Antonio},
  \citenamefont {Frasca}, \citenamefont {Rosa}, \citenamefont {Miller},
  \citenamefont {Palomba},\ and\ \citenamefont {Piccinni}}]{Intini_2020}%
  \BibitemOpen
  \bibfield  {author} {\bibinfo {author} {\bibfnamefont {G}~\bibnamefont
  {Intini}}, \bibinfo {author} {\bibfnamefont {P}~\bibnamefont {Leaci}},
  \bibinfo {author} {\bibfnamefont {P}~\bibnamefont {Astone}}, \bibinfo
  {author} {\bibfnamefont {S~D'}\ \bibnamefont {Antonio}}, \bibinfo {author}
  {\bibfnamefont {S}~\bibnamefont {Frasca}}, \bibinfo {author} {\bibfnamefont
  {I~La}\ \bibnamefont {Rosa}}, \bibinfo {author} {\bibfnamefont
  {A}~\bibnamefont {Miller}}, \bibinfo {author} {\bibfnamefont {C}~\bibnamefont
  {Palomba}}, \ and\ \bibinfo {author} {\bibfnamefont {O}~\bibnamefont
  {Piccinni}},\ }\bibfield  {title} {\enquote {\bibinfo {title} {{A
  Doppler-modulation based veto to discard false continuous gravitational-wave
  candidates}},}\ }\href {\doibase 10.1088/1361-6382/abac43} {\bibfield
  {journal} {\bibinfo  {journal} {Classical and Quantum Gravity}\ }\textbf
  {\bibinfo {volume} {37}},\ \bibinfo {pages} {225007} (\bibinfo {year}
  {2020})}\BibitemShut {NoStop}%
\bibitem [{\citenamefont {Prix}(2007)}]{Prix2007}%
  \BibitemOpen
  \bibfield  {author} {\bibinfo {author} {\bibfnamefont {R.}~\bibnamefont
  {Prix}},\ }\bibfield  {title} {\enquote {\bibinfo {title} {{Search for
  continuous gravitational waves: Metric of the multidetector F-statistic}},}\
  }\href {\doibase 10.1103/PhysRevD.75.023004} {\bibfield  {journal} {\bibinfo
  {journal} {Physical Review D}\ }\textbf {\bibinfo {volume} {75}},\ \bibinfo
  {pages} {023004} (\bibinfo {year} {2007})}\BibitemShut {NoStop}%
\bibitem [{\citenamefont {Cutler}\ and\ \citenamefont
  {Schutz}(2005)}]{Cutler_2005}%
  \BibitemOpen
  \bibfield  {author} {\bibinfo {author} {\bibfnamefont {C.}~\bibnamefont
  {Cutler}}\ and\ \bibinfo {author} {\bibfnamefont {B.~F.}\ \bibnamefont
  {Schutz}},\ }\bibfield  {title} {\enquote {\bibinfo {title} {{Generalized
  F-statistic: Multiple detectors and multiple gravitational wave pulsars}},}\
  }\href {\doibase 10.1103/physrevd.72.063006} {\bibfield  {journal} {\bibinfo
  {journal} {Physical Review D}\ }\textbf {\bibinfo {volume} {72}} (\bibinfo
  {year} {2005}),\ 10.1103/physrevd.72.063006}\BibitemShut {NoStop}%
\bibitem [{\citenamefont {Quinn}\ and\ \citenamefont
  {Hannan}(2001)}]{Quinn2001}%
  \BibitemOpen
  \bibfield  {author} {\bibinfo {author} {\bibfnamefont {B.~G.}\ \bibnamefont
  {Quinn}}\ and\ \bibinfo {author} {\bibfnamefont {E.~J.}\ \bibnamefont
  {Hannan}},\ }\href@noop {} {\emph {\bibinfo {title} {The Estimation and
  Tracking of Frequency}}}\ (\bibinfo  {publisher} {Cambridge University
  Press},\ \bibinfo {year} {2001})\ p.\ \bibinfo {pages} {266}\BibitemShut
  {NoStop}%
\bibitem [{\citenamefont {Viterbi}(1967)}]{Viterbi1967}%
  \BibitemOpen
  \bibfield  {author} {\bibinfo {author} {\bibfnamefont {A.}~\bibnamefont
  {Viterbi}},\ }\bibfield  {title} {\enquote {\bibinfo {title} {Error bounds
  for convolutional codes and an asymptotically optimum decoding algorithm},}\
  }\href {\doibase 10.1109/TIT.1967.1054010} {\bibfield  {journal} {\bibinfo
  {journal} {IEEE Transactions on Information Theory}\ }\textbf {\bibinfo
  {volume} {13}},\ \bibinfo {pages} {260--269} (\bibinfo {year}
  {1967})}\BibitemShut {NoStop}%
\bibitem [{\citenamefont {Middleton}\ \emph {et~al.}(2020)\citenamefont
  {Middleton}, \citenamefont {Clearwater}, \citenamefont {Melatos},\ and\
  \citenamefont {Dunn}}]{Middleton2020}%
  \BibitemOpen
  \bibfield  {author} {\bibinfo {author} {\bibfnamefont {H.}~\bibnamefont
  {Middleton}}, \bibinfo {author} {\bibfnamefont {P.}~\bibnamefont
  {Clearwater}}, \bibinfo {author} {\bibfnamefont {A.}~\bibnamefont {Melatos}},
  \ and\ \bibinfo {author} {\bibfnamefont {L.}~\bibnamefont {Dunn}},\
  }\bibfield  {title} {\enquote {\bibinfo {title} {{Search for gravitational
  waves from five low mass x-ray binaries in the second Advanced LIGO observing
  run with an improved hidden Markov model}},}\ }\href {\doibase
  10.1103/PhysRevD.102.023006} {\bibfield  {journal} {\bibinfo  {journal}
  {Phys. Rev. D}\ }\textbf {\bibinfo {volume} {102}},\ \bibinfo {pages}
  {023006} (\bibinfo {year} {2020})}\BibitemShut {NoStop}%
\bibitem [{Note1()}]{Note1}%
  \BibitemOpen
  \bibinfo {note} {\protect \leavevmode {\color {black}We use the software
  package LALSuite for all the simulations in this paper~\cite
  {lalsuite}.}}\BibitemShut {Stop}%
\bibitem [{\citenamefont {Abbott}\ \emph
  {et~al.}(2021{\natexlab{d}})\citenamefont {Abbott} \emph {et~al.}}]{O3aINS}%
  \BibitemOpen
  \bibfield  {author} {\bibinfo {author} {\bibfnamefont {R.}~\bibnamefont
  {Abbott}} \emph {et~al.} (\bibinfo {collaboration} {LIGO Scientific
  Collaboration, Virgo Collaboration, and KAGRA Collaboration}),\ }\bibfield
  {title} {\enquote {\bibinfo {title} {{All-sky search for continuous
  gravitational waves from isolated neutron stars in the early O3 LIGO
  data}},}\ }\href {\doibase 10.1103/PhysRevD.104.082004} {\bibfield  {journal}
  {\bibinfo  {journal} {Phys. Rev. D}\ }\textbf {\bibinfo {volume} {104}},\
  \bibinfo {pages} {082004} (\bibinfo {year} {2021}{\natexlab{d}})}\BibitemShut
  {NoStop}%
\bibitem [{\citenamefont {Abbott}\ \emph
  {et~al.}(2021{\natexlab{e}})\citenamefont {Abbott} \emph {et~al.}}]{O3aBNS}%
  \BibitemOpen
  \bibfield  {author} {\bibinfo {author} {\bibfnamefont {R.}~\bibnamefont
  {Abbott}} \emph {et~al.} (\bibinfo {collaboration} {The LIGO Scientific
  Collaboration and the Virgo Collaboration}),\ }\bibfield  {title} {\enquote
  {\bibinfo {title} {{All-sky search in early O3 LIGO data for continuous
  gravitational-wave signals from unknown neutron stars in binary systems}},}\
  }\href {\doibase 10.1103/PhysRevD.103.064017} {\bibfield  {journal} {\bibinfo
   {journal} {Phys. Rev. D}\ }\textbf {\bibinfo {volume} {103}},\ \bibinfo
  {pages} {064017} (\bibinfo {year} {2021}{\natexlab{e}})}\BibitemShut
  {NoStop}%
\bibitem [{\citenamefont {Abbott}\ \emph
  {et~al.}(2022{\natexlab{d}})\citenamefont {Abbott} \emph
  {et~al.}}]{O3aCassVela}%
  \BibitemOpen
  \bibfield  {author} {\bibinfo {author} {\bibfnamefont {R.}~\bibnamefont
  {Abbott}} \emph {et~al.} (\bibinfo {collaboration} {LIGO Scientific
  Collaboration and Virgo Collaboration}),\ }\bibfield  {title} {\enquote
  {\bibinfo {title} {{Search of the early O3 LIGO data for continuous
  gravitational waves from the Cassiopeia A and Vela Jr. supernova
  remnants}},}\ }\href {\doibase 10.1103/PhysRevD.105.082005} {\bibfield
  {journal} {\bibinfo  {journal} {Phys. Rev. D}\ }\textbf {\bibinfo {volume}
  {105}},\ \bibinfo {pages} {082005} (\bibinfo {year}
  {2022}{\natexlab{d}})}\BibitemShut {NoStop}%
\bibitem [{\citenamefont {Mastrogiovanni}\ \emph {et~al.}(2018)\citenamefont
  {Mastrogiovanni}, \citenamefont {Astone}, \citenamefont {Antonio},
  \citenamefont {Frasca}, \citenamefont {Intini}, \citenamefont {La~Rosa},
  \citenamefont {Leaci}, \citenamefont {Miller}, \citenamefont {Muciaccia},
  \citenamefont {Palomba}, \citenamefont {Piccinni},\ and\ \citenamefont
  {Singhal}}]{Mastrogiovanni_2018}%
  \BibitemOpen
  \bibfield  {author} {\bibinfo {author} {\bibfnamefont {S.}~\bibnamefont
  {Mastrogiovanni}}, \bibinfo {author} {\bibfnamefont {P.}~\bibnamefont
  {Astone}}, \bibinfo {author} {\bibfnamefont {S.~D.}\ \bibnamefont {Antonio}},
  \bibinfo {author} {\bibfnamefont {S.}~\bibnamefont {Frasca}}, \bibinfo
  {author} {\bibfnamefont {G.}~\bibnamefont {Intini}}, \bibinfo {author}
  {\bibfnamefont {I.}~\bibnamefont {La~Rosa}}, \bibinfo {author} {\bibfnamefont
  {P.}~\bibnamefont {Leaci}}, \bibinfo {author} {\bibfnamefont
  {A.}~\bibnamefont {Miller}}, \bibinfo {author} {\bibfnamefont
  {F.}~\bibnamefont {Muciaccia}}, \bibinfo {author} {\bibfnamefont
  {C.}~\bibnamefont {Palomba}}, \bibinfo {author} {\bibfnamefont {O.~J.}\
  \bibnamefont {Piccinni}}, \ and\ \bibinfo {author} {\bibfnamefont
  {A.}~\bibnamefont {Singhal}},\ }\bibfield  {title} {\enquote {\bibinfo
  {title} {Phase decomposition of the template metric for continuous
  gravitational-wave searches},}\ }\href {\doibase 10.1103/PhysRevD.98.102003}
  {\bibfield  {journal} {\bibinfo  {journal} {Phys. Rev. D}\ }\textbf {\bibinfo
  {volume} {98}},\ \bibinfo {pages} {102003} (\bibinfo {year}
  {2018})}\BibitemShut {NoStop}%
\bibitem [{\citenamefont {Isi}\ \emph {et~al.}(2019)\citenamefont {Isi},
  \citenamefont {Sun}, \citenamefont {Brito},\ and\ \citenamefont
  {Melatos}}]{Isi_2019}%
  \BibitemOpen
  \bibfield  {author} {\bibinfo {author} {\bibfnamefont {M.}~\bibnamefont
  {Isi}}, \bibinfo {author} {\bibfnamefont {L.}~\bibnamefont {Sun}}, \bibinfo
  {author} {\bibfnamefont {R.}~\bibnamefont {Brito}}, \ and\ \bibinfo {author}
  {\bibfnamefont {A.}~\bibnamefont {Melatos}},\ }\bibfield  {title} {\enquote
  {\bibinfo {title} {Directed searches for gravitational waves from ultralight
  bosons},}\ }\href {\doibase 10.1103/physrevd.99.084042} {\bibfield  {journal}
  {\bibinfo  {journal} {Physical Review D}\ }\textbf {\bibinfo {volume} {99}}
  (\bibinfo {year} {2019}),\ 10.1103/physrevd.99.084042}\BibitemShut {NoStop}%
\bibitem [{\citenamefont {R.}\ \emph {et~al.}(2021)\citenamefont {R.} \emph
  {et~al.}}]{Abbott-O2}%
  \BibitemOpen
  \bibfield  {author} {\bibinfo {author} {\bibfnamefont {Abbott}\ \bibnamefont
  {R.}} \emph {et~al.},\ }\bibfield  {title} {\enquote {\bibinfo {title} {Open
  data from the first and second observing runs of advanced ligo and advanced
  virgo},}\ }\href {\doibase https://doi.org/10.1016/j.softx.2021.100658}
  {\bibfield  {journal} {\bibinfo  {journal} {SoftwareX}\ }\textbf {\bibinfo
  {volume} {13}},\ \bibinfo {pages} {100658} (\bibinfo {year}
  {2021})}\BibitemShut {NoStop}%
\bibitem [{\citenamefont {Biwer}\ \emph {et~al.}(2017)\citenamefont {Biwer},
  \citenamefont {Barker}, \citenamefont {Batch}, \citenamefont {Betzwieser},
  \citenamefont {Fisher}, \citenamefont {Goetz}, \citenamefont {Kandhasamy},
  \citenamefont {Karki}, \citenamefont {Kissel}, \citenamefont {Lundgren},
  \citenamefont {Macleod}, \citenamefont {Mullavey}, \citenamefont {Riles},
  \citenamefont {Rollins}, \citenamefont {Thorne}, \citenamefont {Thrane},
  \citenamefont {Abbott}, \citenamefont {Allen}, \citenamefont {Brown},
  \citenamefont {Charlton}, \citenamefont {Crowder}, \citenamefont {Fritschel},
  \citenamefont {Kanner}, \citenamefont {Landry}, \citenamefont {Lazzaro},
  \citenamefont {Millhouse}, \citenamefont {Pitkin}, \citenamefont {Savage},
  \citenamefont {Shawhan}, \citenamefont {Shoemaker}, \citenamefont {Smith},
  \citenamefont {Sun}, \citenamefont {Veitch}, \citenamefont {Vitale},
  \citenamefont {Weinstein}, \citenamefont {Cornish}, \citenamefont {Essick},
  \citenamefont {Fays}, \citenamefont {Katsavounidis}, \citenamefont {Lange},
  \citenamefont {Littenberg}, \citenamefont {Lynch}, \citenamefont {Meyers},
  \citenamefont {Pannarale}, \citenamefont {Prix}, \citenamefont
  {O'Shaughnessy},\ and\ \citenamefont {Sigg}}]{Biwer2017}%
  \BibitemOpen
  \bibfield  {author} {\bibinfo {author} {\bibfnamefont {C.}~\bibnamefont
  {Biwer}}, \bibinfo {author} {\bibfnamefont {D.}~\bibnamefont {Barker}},
  \bibinfo {author} {\bibfnamefont {J.~C.}\ \bibnamefont {Batch}}, \bibinfo
  {author} {\bibfnamefont {J.}~\bibnamefont {Betzwieser}}, \bibinfo {author}
  {\bibfnamefont {R.~P.}\ \bibnamefont {Fisher}}, \bibinfo {author}
  {\bibfnamefont {E.}~\bibnamefont {Goetz}}, \bibinfo {author} {\bibfnamefont
  {S.}~\bibnamefont {Kandhasamy}}, \bibinfo {author} {\bibfnamefont
  {S.}~\bibnamefont {Karki}}, \bibinfo {author} {\bibfnamefont {J.~S.}\
  \bibnamefont {Kissel}}, \bibinfo {author} {\bibfnamefont {A.~P.}\
  \bibnamefont {Lundgren}}, \bibinfo {author} {\bibfnamefont {D.~M.}\
  \bibnamefont {Macleod}}, \bibinfo {author} {\bibfnamefont {A.}~\bibnamefont
  {Mullavey}}, \bibinfo {author} {\bibfnamefont {K.}~\bibnamefont {Riles}},
  \bibinfo {author} {\bibfnamefont {J.~G.}\ \bibnamefont {Rollins}}, \bibinfo
  {author} {\bibfnamefont {K.~A.}\ \bibnamefont {Thorne}}, \bibinfo {author}
  {\bibfnamefont {E.}~\bibnamefont {Thrane}}, \bibinfo {author} {\bibfnamefont
  {T.~D.}\ \bibnamefont {Abbott}}, \bibinfo {author} {\bibfnamefont
  {B.}~\bibnamefont {Allen}}, \bibinfo {author} {\bibfnamefont {D.~A.}\
  \bibnamefont {Brown}}, \bibinfo {author} {\bibfnamefont {P.}~\bibnamefont
  {Charlton}}, \bibinfo {author} {\bibfnamefont {S.~G.}\ \bibnamefont
  {Crowder}}, \bibinfo {author} {\bibfnamefont {P.}~\bibnamefont {Fritschel}},
  \bibinfo {author} {\bibfnamefont {J.~B.}\ \bibnamefont {Kanner}}, \bibinfo
  {author} {\bibfnamefont {M.}~\bibnamefont {Landry}}, \bibinfo {author}
  {\bibfnamefont {C.}~\bibnamefont {Lazzaro}}, \bibinfo {author} {\bibfnamefont
  {M.}~\bibnamefont {Millhouse}}, \bibinfo {author} {\bibfnamefont
  {M.}~\bibnamefont {Pitkin}}, \bibinfo {author} {\bibfnamefont {R.~L.}\
  \bibnamefont {Savage}}, \bibinfo {author} {\bibfnamefont {P.}~\bibnamefont
  {Shawhan}}, \bibinfo {author} {\bibfnamefont {D.~H.}\ \bibnamefont
  {Shoemaker}}, \bibinfo {author} {\bibfnamefont {J.~R.}\ \bibnamefont
  {Smith}}, \bibinfo {author} {\bibfnamefont {L.}~\bibnamefont {Sun}}, \bibinfo
  {author} {\bibfnamefont {J.}~\bibnamefont {Veitch}}, \bibinfo {author}
  {\bibfnamefont {S.}~\bibnamefont {Vitale}}, \bibinfo {author} {\bibfnamefont
  {A.~J.}\ \bibnamefont {Weinstein}}, \bibinfo {author} {\bibfnamefont
  {N.}~\bibnamefont {Cornish}}, \bibinfo {author} {\bibfnamefont {R.~C.}\
  \bibnamefont {Essick}}, \bibinfo {author} {\bibfnamefont {M.}~\bibnamefont
  {Fays}}, \bibinfo {author} {\bibfnamefont {E.}~\bibnamefont {Katsavounidis}},
  \bibinfo {author} {\bibfnamefont {J.}~\bibnamefont {Lange}}, \bibinfo
  {author} {\bibfnamefont {T.~B.}\ \bibnamefont {Littenberg}}, \bibinfo
  {author} {\bibfnamefont {R.}~\bibnamefont {Lynch}}, \bibinfo {author}
  {\bibfnamefont {P.~M.}\ \bibnamefont {Meyers}}, \bibinfo {author}
  {\bibfnamefont {F.}~\bibnamefont {Pannarale}}, \bibinfo {author}
  {\bibfnamefont {R.}~\bibnamefont {Prix}}, \bibinfo {author} {\bibfnamefont
  {R.}~\bibnamefont {O'Shaughnessy}}, \ and\ \bibinfo {author} {\bibfnamefont
  {D.}~\bibnamefont {Sigg}},\ }\bibfield  {title} {\enquote {\bibinfo {title}
  {{Validating gravitational-wave detections: The Advanced LIGO hardware
  injection system}},}\ }\href {\doibase 10.1103/PhysRevD.95.062002} {\bibfield
   {journal} {\bibinfo  {journal} {Phys. Rev. D}\ }\textbf {\bibinfo {volume}
  {95}},\ \bibinfo {pages} {062002} (\bibinfo {year} {2017})}\BibitemShut
  {NoStop}%
\bibitem [{\citenamefont {Tenorio}\ \emph
  {et~al.}(2021{\natexlab{b}})\citenamefont {Tenorio}, \citenamefont {Keitel},\
  and\ \citenamefont {Sintes}}]{Tenorio2021_2}%
  \BibitemOpen
  \bibfield  {author} {\bibinfo {author} {\bibfnamefont {R.}~\bibnamefont
  {Tenorio}}, \bibinfo {author} {\bibfnamefont {D.}~\bibnamefont {Keitel}}, \
  and\ \bibinfo {author} {\bibfnamefont {A.~M.}\ \bibnamefont {Sintes}},\
  }\bibfield  {title} {\enquote {\bibinfo {title} {Time-frequency track
  distance for comparing continuous gravitational wave signals},}\ }\href
  {\doibase 10.1103/PhysRevD.103.064053} {\bibfield  {journal} {\bibinfo
  {journal} {Phys. Rev. D}\ }\textbf {\bibinfo {volume} {103}},\ \bibinfo
  {pages} {064053} (\bibinfo {year} {2021}{\natexlab{b}})}\BibitemShut
  {NoStop}%
\bibitem [{\citenamefont {Dhurandhar}\ \emph {et~al.}(2008)\citenamefont
  {Dhurandhar}, \citenamefont {Krishnan}, \citenamefont {Mukhopadhyay},\ and\
  \citenamefont {Whelan}}]{Dhurandhar2008}%
  \BibitemOpen
  \bibfield  {author} {\bibinfo {author} {\bibfnamefont {Sanjeev}\ \bibnamefont
  {Dhurandhar}}, \bibinfo {author} {\bibfnamefont {Badri}\ \bibnamefont
  {Krishnan}}, \bibinfo {author} {\bibfnamefont {Himan}\ \bibnamefont
  {Mukhopadhyay}}, \ and\ \bibinfo {author} {\bibfnamefont {John~T.}\
  \bibnamefont {Whelan}},\ }\bibfield  {title} {\enquote {\bibinfo {title}
  {Cross-correlation search for periodic gravitational waves},}\ }\href
  {\doibase 10.1103/PhysRevD.77.082001} {\bibfield  {journal} {\bibinfo
  {journal} {Phys. Rev. D}\ }\textbf {\bibinfo {volume} {77}},\ \bibinfo
  {pages} {082001} (\bibinfo {year} {2008})}\BibitemShut {NoStop}%
\bibitem [{\citenamefont {Chung}\ \emph {et~al.}(2011)\citenamefont {Chung},
  \citenamefont {Melatos}, \citenamefont {Krishnan},\ and\ \citenamefont
  {Whelan}}]{Chung2011}%
  \BibitemOpen
  \bibfield  {author} {\bibinfo {author} {\bibfnamefont {C.~T.~Y.}\
  \bibnamefont {Chung}}, \bibinfo {author} {\bibfnamefont {A.}~\bibnamefont
  {Melatos}}, \bibinfo {author} {\bibfnamefont {B.}~\bibnamefont {Krishnan}}, \
  and\ \bibinfo {author} {\bibfnamefont {J.~T.}\ \bibnamefont {Whelan}},\
  }\bibfield  {title} {\enquote {\bibinfo {title} {{Designing a
  cross-correlation search for continuous-wave gravitational radiation from a
  neutron star in the supernova remnant SNR 1987A*}},}\ }\href {\doibase
  10.1111/j.1365-2966.2011.18585.x} {\bibfield  {journal} {\bibinfo  {journal}
  {Monthly Notices of the Royal Astronomical Society}\ }\textbf {\bibinfo
  {volume} {414}},\ \bibinfo {pages} {2650--2663} (\bibinfo {year}
  {2011})}\BibitemShut {NoStop}%
\bibitem [{\citenamefont {Sun}\ \emph {et~al.}(2016)\citenamefont {Sun},
  \citenamefont {Melatos}, \citenamefont {Lasky}, \citenamefont {Chung},\ and\
  \citenamefont {Darman}}]{Crosscorr}%
  \BibitemOpen
  \bibfield  {author} {\bibinfo {author} {\bibfnamefont {L.}~\bibnamefont
  {Sun}}, \bibinfo {author} {\bibfnamefont {A.}~\bibnamefont {Melatos}},
  \bibinfo {author} {\bibfnamefont {P.~D.}\ \bibnamefont {Lasky}}, \bibinfo
  {author} {\bibfnamefont {C.~T.~Y.}\ \bibnamefont {Chung}}, \ and\ \bibinfo
  {author} {\bibfnamefont {N.~S.}\ \bibnamefont {Darman}},\ }\bibfield  {title}
  {\enquote {\bibinfo {title} {{Cross-correlation search for continuous
  gravitational waves from a compact object in SNR 1987A in LIGO Science run
  5}},}\ }\href {\doibase 10.1103/PhysRevD.94.082004} {\bibfield  {journal}
  {\bibinfo  {journal} {Phys. Rev. D}\ }\textbf {\bibinfo {volume} {94}},\
  \bibinfo {pages} {082004} (\bibinfo {year} {2016})}\BibitemShut {NoStop}%
\bibitem [{\citenamefont {{LIGO Scientific Collaboration}}(2018)}]{lalsuite}%
  \BibitemOpen
  \bibfield  {author} {\bibinfo {author} {\bibnamefont {{LIGO Scientific
  Collaboration}}},\ }\href {\doibase 10.7935/GT1W-FZ16} {\enquote {\bibinfo
  {title} {{LIGO Algorithm Library - LALSuite}},}\ }\bibinfo {howpublished}
  {free software (GPL)} (\bibinfo {year} {2018})\BibitemShut {NoStop}%
\end{thebibliography}%

\end{document}